\documentclass[aps,preprintnumbers,floatfix,article,amsmath,amssymb,floatfix,10pt,prd,superscriptaddress,nofootinbib,natbib]{revtex4-2}
\usepackage{bm}
\usepackage{amsfonts}
\usepackage{latexsym}
\usepackage[latin1]{inputenc}
\usepackage{graphicx}
\usepackage{amsmath}
\usepackage{palatino}
\usepackage{mathpazo}
\usepackage{textcomp}
\usepackage{float}
\usepackage{booktabs}
\usepackage{dcolumn}
\usepackage{ragged2e}
\usepackage{hyperref}
\hypersetup{colorlinks,citecolor=purple}
\hypersetup{colorlinks=true,linkcolor=red,filecolor=magenta,    urlcolor=blue}
\usepackage{amsmath}
\usepackage{xcolor}
\usepackage{orcidlink}
\usepackage{epsfig}
\usepackage{caption}
\usepackage{subcaption}
\usepackage{commath}
\usepackage{color,soul}
\usepackage{titlesec}

\titleformat{\paragraph}
{\normalfont\normalsize\bfseries}{\theparagraph}{1em}{}
\titlespacing*{\paragraph}
{0pt}{3.25ex plus 1ex minus .2ex}{1.5ex plus .2ex}

\def\jnl@style{\it}
\def\aaref@jnl#1{{\jnl@style#1}}

\def\aaref@jnl#1{{\jnl@style#1}}

\def\aj{\aaref@jnl{AJ}}                   % Astronomical Journal
\def\apj{\aaref@jnl{ApJ}}                 % Astrophysical Journal
\def\apjl{\aaref@jnl{ApJ}}                % Astrophysical Journal, Letters
\def\apjs{\aaref@jnl{ApJS}}               % Astrophysical Journal, Supplement
\def\apss{\aaref@jnl{Ap\&SS}}             % Astrophysics and Space Science
\def\aap{\aaref@jnl{A\&A}}                % Astronomy and Astrophysics
\def\aapr{\aaref@jnl{A\&A~Rev.}}          % Astronomy and Astrophysics Reviews
\def\aaps{\aaref@jnl{A\&AS}}              % Astronomy and Astrophysics, Supplement
\def\mnras{\aaref@jnl{Mon.~Not.~Roy.~Astron.~Soc.}}             % Monthly Notices of the RAS
\def\prd{\aaref@jnl{Phys.~Rev.~D}}        % Physical Review D
\def\prc{\aaref@jnl{Phys.~Rev.~C}}  % Physical Review C
\def\prl{\aaref@jnl{Phys.~Rev.~Lett.}}    % Physical Review Letters
\def\qjras{\aaref@jnl{QJRAS}}             % Quarterly Journal of the RAS
\def\skytel{\aaref@jnl{S\&T}}             % Sky and Telescope
\def\ssr{\aaref@jnl{Space~Sci.~Rev.}}     % Space Science Reviews
\def\zap{\aaref@jnl{ZAp}}                 % Zeitschrift fuer Astrophysik
\def\nat{\aaref@jnl{Nature}}              % Nature
\def\aplett{\aaref@jnl{Astrophys.~Lett.}} % Astrophysics Letters
\def\apspr{\aaref@jnl{Astrophys.~Space~Phys.~Res.}} % Astrophysics Space Physics Research
\def\physrep{\aaref@jnl{Phys.~Rep.}}      % Physics Reports
\def\physscr{\aaref@jnl{Phys.~Scr}}       % Physica Scripta
\def\commat{\aaref@jnl{Comm.~Math.~Phys.}}              % Communications in Mathematical Physics
\def\science{\aaref@jnl{Science}}               % Science
\def\cqg{\aaref@jnl{Classical Quant.~Grav.}}            % Classical and Quantum Gravity
\def\jpcs{\aaref@jnl{JPCS}}                                     % Journal of Physics Conference Series
\def\ijmpd{\aaref@jnl{Int.~J.~Mod.~Phys.~D}}                    % International Journal of Modern Physics D
\def\grg{\aaref@jnl{Gen.~Relat.~Gravit.}}               % General Relativity and Gravitation
\def\rpp{\aaref@jnl{Rep.~Prog.~Phys.}}          % Reports on Progress in Physics
\def\npa{\aaref@jnl{Nucl.~Phys.~A}}        % Nuclear Physics A
\def\lrr{\aaref@jnl{Living Rev.~Rel.}}                   % Living reviews in relativity
\def\jcap{\aaref@jnl{J.~Cosmology Astropart.~Phys.}}    % Journal of cosmology and astroparticle physics
\def\rmp{\aaref@jnl{Rev.~Mod.~Phys.}}   %Reviews of modern physics
\def\epjc{\aaref@jnl{Eur.~Phys.~J.~C}} 
\def\plb{\aaref@jnl{~Phy.~Lett.~B}} 
\def\mpla{\aaref@jnl{Mod.~Phy.~Lett.~A}} 
\def\arxiv{\aaref@jnl{arxiv.org}}

\allowdisplaybreaks[1]
\DeclareMathOperator{\sgn}{sgn}
\begin{document}
	\title{\bf Global phase space analysis for a class of single scalar field bouncing solutions in general relativity}
	
	\author{A. S. Agrawal\orcidlink{0000-0003-4976-8769}}
	\email{agrawalamar61@gmail.com}
	\affiliation{Department of Mathematics,
		Birla Institute of Technology and Science-Pilani, Hyderabad Campus,
		Hyderabad-500078, India.}
	\author{Saikat Chakraborty\orcidlink{0000-0002-5472-304X}}
	\email{saikat.ch@nu.ac.th}
	\affiliation{The Institute for Fundamental Study ``The Tah Poe Academia Institute", Naresuan University, Phitsanulok 65000, Thailand}
	\affiliation{Center for Space Research, North-West University, Mahikeng 2745, South Africa}
	\author{B. Mishra\orcidlink{0000-0001-5527-3565}}
	\email{bivu@hyderabad.bits-pilani.ac.in}
	\affiliation{Department of Mathematics,
		Birla Institute of Technology and Science-Pilani, Hyderabad Campus,
		Hyderabad-500078, India.}
	\author{Jibitesh Dutta\orcidlink{0000-0002-6097-454X}}
	\email{jibitesh@nehu.ac.in}
	\affiliation{Mathematics Division, Department of Basic Sciences and Social Sciences, North Eastern Hill University,  Shillong, Meghalaya 793022, India.}
	\affiliation{Inter University Centre for Astronomy and Astrophysics, Pune 411 007, India}
	\author{Wompherdeiki Khyllep\orcidlink{0000-0003-3930-4231}}
	\email{sjwomkhyllep@gmail.com}
	\affiliation{Department of Mathematics, St.\ Anthony's College, Shillong, Meghalaya 793001, India}

	\begin{abstract}
		We carry out a compact phase space analysis of a non-canonical scalar field theory whose Lagrangian is of the form $F(X)-V(\phi)$ within general relativity. In particular, we focus on a kinetic term of the form $F(X)=\beta X^m$ with power law potential $V_0 \phi^n$ and exponential potential $V_0 e^{-\lambda\phi/M_{Pl}}$ of the scalar field. The main aim of this work is to investigate the genericity of nonsingular bounce in these models and to investigate the cosmic future of the bouncing cosmologies when they are generic. A global dynamical system formulation that is particularly suitable for investigating nonsingular bouncing cosmologies is used to carry out the analysis. We show that when $F(X)=\beta X^m$ ($\beta<0$), nonsingular bounce is generic for a power law potential $V(\phi) = V_0 \phi^n$ only within the parameter range $\left\lbrace \frac{1}{2}<m<1,\,n<\frac{2m}{m-1}\right\rbrace$ and for an exponential potential $V(\phi) = V_0 e^{-\lambda\phi/M_{Pl}}$ only within the parameter range $\left\lbrace\frac{1}{2}<m\leq1\right\rbrace$. Except in these cases, nonsingular bounce in these models is not generic due to the non-existence of global past or future attractors. Our analysis serves to show the importance of a global phase space analysis to address important questions about nonsingular bouncing solutions, an idea that may and must be adopted for such solutions even in other theories.
	\end{abstract}
	
	\maketitle
	%\textbf{PACS number}:\\
	
	\section{Introduction} 
	
	The nonsingular bouncing paradigm has been under study for quite a long time now as an alternative to the inflationary paradigm. The main problem with the inflationary paradigm is that it is inherently singular; timelike and null geodesics cannot be extended indefinitely towards the past in a nonsingular way \cite{Borde:2001nh}. Studies on the nonsingular bouncing paradigm were fuelled mainly by the theoretical necessity of avoiding this singularity problem. Nonsingular bouncing cosmologies consist of a pre-bounce contracting phase ($\dot{a}(t)<0$, $a(t)$ being the scale factor of the Universe) connecting smoothly through a nonsingular bounce ($\dot{a}=0,\,\Ddot{a}>0$ or $H=0,\,\dot{H}>0$, $H(t)$ being the Hubble parameter) to the post bounce expanding phase ($\dot{a}(t)>0$) that we live in. Various aspects of the nonsingular bouncing solutions have been studied in the literature, for example, the anisotropy problem during the contraction phase and the evolution of cosmological perturbations through a bounce. For a critical review of the bouncing paradigm, one can see Refs. \cite{Xue:2013iqy,Battefeld:2014uga}. An aspect that is rarely addressed in relevant literature is, even if a theory admits nonsingular bouncing solutions, how generic these bouncing solutions are i.e. how sensitive are these solutions to small changes in the initial conditions of the universe? This is precisely the question that we will address in this article.
	
	 In this article, we will be dealing with a single scalar field model whose Lagrangian density can be written as $\mathcal{L}=F(X)-V(\phi)$. $F(X)-V(\phi)$ model can produce nonsingular bouncing solutions \cite{Cai:2012va,Cai:2013kja} with a phantom field. It is known that if one tries to achieve nonsingular bouncing solutions within general relativity (GR), one needs to allow for the violation of the null energy condition (NEC) and/or the strong energy condition (SEC) \cite{Novello:2008ra}. For the spatially flat case like we consider here, NEC violation is necessary. When the dynamics of inhomogeneous cosmological perturbations are considered, NEC-violating scalar fields may exhibit ghost instability \cite{Carroll_2003,Garriga_2013,Sawicki_2013} and gradient instability \cite{Vikman_2005}. We must mention here that attempts to construct stable nonsingular bouncing models in spite of NEC violation has given birth to more sophisticated and involved models that involve, instead of a usual scalar field, either ghost condensates \cite{Buchbinder_2007,Arkani-Hamed_2004} and Galileons \cite{Creminelli_2010_2010,Easson_2011,Qiu_2011_2011}. However, since the main focus of this paper is not an analysis of the dynamics of inhomogeneous perturbations but to address the question of stability of a bouncing solution with slight perturbation in the initial condition, we stick to the simpler $F(X)$-$V(\phi)$ model here to explain our method. The analysis can, in principle, of course be generalized to the more sophisticated bouncing models mentioned above.
 
    Considering a simple form of the kinetic term $F(X)$ and two particular examples for the potential $V(\phi)$, we will attempt to address the question of the generality of nonsingular bouncing solutions. The canonical field, often referred to as the quintessence field with potential, is the most basic type of scalar field \cite{Fang_2007}. However, there are several complex cosmological dynamics in the universe that the canonical scalar field is unable to fully explain. For instance, the crossing of the phantom divide line and the bouncing solution cannot be explained by the quintessence field model. This leads to a broader explanation of the non-canonical scalar field, a type of scalar field. The coincidence problem may be solved in the non-canonical scenario without causing any fine-tuning problems, which is another benefit over the canonical setup. In addition, the tensor-to-scalar ratio in non-canonical models is smaller than in canonical models, resulting in better agreement with CMB measurements. We are inspired to learn more about the cosmic dynamics of non-canonical scalar fields by their intriguing properties \cite{Unnikrishnan_2012}.

	Our approach to tackling the problem of genericity of bouncing solutions  will be to carry out a dynamical system analysis of the models and investigate the nonsingular bouncing solutions within the phase space. The dynamical systems technique has been a very useful tool in understanding the qualitative behaviors of cosmological models without analytically solving the system of differential equations \cite{Ellis,Coley:2003mj}. With this technique, one can reframe cosmological dynamics as the phase flow in a suitably defined phase space. Although this technique is now extensively used in investigating inflationary and dark energy models \cite{Bahamonde:2017ize}, it is rarely used to investigate nonsingular bouncing cosmologies. The main reason seems to be that the Hubble normalized dimensionless dynamical variables that are mostly used for the cosmological phase space analysis diverge at a nonsingular bounce. On the other hand, the dynamical system formulation seems to be perfectly fitted to answer qualitative questions like the genericity of a cosmological solution. By genericity of the bouncing solutions in a theory, we mean how sensitive are they to the initial conditions. Suppose one sets some initial conditions at a random moment during the contracting phase of the evolution, numerically evolves the field equations and obtains a bounce. The question is whether one would also get a bounce if (s)he applies a random slight change in the initial condition. To investigate nonsingular bouncing solutions in the phase space picture, one needs to either define an alternative set of dimensionless dynamical variables or compactify the infinite phase space into a finite domain using a specific compactification prescription. Apart from the question of genericity, the phase space picture also provides information about the past and future asymptotics of a nonsingular bouncing cosmology. The future asymptotic is really important, as they determine the end states of the bouncing cosmologies. Do they end up being asymptotically De-Sitter like in $\Lambda$CDM, or lead to a big-rip?  These are our main motivations behind this work.
 
	Cosmological phase space of $F(X)-V(\phi)$ models have been investigated in Ref. \cite{De-Santiago:2012ibi}, where the authors also attempt to touch upon the bouncing solutions from a phase space point of view by defining an alternative set of dynamical variables. Ref. \cite{Panda:2015wya} extends the same analysis to the case of Bianchi-I. However, none of the above works carries out a compact phase space analysis. The system, in general, may contain interesting cosmological scenarios described by fixed points hidden at infinity; in that case, compact analysis is required to extract the global dynamics of the system. For example, in Ref. \cite{De-Santiago:2012ibi}, even though the authors could show phase trajectories that represent nonsingular bounce, one cannot really answer the question of genericity and future asymptotics. In the present work, we employ the same dynamical system formulation as was used in Refs.\cite{De-Santiago:2012ibi,Panda:2015wya} but complement these earlier works by carrying out a compact phase space analysis and hence answering the aforementioned questions. For the $F(X)$ and $V(\phi)$ we consider here, we explicitly single out the cases where the bounce is generic and also find out when they are asymptotically De-Sitter and when they lead to a big-rip. We must mention, however, that bouncing solutions not being generic does not mean one cannot obtain bouncing solutions. One could of course still get a bounce for a set of initial conditions. However, unlike the case when the bounce is generic, for a chosen initial condition that leads to a bounce, it cannot be guaranteed that a random slight change of it will still lead to a bounce. In this sense, the cases for which bounce is generic are more interesting than the cases when they are not, as long as constructing a bouncing model is concerned.
 % Some more research work can be seen in Refs. \cite{Lilley:2011ag,Koehn:2016,Chen:2017,Klinkhamer:2019,Battista:2021,Gungor:2021}
	
	The paper is structured as follows. In section \ref{sec:basics}, we present the basic equations for $F(X)-V(\phi)$ model considering a spatially flat homogeneous and isotropic FLRW Universe. In section \ref{sec:dsa}, we present the dynamical system formulation suitable for investigating nonsingular bouncing cosmologies and carry out the phase space analysis for two particular models. In section \ref{sec:summary}, we summarize the take-home results from our phase space analysis. Finally, we conclude in section \ref{sec:concl}.
	
	%%%%%%%%%%%%%%%%%%%%%%%%%%%%%%%%%%%%%%%%%%%%%%%%%%%%%%%%%%%%%%%%%%%%
	
	\section{Basic cosmological equations for \texorpdfstring{$\mathcal{L}=F(X)-V(\phi)$}{}}\label{sec:basics}
	
	The most general action of a minimally coupled scalar field theory
	is given by
	\begin{equation}\label{eq:action}
		S=\int d^4x \sqrt{-g}\left(\frac{M_{Pl}^2}{2} R+\mathcal{L}(\phi,X)\right)+S_m\,,
	\end{equation}
	where $M_{Pl}$ is the reduced Planck mass, $R$ is the Ricci scalar, $g$ is the metric determinant, $\mathcal{L}(\phi, X)$ is the  Lagrangian density of scalar field $\phi$ whose kinetic component is denoted by $X$  (i.e., $X = -\frac{1}{2}\partial_{\mu} \phi \partial^{\mu} \phi$) and the last term $S_m$ is the action corresponds to the matter component taken to be a perfect fluid.
	
	Variation of  \eqref{eq:action} with respect to the metric $g_{\mu\nu}$ yields the Einstein field equations given by
	\begin{equation}
		G_{\mu\nu}= T_{\mu\nu}^{(\phi)} + T_{\mu\nu}^{(m)}\,,
	\end{equation}
	where $G_{\mu\nu}$ denotes the Einstein tensor, $T_{\mu\nu}^{(\phi)}$ is the scalar field  energy-momentum tensor given by
	\begin{equation}
		T_{\mu\nu}^{(\phi)}=\frac{\partial \mathcal{L}}{\partial X} \partial_{\mu} \phi \partial_{\mu} \phi-g_{\mu\nu} \mathcal{L}\,,
	\end{equation}
	and the matter energy-momentum tensor $T_{\mu\nu}^{(m)}$ is given by
	\begin{equation}\label{eq:EMT_mat}
		T_{\mu\nu}^{(m)}=(\rho_m+P_m) u_\mu u_\nu+P_m g_{\mu\nu}\,.
	\end{equation}
	Here  $\rho_{m}$ and $P_m$ denote the energy density and pressure of the matter component, respectively, with the four-velocity vector $u_\mu$. In this paper, we consider a spatially flat Friedmann-Lema\^itre-Robertson-Walker (FLRW) cosmology described by the metric
	\begin{equation}
		ds^2=-dt^2+a(t)^2(dx^2+dy^2+dz^2)\,,
	\end{equation} 
	where $a(t)$ is a scale factor, $t$ is cosmic time and $x, y, z$ are the Cartesian coordinates. We also focus on a scalar field model whose generic form of the Lagrangian is given by 
	\begin{equation}
		\mathcal{L}(\phi,X) = F(X) - V(\phi)\,,
		%\qquad \left(X=-\frac{1}{2}\partial_{\mu}\phi\partial^{\mu}\phi\right),
	\end{equation}
	where $V(\phi)$  is a scalar field potential and $F(X)$ is an arbitrary  function of $X$.
	
	The energy-momentum tensor \eqref{eq:EMT_mat} of  a perfect fluid under the FLRW cosmology is
	\begin{equation}
		T^{\mu (m)}_{\nu}= \text{diag}(-\rho_m, P_m, P_m, P_m) %=\text{diag}(-\rho_m, P_m, P_m, P_m). 
	\end{equation}
	The spatially flat FLRW space time transforms the above Einstein field equations to the following cosmological field equations,
	\begin{subequations}
		\begin{eqnarray}
			&& H^{2} = \frac{1}{3M_{Pl}^{2}}\left[2XF_X - F + V + \rho_{m}\right], \\
			&& \dot{H} = -\frac{1}{2M_{Pl}^{2}}\left[2XF_X + (1+\omega_{m})\rho_{m}\right]\,, \label{eq:Raychaudhuri}
		\end{eqnarray}
	\end{subequations}
	where $H=\frac{\dot{a}}{a}$ is the Hubble parameter and an over dot denote derivative with respect to $t$, the subscript $X$ denotes derivative with respect to $X$ and $\omega_{m}$ is the perfect fluid equation of state parameter defined as $P_m=\omega_{m} \rho_{m}$. When there is no energy exchange between the fluid and the field, the fluid component scales according to the ordinary continuity equation
	\begin{equation}
		\dot{\rho}_m + 3H(1+\omega_m)\rho_m = 0 \quad \Rightarrow \quad \rho_{m}\propto a^{-3(1+\omega_{m})}\,,
	\end{equation}
	and the scalar field satisfies the generic Klein-Gordon equation
	\begin{equation}
		\frac{d}{dN}(2XF_X - F + V) + 6XF_X = 0,
	\end{equation}
	% \begin{equation}
	% \frac{d}{d\tau}(2XF_X - F + V) + \frac{\sqrt{3}M_{Pl}H}{\sqrt{|2XF_X - F|}}(6XF_X)=0,
	% \end{equation}
	% The time differentiation here has been changed to $d\tau=\sqrt{\frac{\vert \rho_{k}\vert}{3M_{PI}^{2}}}dt$, a variable that for an expanding FLRW model can be used as the independent variable instead of the cosmological time, with the relation $d\tau=\sqrt{\frac{\vert 2XF_{X}-F\vert}{3M_{PI}^{2}}}dt$.
	where $N=\ln a$. 
	
	In the next section, we shall construct the corresponding 
	autonomous system for the above cosmological equations and then we investigate the bouncing scenarios via a global dynamical system analysis.
	
	%%%%%%%%%%%%%%%%%%%%%%%%%%%%%%%%%%%%%%%%%%%%%%%%%%%%%%%%%%%%%%%%%%%%
	
	\section{\bf{Dynamical system formulation of \texorpdfstring{$\mathcal{L}=F(X)-V(\phi)$}{} models suitable for investigating nonsingular bounces}}\label{sec:dsa}
	
	In order to obtain an autonomous system of equations that is suitable to investigate nonsingular bouncing solutions through a phase space point of view, we follow a dynamical system construction that was presented in Ref.\cite{De-Santiago:2012ibi}. We define the dynamical variables
	\begin{equation}\label{dynvar_def}
		\begin{aligned}
			& x = \frac{\sqrt{3}M_{Pl}H}{\sqrt{|\rho_k|}}, \quad 
			y = \sqrt{\frac{|V|}{|\rho_k|}}\sgn(V), \quad 
			\Omega_{m} = \frac{\rho_{m}}{\vert \rho_{k}\vert},\\
			& \sigma = -\frac{M_{Pl}V_{\phi}}{V}\sqrt{\frac{2X}{3|\rho_k|}}\sgn(\dot{\phi}) = -\frac{M_{Pl}}{\sqrt{3|\rho_k|}}\frac{dlog{V}}{dt}
		\end{aligned}
	\end{equation}
	where we have denoted the kinetic part of the energy density by
	\begin{equation}
		\rho_{k} = 2XF_{X} - F.
	\end{equation}
	The kinetic part of the pressure is just
	\begin{equation}
		P_k = F.
	\end{equation}
	Motivated with this, one can define 
	\begin{equation}
		\omega_{k} \equiv \frac{P_k}{\rho_k} = \frac{F}{2XF_{X}-F},
	\end{equation}
	which can be interpreted as the equation of state parameter for the kinetic part of the Lagrangian. The equation of state of the scalar field can be obtained in terms of the new variable as,  
	\begin{equation}
		\omega_{\phi}=\frac{p_{\phi}}{\rho_{\phi}}=\frac{\omega_{k}x^{2}-y^{2}}{x^{2}+y^{2}}\,.
	\end{equation} 
	Next, we define two auxiliary variables 
	\begin{equation}\label{auxvar_def}
		\Xi=\frac{X F_{XX}}{F_{X}}, \qquad \Gamma=\frac{V V_{\phi \phi}}{V_{\phi}^{2}},
	\end{equation}
	which will be required to write the dynamical system. Lastly, we define the phase space time variable \cite{De-Santiago:2012ibi}
	\begin{equation}
		d\tau = \sqrt{\frac{|\rho_{k}|}{3M_{Pl}^{2}}}dt.
	\end{equation}
	
	With respect to the dynamical variables and auxiliary variables defined in Eq.\eqref{dynvar_def} and Eq.\eqref{auxvar_def}, the Friedmann constraint and the dynamical equations become
	\begin{subequations}\label{dynsys_0}
		\begin{eqnarray}
			&& x^2 - y|y| - \Omega_{m} = 1\times \sgn(\rho_{k}),\label{constr_a}\\
			&& \frac{dx}{d\tau} = \frac{3}{2}x\left[(\omega_{k}+1)x-\sigma y|y|\sgn(\rho_{k})\right] - \frac{3}{2}\left[(\omega_{k}-\omega_{m})\sgn(\rho_{k})+(1+\omega_{m})(x^{2}-y|y|)\right],\\
			&& \frac{dy}{d\tau} = \frac{3}{2}y\left[-\sigma+(\omega_{k}+1)x-\sigma y|y|\sgn(\rho_{k})\right],\\
			&& \frac{d\sigma}{d\tau} = -3\sigma^{2}(\Gamma-1) + \frac{3\sigma[2\Xi(\omega_{k}+1)+\omega_{k}-1]}{2(4\Xi+1)(\omega_{k}+1)}\left((\omega_{k}+1)x-\sigma y^{2}\right)\,.
		\end{eqnarray}
	\end{subequations}
	Because of the definition of the dynamical variable $y$, $y|y|$ can also be written as $y^{2}\sgn(V)$. Therefore, the dynamical system \eqref{dynsys_0} can also be written as
	\begin{subequations}\label{dynsys}
		\begin{eqnarray}
			&& x^2 - y^{2}\sgn(V) - \Omega_{m} = 1\times \sgn(\rho_{k}),\label{constr}\\
			&& \frac{dx}{d\tau} = \frac{3}{2}x\left[(\omega_{k}+1)x-\sigma y^{2}\sgn(V)\sgn(\rho_{k})\right] - \frac{3}{2}\left[(\omega_{k}-\omega_{m})\sgn(\rho_{k})+(1+\omega_{m})(x^{2}-y^{2}\sgn(V))\right],\label{eq:ds_x_gen}\\
			&& \frac{dy}{d\tau} = \frac{3}{2}y\left[-\sigma+(\omega_{k}+1)x-\sigma y^{2}\sgn(V)\sgn(\rho_{k})\right],\label{eq:ds_y_gen}\\
			&& \frac{d\sigma}{d\tau} = -3\sigma^{2}(\Gamma-1) + \frac{3\sigma[2\Xi(\omega_{k}+1)+\omega_{k}-1]}{2(2\Xi+1)(\omega_{k}+1)}\left((\omega_{k}+1)x-\sigma y^{2}\right)\,.\label{eq:ds_s_gen}
		\end{eqnarray}
	\end{subequations}
	
	Since $\Omega_m\geq0$, the phase space of the system \eqref{dynsys} is given by
	\begin{equation}
		\lbrace (x,y,\sigma) \in \mathbb{R}^{3}: x^2-y^2\sgn(V)-\sgn(\rho_k)\geq 0\rbrace\,.
	\end{equation}
	It is important to express the quantity $\frac{\dot{H}}{H^2}$ in terms of the dynamical variables as this will help us to find the cosmological evolution corresponding to a fixed point
	\begin{equation}\label{H_eq}
		\frac{\dot{H}}{H^{2}}=-\frac{3}{2x^{2}}\left[(1+\omega_{m})(x^{2}-y\vert y\vert)+(\omega_{k}-\omega_{m})\sgn(\rho_{k})\right]\,.
	\end{equation}
	
	Additionally, one can introduce the effective equation of state $\omega_{\rm eff}$ given by
	\begin{equation}\label{eq:weff}
		\omega_{\rm eff}=-1-\frac{2}{3}\frac{\dot{H}}{H^2}=-1+\frac{1}{x^{2}}\left[(1+\omega_{m})(x^{2}-y\vert y\vert)+(\omega_{k}-\omega_{m})\sgn(\rho_{k})\right]\,.
	\end{equation}
	
	% Before analyzing any specific case, another important point is worth mentioning here. 
	% Writing this way one can notice that the dynamical system \eqref{dynsys} is symmetric under reflection against the $y=0$ line, i.e. under the transformation $y\to-y$. This implies that it suffices to take into consideration only the $y>0$ region of the phase space, since the phase portrait in the $y<0$ region will just be a reflection of that against $y=0$. Since the region $y>0$ corresponds to the requirement $\sgn(V)>0$, this means that. Therefore, from the physical point of view,  it means that the qualitative behavior of the model is independent of the signature of the potential. Further, for physical viability of the model, one requires the condition $\Omega_{m}\geq 0$. 
	Below we consider some specific examples of $F(X)-V(\phi)$ models and look for nonsingular bouncing solutions in the phase space.
	
\subsection{\bf{Specific case I: \quad $F(X)=\beta X$}}

In this section we consider scalar field Lagrangians of the form
\begin{equation}
\mathcal{L}(\phi,X) = \beta X - V(\phi)\,,  \label{langragian}  
\end{equation}
Since $\rho_{\phi}+P_{\phi}=2XF_X=2m\beta X^m$ and $X\geq0$, it reduces to a canonical scalar field when $\beta>0$ and to a phantom scalar field when $\beta <0$. One can notice from equation \eqref{eq:Raychaudhuri} that for $\dot{H}>0$ near the bounce one must necessarily require $\beta<0$, i.e. a phantom scalar field.
 
\subsubsection{\bf{Power Law Potential $V(\phi)=V_0 \phi^n$}}\label{subsec:modelI}

 The first example we consider is the specific case given by $F(X)=\beta X,\,V(\phi)=V_0 \phi^n$ where $\beta, V_0$  are constants with suitable dimension and $n$ is a dimensionless constant. For this choice, we have
 
	\begin{equation}
		\Xi=0,\qquad 
	  \Gamma=1-\frac{1}{n}, \qquad \omega_{k}=1, \qquad \rho_{k}=\beta X.
	\end{equation}

 \paragraph{Finite Fixed Point Analysis}
 
	In this case, the system \eqref{dynsys} reduces to
	 \begin{subequations}\label{dynsys_pow}
 	\begin{eqnarray}
	 		&& \frac{dx}{d\tau} = \frac{3}{2}x\left[2x - \sigma y^2 \sgn(V)\sgn(\beta)\right] - \frac{3}{2}\left[(1-\omega_{m})\sgn(\beta) + (1+\omega_{m})(x^{2}-y^2 \sgn(V))\right],\label{eq:ds_x_pow}\\
 		&& \frac{dy}{d\tau} = \frac{3}{2}y\left[-\sigma + 2x - \sigma y^2 \sgn(V)\sgn(\beta)\right],\label{eq:ds_y_pow}\\
			&& \frac{d\sigma}{d\tau} = \frac{3}{n}\sigma^{2}\,.\label{eq:ds_s_pow}
		\end{eqnarray}
	 \end{subequations}
       One can notice that the dynamical system \eqref{dynsys_pow} is symmetric under reflection against the $y=0$ plane, i.e. under the transformation $y\to-y$. This implies that it suffices to take into consideration only the $y>0$ region of the phase space, since the phase portrait in the $y<0$ region will just be a reflection of that against $y=0$. From the physical point of view, this means that the qualitative behaviour of the  model is independent of the signature of the potential and we can confine our attention to $\sgn(V)=1$. Also, the $y=0$ line acts as an invariant submanifold, meaning that dynamics taking place in the positive branch of the potential can never cross into the negative branch of the potential, and vice versa. Furthermore, for the physical viability of the model, one requires the condition $\Omega_{m}\geq 0$. 
	The phase space of the system \eqref{dynsys_pow} is therefore constrained within the region given by
	\begin{equation}
		\lbrace (x,y,\sigma) \in \mathbb{R}^3: y \geq 0, x^2-y^2-\sgn(\beta)\geq 0\rbrace\,.
	\end{equation}
	
	%%%%%%%%%%%%%%%%%%%%%%%%%%
	
	\begin{table}[H]
		\begin{center}
			\begin{tabular}{ |c|c|c|c| } 
				\hline
				Point & Co-ordinate $(x,y,\sigma)$ &  Existence &  Physical viability $(\Omega_{m}\geq0)$\\ \hline
				$A_{1+}$ & $(1,0,0)$ & $\beta>0$ & Always \\ \hline
				$A_{1-}$ & $(-1,0,0)$ & $\beta>0$ & Always \\ \hline
			\end{tabular}
		\end{center}
		\caption{Existence and physical viability conditions for finite fixed points for a scalar field with kinetic term $F(X)=\beta X$ and potential $V(\phi)=V_0 \phi^n$, calculated from the system \eqref{dynsys_pow}.}
		\label{tab:finite_fixed_pts_pow}
	\end{table}
	
	%%%%%%%%%%%%%%%%%%%%%%%%%%
	
	\begin{table}[H]
		\begin{center}
			\begin{tabular}{ |c|c|c|c| } 
				\hline
				Point   & Co-ordinates $(x,y,\sigma)$ & Stability & Cosmology \\ \hline
				$A_{1+}$ & $(1,0,0)$ &  \begin{tabular}{@{}c@{}}unstable\end{tabular} & $a(t)=(t-t_{*})^{\frac{1}{3}}, ~t\geq t_{*}$ \\ \hline
				$A_{1-}$ & $(-1,0,0)$ & \begin{tabular}{@{}c@{}} saddle (NH) \end{tabular} & $a(t)=(t_{*}-t)^{\frac{1}{3}}, ~t\leq t_{*}$ \\ \hline
			\end{tabular}
		\end{center}
		\caption{Stability condition of physically viable fixed points given in Table \ref{tab:finite_fixed_pts_pow} along with their cosmological behavior. Here and throughout the paper, NH stands for nonhyperbolic.}
		\label{tab:stab_finite_fixed_pts_pow}
	\end{table}
	
	%%%%%%%%%%%%%%%%%%%%%%%%%%
	
\begin{figure}[H]
		\centering
		\includegraphics[width=7cm, height=7.2cm]{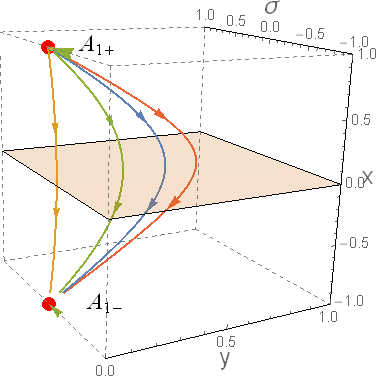}
		\caption{Phase portrait of system \eqref{dynsys_pow} for $\omega_{m}=0$, $n=4$ and $\beta<0$.}\label{fig:finte_pow}
	\end{figure}

	The system \eqref{dynsys_pow} contains two finite fixed points viz. $A_{1+}$ and $A_{1-}$ (see Table \ref{tab:finite_fixed_pts_pow}). Both points exist only in the case of the canonical scalar field, i.e., $\beta>0$. From the value of $x$-coordinate and the nature of scale factor $a(t)$ (see Table \ref{tab:stab_finite_fixed_pts_pow}), we see that while point $A_{1+}$ corresponds to a decelerated expansion of the universe, point $A_{1-}$ corresponds to a decelerated contraction of the universe. We note here that point $A_{1-}$ is non-hyperbolic with an empty unstable subspace near a point and therefore, we refer to the center manifold theory for further analysis.  On performing the analysis, we obtained that the point $A_{1-}$ is a  saddle point (see Appendix \ref{app:cmt}). We also find that the trajectories flow from an unstable point $A_{1+}$ towards a saddle point $A_{1-}$. Therefore, from the finite analysis one can not extract any bouncing solution but it shows recollapsing solutions (see Fig. \ref{fig:finte_pow}).
	
	%On performing the analysis, we obtained that the dynamics on the center manifold of a point is completely determined by the equation
	%	\begin{equation}
	%		\frac{d\sigma}{d\tau} = \frac{3}{n}\sigma^{2}\,.
	%	\end{equation}
	% The above equation shows that the point $A_{1-}$ is a  saddle point. 

	%%%%%%%%%%%%%%%%%%%%%%%%%%
	
	%%%%%%%%%%%%%%%%%%%%%%%
	
	\paragraph{\bf{Fixed Points at Infinity}}
	
	To get a global picture of the phase space, we introduce  the following compact dynamical variables
	\begin{equation}\label{3D_dynvar_comp_def}
		\Bar{x}=\frac{x}{\sqrt{1+x^{2}}}\,, \qquad \Bar{y}=\frac{y}{\sqrt{1+y^{2}}} \,, \qquad  \Bar{\sigma}=\frac{\sigma}{\sqrt{1+\sigma^{2}}}\,.
	\end{equation}
	The evolution equation \eqref{dynsys} can be converted to the following system of equations
	\begin{subequations}\label{dynsys_3d}
		\begin{eqnarray}
			\frac{d\Bar{x}}{d\tau}&=&\frac{3}{2}(1-\Bar{x}^{2})^{\frac{3}{2}}\left[(\omega_{k}-\omega_{m})\left(\frac{\Bar{x}^{2}}{1-\Bar{x}^{2}}-\sgn(\beta)\right)+\left(1+\omega_{m}-\frac{\Bar{\sigma}\Bar{x}\sgn(\beta)}{\sqrt{1-\Bar{\sigma}^2}\sqrt{1-\Bar{x}^{2}}}\right)\frac{\Bar{y}^{2}\sgn(V)}{1-\Bar{y}^{2}}\right],\\    
			\frac{d\Bar{y}}{d\tau}&=&\frac{3}{2}\Bar{y}(1-\Bar{y}^{2})\left[-\frac{\Bar{\sigma}}{\sqrt{1-\Bar{\sigma}^2}} +\frac{(\omega_{k}+1)\Bar{x}}{\sqrt{1-\Bar{x}^{2}}}-\frac{\Bar{\sigma}\Bar{y}^{2}\sgn(V)}{\sqrt{1-\Bar{\sigma}^2}(1-\Bar{y}^{2})}\sgn(\beta)\right] \,,\\
			\frac{d\Bar{\sigma}}{d\tau}&=&\frac{3}{n} \Bar{\sigma}^2\sqrt{1-\Bar{\sigma}^2}\,. %+\frac{3\Bar{\sigma}(1-\Bar{\sigma}^2)\left[2\Xi (\omega_{k}+1)+(\omega_{k}-1) \right]}{2(2\Xi+1)(\omega_{k}+1)}\left[(\omega_{k}+1)\frac{\Bar{x}}{\sqrt{1-\Bar{x}^2}}-\frac{\Bar{\sigma}\Bar{y}^2}{\sqrt{1-\sigma^2}(1-\Bar{y}^2)}\right]\,.
		\end{eqnarray}
	\end{subequations}
	Note that the variables $\Bar{x}, \Bar{y}, \Bar{\sigma}$ are bounded between $-1$ and 1. The dynamical system in Eq.\eqref{dynsys_3d} is however not regular at the boundaries of the compact phase space $\Bar{x}^2=1,\,\Bar{y}^2=1$ and $\Bar{\sigma}^2=1$, it has a pole of order $\frac{1}{2}$ at $\Bar{x}^2=1$ and $\Bar{\sigma}^2=1$ and a pole of order $1$ at $\Bar{y}^2=1$. This can be regularized following a prescription from Ref.\cite{Bouhmadi-Lopez:2016dzw}. The idea is to redefine the phase space time variable as
	\begin{equation}\label{time_redef_pow}
		d\tau \rightarrow d\Bar{\tau} = \frac{d\tau}{(1-\Bar{x}^{2})^{\frac{1}{2}}(1-\Bar{y}^{2})(1-\Bar{\sigma}^2)^\frac{1}{2}}.
	\end{equation}
	%\begin{equation*}
	%\frac{d\Bar{x}}{d\Bar{N}}=\frac{3}{2}(1-\Bar{x}^{2})\left[(\omega_{k}-\omega_{m})\left(\Bar{x}^{2}-(1-\Bar{x}^{2})\sgn(\beta)\right)(1-\Bar{y}^{2})+\left((1+\omega_{m})(1-\Bar{x}^{2})-\sigma\Bar{x}\sgn(\beta)\sqrt{1-\Bar{x}^{2}}\right)\Bar{y}^{2}\sgn(V_{0})\right]    
	%\end{equation*}
	With respect to this redefined time variable, the dynamical system \eqref{dynsys_3d} can be rewritten as
	\begin{subequations}\label{dynsys_3d_new}
		\begin{eqnarray}
			\frac{d\Bar{x}}{d\Bar{\tau}} &=& \frac{3}{2}(\omega_{k}-\omega_{m})\left(\Bar{x}^{2}-(1-\Bar{x}^{2})\sgn(\beta)\right)(1-\Bar{x}^{2})(1-\Bar{y}^{2}) \sqrt{1-\Bar{\sigma}^2} \nonumber\\&& + \frac{3}{2}\left((1+\omega_{m})(1-\Bar{x}^{2}) \sqrt{1-\Bar{\sigma}^2}-\Bar{\sigma}\Bar{x}\sgn(\beta)\sqrt{1-\Bar{x}^{2}}\right)(1-\Bar{x}^{2})\Bar{y}^{2},\label{dynsys_3d_new_x}\\
			\frac{d\Bar{y}}{d\Bar{\tau}}& =& \frac{3}{2}\Bar{y}(1-\Bar{y}^{2})\left[(-\Bar{\sigma}\sqrt{1-\Bar{x}^{2}} +(\omega_{k}+1)\Bar{x}\sqrt{1-\Bar{\sigma}^2})(1-\Bar{y}^{2})-\Bar{\sigma}\Bar{y}^{2}\sqrt{1-\Bar{x}^{2}}\sgn(\beta)\right]\,, \label{dynsys_3d_new_y}\\
			\frac{d\Bar{\sigma}}{d\Bar{\tau}}&=&\frac{3}{n}\Bar{\sigma}^2(1-\Bar{\sigma}^2) \sqrt{1-\Bar{x}^2}(1-\Bar{y}^2)\,.\label{dynsys_3d_new_s}
			%+\frac{3\Bar{\sigma}(1-\Bar{\sigma}^2)\left[2\Xi (\omega_{k}+1)+(\omega_{k}-1) \right]}{2(2\Xi+1)(\omega_{k}+1)}\nonumber\\ &&\left[(\omega_{k}+1)\Bar{x}(1-\Bar{y}^2)\sqrt{1-\Bar{\sigma}^2}-\Bar{\sigma}\Bar{y}^2\sqrt{1-\Bar{x}^2}\right]\,.\label{dynsys_3d_new_s}
		\end{eqnarray}
	\end{subequations}
	%%%%%%%%%%%%%%%%%%%%%%%
	
	\begin{table}[H]
		\begin{center}
                %\small\addtolength{\tabcolsep}{-15pt}
                \resizebox{\textwidth}{!}{
			\begin{tabular}{|*{4}{c|}} 
				\hline
				Point & Co-ordinate $(\Bar{x},\Bar{y},\Bar{\sigma})$ &  Existence &  Physical viability $(\Omega_{m}\geq0)$\\ \hline
				$B_{1+}$ & $(1,1,\Bar{\sigma})$ & Always & Always \\ \hline
			$B_{1-}$ & $(-1,1,\Bar{\sigma})$ & Always &          Always\\ \hline
				$B_{2+}$ & $(1,0,\Bar{\sigma})$ & Always & Always \\ \hline
				\parbox[c][0.6cm]{0.5cm}{$B_{2-}$} & $(-1,0,\Bar{\sigma})$ & Always & Always \\ \hline
				\parbox[c][1cm]{0.5cm}{$B_{3+}$} & $\left(\Bar{x},0,1\right)$ & Always & \begin{tabular}{@{}c@{}}$\frac{1}{2}\leq \Bar{x}^2 \leq 1$ \& $\Bar{x}\neq 0$ if $\sgn(\beta)=1$\\$ \Bar{x}^2 \leq 1$ if $\sgn(\beta)=-1$ \end{tabular}  \\ \hline
				\parbox[c][1cm]{0.5cm}{$B_{3-}$} & $\left(\Bar{x},0,-1\right)$ & Always &\begin{tabular}{@{}c@{}}$\frac{1}{2}\leq \Bar{x}^2 \leq 1$ \& $\Bar{x}\neq 0$  if $\sgn(\beta)=1$\\$ \Bar{x}^2 \leq 1$ if $\sgn(\beta)=-1$ \end{tabular} \\ \hline
				$B_{4+}$ & $\left(\frac{(1+\omega_{m})\sqrt{1-\Bar{\sigma}^2}}{\sqrt{(1+\omega_{{m}})^2-\Bar{\sigma}^2\omega_{m}(\omega_{m}+2)}},1,\Bar{\sigma}\right)$ & \begin{tabular}{@{}c@{}}$\left(\left(\Bar{\sigma}^2\leq \frac{(1+\omega_{m})^2}{\omega_{m}(\omega_{m}+2)}\right)\wedge  \left((\beta<0)\wedge\left[ (\Bar{\sigma}\leq 0) \vee (\Bar{\sigma}=1)\right]\right)\right)\vee$ \\ $\left(\left(\Bar{\sigma}^2\leq \frac{(1+\omega_{m})^2}{\omega_{m}(\omega_{m}+2)}\right) \wedge\left((\beta>0)\wedge \left[(\Bar{\sigma}\geq 0)\vee(\Bar{\sigma}=-1)\right]\right)\right)$\end{tabular}& $\Bar{\sigma}=0$ \\ \hline
                $B_{4-}$ & $\left(-\frac{(1+\omega_{m})\sqrt{1-\Bar{\sigma}^2}}{\sqrt{(1+\omega_{{m}})^2-\Bar{\sigma}^2\omega_{m}(\omega_{m}+2)}},1,\Bar{\sigma}\right)$ & \begin{tabular}{@{}c@{}}$\left(\left(\Bar{\sigma}^2\leq \frac{(1+\omega_{m})^2}{\omega_{m}(\omega_{m}+2)}\right)\wedge \left((\beta<0)\wedge \left[(\Bar{\sigma}\geq 0)\vee(\Bar{\sigma}=-1)\right]\right)\right)\vee$\\ 
                $\left(\left(\Bar{\sigma}^2\leq \frac{(1+\omega_{m})^2}{\omega_{m}(\omega_{m}+2)}\right)\wedge\left((\beta>0)\wedge \left[(\Bar{\sigma}\leq 0)\vee (\Bar{\sigma}=1)\right]\right)\right)$\end{tabular} & $\Bar{\sigma}=0$ \\ \hline
				$B_{5+}$ & $\left(0,\frac{1}{\sqrt{2}},1\right)$ & $\beta <0$ & Always \\ \hline
				$B_{5-}$ &$\left(0,\frac{1}{\sqrt{2}},-1\right)$ & $\beta <0$ & Always \\ \hline

				%	$B_{3+}$ & $\left(\frac{1}{\sqrt{2}},0,1\right)$ & $\beta >0$ & Always \\ \hline
				%	$B_{3-}$ & $\left(-\frac{1}{\sqrt{2}},0,1\right)$ & $\beta >0$ & Always \\ \hline
				%	$B_{4+}$ & $\left(\frac{1}{\sqrt{2}},0,-1\right)$ & $\beta >0$ & Always \\ \hline
				%	$B_{4-}$ & $\left(-\frac{1}{\sqrt{2}},0,-1\right)$ & $\beta >0$ & Always \\ \hline
				%		$B_{5+}$ & $\left(\frac{2}{\sqrt{5}},\frac{\sqrt{3}}{2},1\right)$  & $\beta>0$ & Always \\ \hline
				%	$B_{5-}$ & $\left(-\frac{2}{\sqrt{5}},\frac{\sqrt{3}}{2},-1\right)$& $\beta>0$ & Always  \\ \hline
				%	$B_{6+}$ & $\left(\frac{1}{\sqrt{(1+\omega_{m})^{2}+1}},\sqrt{\frac{{1-\omega_{m}}}{2}},1\right)$ & $\beta>0$ & $\omega_{m}\leq -\frac{1}{2}$ \\ \hline
				%		$B_{6-}$ & $\left(\frac{1}{\sqrt{(1+\omega_{m})^{2}+1}},\sqrt{\frac{{1-\omega_{m}}}{2}},-1\right)$ & $\beta>0$ & $\omega_{m}\leq -\frac{1}{2}$ \\ \hline
				%		$B_{7+}$ & $\left(-\frac{1}{\sqrt{(1+\omega_{m})^{2}+1}},\sqrt{\frac{{1-\omega_{m}}}{2}},1\right)$ & $\beta>0$ & $\omega_{m}\leq -\frac{1}{2}$ \\ \hline
				%		$B_{7-}$ & $\left(-\frac{1}{\sqrt{(1+\omega_{m})^{2}+1}},\sqrt{\frac{{1-\omega_{m}}}{2}},-1\right)$ & $\beta>0$ & $\omega_{m}\leq -\frac{1}{2}$ \\ \hline
				%		$B_{8+}$ & $\left(\frac{2}{\sqrt{5}},\frac{5}{\sqrt{6}},-1\right)$ & $\beta<0$ & Always \\ \hline
				%		$B_{8-}$ & $\left(-\frac{2}{\sqrt{5}},\frac{5}{\sqrt{6}},1\right)$ & $\beta<0$ & Always \\ \hline

			\end{tabular}}
		\end{center}
		\caption{Existence and physical viability condition for fixed points at infinity for a scalar field with kinetic term $F(X)=\beta X$ and potential $V(\phi)=V_0 \phi^n$ calculated from the system \eqref{dynsys_3d_new}. }   
		\label{tab:infinite_fixed_pts_pow}
	\end{table}
	
	%%%%%%%%%%%%%%%%%%%%%%%
	
	\begin{table}[H]
		\begin{center}
			\begin{tabular}{|*{4}{c|}}
				\hline
				Point   & Co-ordinates $(\Bar{x},\Bar{y},\Bar{\sigma})$ & Stability & Cosmology \\ \hline
				\parbox[c][1.2cm]{0.5cm}{$B_{1+}$} & $(1,1,\Bar{\sigma})$ & \begin{tabular}{@{}c@{}}stable for $\Bar{\sigma}=0 ~\text{or} ~\{\Bar{\sigma} \neq 0, \sgn(\beta)\neq \sgn(\Bar{\sigma})\}$ \\ \& saddle for $\{\Bar{\sigma} \neq 0, \sgn(\beta)= \sgn(\Bar{\sigma})\}$\end{tabular}& De Sitter\\ \hline
				\parbox[c][1.2cm]{0.5cm}{$B_{1-}$} &  $(-1,1,\Bar{\sigma})$& \begin{tabular}{@{}c@{}} unstable for $\Bar{\sigma}=0 ~\text{or} ~\{\Bar{\sigma}\neq 0, \sgn(\beta)= \sgn(\Bar{\sigma})\}$  \\ \& saddle for $\{\Bar{\sigma} \neq 0, \sgn(\beta)\neq \sgn(\Bar{\sigma})\}$\end{tabular} & De Sitter \\ \hline
				\parbox[c][0.8cm]{0.5cm}{$B_{2+}$}   & $(1,0,\Bar{\sigma})$ & saddle always & $a(t)=(t-t_{*})^{\frac{2}{3(1+\omega_{m})}}, ~t\geq t_{*}$\\ \hline
				\parbox[c][0.8cm]{0.5cm}{$B_{2-}$}   & $(-1,0,\Bar{\sigma})$ & saddle always & $a(t)=(t_{*}-t)^{\frac{2}{3(1+\omega_{m})}}, ~t\leq t_{*}$\\ \hline
				$B_{3+}$ & $\left(\Bar{x},0,1\right)$ &  \begin{tabular}{@{}c@{}}stable if $\sgn(n)>0$\\saddle otherwise \end{tabular}  & depending on $\Bar{x}$ and $\beta$\\ \hline
				$B_{3-}$ & $\left(\Bar{x},0,-1\right)$  &  \begin{tabular}{@{}c@{}}unstable if $\sgn(n)>0$\\saddle otherwise \end{tabular}  &  depending on $\Bar{x}$ and $\beta$ \\ \hline
				
				%	$B_{4+}$   & $\left(\frac{(1+\omega_{m})\sqrt{1-\Bar{\sigma}^2}}{\sqrt{(1+\omega_{{m}})^2-\Bar{\sigma}^2\omega_{m}(\omega_{m}+2)}},1,\Bar{\sigma}\right)$ & saddle & To be obtained \\ \hline
				%		$B_{4-}$   &$\left(-\frac{(1+\omega_{m})\sqrt{1-\Bar{\sigma}^2}}{\sqrt{(1+\omega_{{m}})^2-\Bar{\sigma}^2\omega_{m}(\omega_{m}+2)}},1,\Bar{\sigma}\right)$ & saddle  & To be obtained \\ \hline
				$B_{5+}$ & $\left(0,\frac{1}{\sqrt{2}},1\right)$ &  \begin{tabular}{@{}c@{}}unstable if $\sgn(n)<0$\\saddle otherwise \end{tabular}  &$a(t)=$constant\\ \hline
				$B_{5-}$ & $\left(0,\frac{1}{\sqrt{2}},-1\right)$ & \begin{tabular}{@{}c@{}}stable if $\sgn(n)<0$\\saddle otherwise \end{tabular} & $a(t)=$constant\\ \hline

			\end{tabular}
		\end{center}
		\caption{Stability condition of physically viable fixed points given in Table \ref{tab:infinite_fixed_pts_pow} along with their cosmological behavior. Stability of the lines of fixed points $B_{1\pm},\,B_{2\pm}$ and $B_{3\pm}$ can be determined by investigating the stability of the invariant submanifolds $\Bar{x}=\pm1,\,\Bar{y}=0,1$ and $\Bar{\sigma}=\pm1$ (see appendix \ref{app:stab_inv_sub}).}
		\label{tab:stab_infinite_fixed_pts_pow}
	\end{table}
	%%%%%%%%%%%%%%%%%%%%%%%
	
	In terms of the compact variables, the constraint equation \eqref{constr} can be rewritten for this model as
	\begin{equation}\label{constr_comp}
		\frac{\Bar{x}^{2}}{1-\Bar{x}^{2}} - \frac{\Bar{y}^{2}}{1-\Bar{y}^{2}} - \sgn(\beta) = \Omega_{m} \,.   
	\end{equation}
	Therefore, for a canonical scalar field where $\sgn(\beta)=1$, the physical viability condition requires 
	\begin{equation} \label{physviab_can_1}
		\frac{\Bar{x}^{2}}{1-\Bar{x}^{2}}-\frac{\Bar{y}^{2}}{1-\Bar{y}^{2}}-1=\Omega_{m}\geq 0 \,.  
	\end{equation}
	It can be checked that the necessary and sufficient conditions for physical viability of canonical scalar field are given respectively as follows
	\begin{eqnarray}\label{physviab_can}
		\Bar{y}^{2}\leq 2-\frac{1}{\Bar{x}^{2}}, \hspace{1cm} \Bar{x}^{2}\geq \frac{1}{2}.
	\end{eqnarray}
	For a phantom scalar field where $\sgn(\beta)=-1$, the physical viability condition requires
	\begin{equation}
		\frac{\Bar{x}^{2}}{1-\Bar{x}^{2}}-\frac{\Bar{y}^{2}}{1-\Bar{y}^{2}}+1=\Omega_{m}\geq 0\,.
	\end{equation}
	The necessary \emph{and} sufficient condition for the above to be satisfied is
	\begin{eqnarray}\label{physviab_noncan}
		\Bar{y}^{2}\leq \frac{1}{2-\Bar{x}^{2}}.
	\end{eqnarray}
	The constraints in Eq.\eqref{physviab_can} and Eq.\eqref{physviab_noncan} specify the physically viable region of the entire 3-dimensional compact phase space for canonical and phantom scalar fields respectively. One can notice that the line $\Bar{x}=0$ is physical only for the case of a phantom scalar field ($\beta<0$), which is consistent with the fact that a phantom scalar field is necessarily required to achieve a nonsingular bounce.
	
	The system \eqref{dynsys_3d_new} presents a total of six invariant submanifolds $\bar{x}=\pm1,\,\Bar{y}=0,1$ and $\bar{\sigma}=\pm1$. Their stability is calculated in appendix \ref{app:stab_inv_sub}. The fixed points for system \eqref{dynsys_3d_new} are presented in Table \ref{tab:infinite_fixed_pts_pow}, along with their stability conditions in Table \ref{tab:stab_infinite_fixed_pts_pow}. Although from Table \ref{tab:infinite_fixed_pts_pow} it might appear that there are five different pairs of fixed points at the infinity of the phase space, the pair $B_{4\pm}$ is physically viable only for $\Bar{\sigma}=0$, in which case they already lie on the line of fixed points $B_{1\pm}$. Therefore there are only four different pairs of fixed points at infinity. $B_{1\pm}$ are De-Sitter solutions, whereas $B_{2\pm}$ correspond to cosmological phases dominated by the hydrodynamic matter component. For $\omega_{m}=0$, the fixed point $B_{2+}$ can be interpreted as the matter-dominated epoch since this is a saddle and therefore, always represents an intermediate phase of evolution. Fixed points on the line $B_{3\pm}$ can exhibit various cosmological scenarios. For instance, they can describe a matter-dominated universe for $\Bar{x}=1,\,\beta>0$, static universe for $\Bar{x}=0,\,\beta<0$. Finally, the fixed points $B_{5\pm}$ correspond to a static universe.
	
\begin{figure}[H]
		\centering
		\minipage{0.40\textwidth}
		\includegraphics[width=\textwidth]{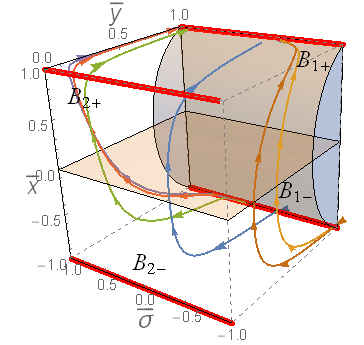}
		\endminipage
		\minipage{0.40\textwidth}
		\includegraphics[width=\textwidth]{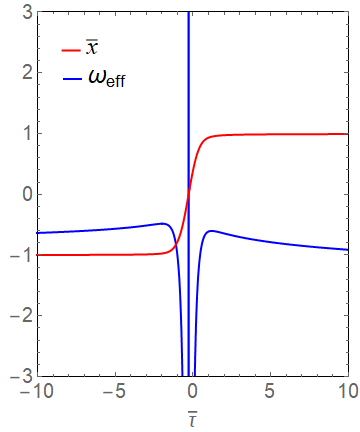}
		\endminipage\\
		\minipage{0.40\textwidth}
		\includegraphics[width=\textwidth]{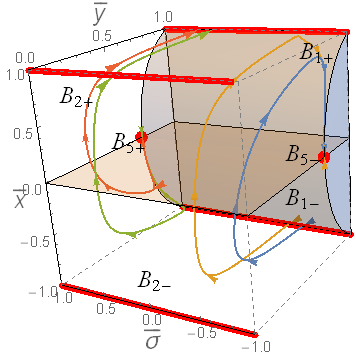}
  \endminipage
		\minipage{0.40\textwidth}
		\includegraphics[width=\textwidth]{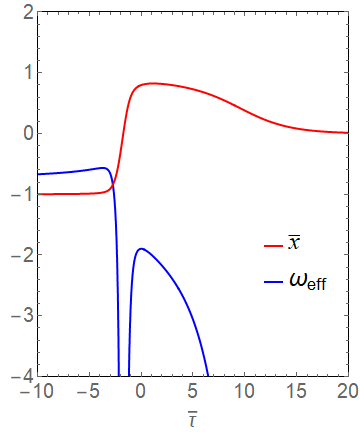}
		\endminipage
		\caption{Phase portrait of the system \eqref{dynsys_3d_new} with the shaded region representing the non-physical region of the phase space (upper and lower left panels). Here $\omega_{m}=0$, $\beta<0$ with $n=4$ in the upper left panel and $n=-4$ in the lower panel. A plot of variables $\Bar{x}$ and effective equation of state $\omega_{\rm eff}$ is shown in the upper and lower right panel for the blue trajectories of the upper and lower left panel, which represents two characteristic types of nonsingular bouncing solutions, namely, asymptotically De-Sitter and asymptotically static in future.}
    \label{fig:infinte_pow}
	\end{figure}

	In the left panel of Fig.\ref{fig:infinte_pow}, we present the phase portrait of the system  \eqref{dynsys_3d_new} in the compact phase space that corresponds to physically viable nonsingular bouncing solutions. We also observe that there is a violation of the null energy condition (NEC) during bounce (see the right panel of Fig.\ref{fig:infinte_pow}), as expected. The plots also clearly show the matter-dominated fixed points $B_{2\pm}$ are saddles, i.e. intermediate epochs of evolution, as expected. 
 
   The phase trajectories in the upper left panel, which corresponds to $F(X)=\beta X\,(\beta<0),\,V(\phi)=V_{0}\phi^4$, show nonsingular bouncing solutions connecting the contracting and expanding De-Sitter phases $B_{1-}$ and $B_{1+}$ respectively. However, one cannot say this is the generic behaviour of phase trajectories as $B_{1-}$ and $B_{1+}$ are not global repellers and attractors. Apart from  $B_{1-}$ and $B_{1+}$  there is also  another repeller $B_{3-}$ and another attractor $B_{3+}$. Therefore, there can exist heteroclinic trajectories connecting $B_{3-}$ to $B_{1+}$, $B_{1-}$ to $B_{3+}$ and $B_{3-}$ to $B_{3+}$, which may or may not correspond to a nonsingular bouncing cosmology, depending on how the trajectories evolve.

   The phase trajectories in the lower left panel, which corresponds to $F(X)=\beta X\,(\beta<0),\,V(\phi)=V_{0}\phi^{-4}$, show nonsingular bouncing solutions connecting two De-Sitter phases ($B_{1-} \rightarrow B_{1+}$) or a De-Sitter phase with a static universe ($B_{1-} \rightarrow B_{5-}$ or $B_{5+} \rightarrow B_{1+}$). However, again, one cannot say that such behaviour is  generic. For $n=-4<0$, the fixed points $B_{3\pm}$ are saddles, but $B_{5\pm}$ becomes an attractor/repeller pair. Thus, there can exist heteroclinic trajectories connecting $B_{5-}$ to $B_{5+}$, which, again, may or may not correspond to a nonsingular bouncing cosmology, depending on how the trajectories evolve.

	%%%%%%%%%%%%%%%%%%%%%%%  
	
 %%%%%%%%%%%%%%%%%%%%%%%%%%%%%%%%%%%%%%%%%
		
    \subsubsection{\bf{Exponential potential $V(\phi)=V_{0}e^{-\lambda \phi/M_{Pl}}$: }}\label{subsec:modelII}
    
    The second example we consider is the specific case given by $F(X)=\beta X,\,V(\phi)=V_{0}e^{-\lambda \phi/M_{Pl}}$ where $\beta, V_0$ are constants with suitable dimension, $\lambda$ is a dimensionless constant. For this choice, we have
        \begin{equation}
	\Xi=0, \qquad \Gamma=1, \qquad \omega_{k}=1, \qquad \rho_{k}=\beta X, \qquad \sigma = \frac{\sqrt{{2}/{3}}\lambda}{\sqrt{|\beta|}}.
	\end{equation}

 \paragraph{Finite Fixed Point Analysis}
	
	In this case, since $\sigma$ is a constant, the system \eqref{dynsys} reduces to a 2-dimensional dynamical system
	\begin{subequations}\label{dynsys_exp}
		\begin{eqnarray}
			&& \frac{dx}{d\tau} = \frac{3}{2}x\left[2x-\sigma y^{2}\sgn(V_{0})\sgn(\beta)\right] - \frac{3}{2}\left[(1-\omega_{m})\sgn(\beta)+(1+\omega_{m})(x^{2}-y^{2}\sgn(V_{0}))\right],\\
			&& \frac{dy}{d\tau} = \frac{3}{2}y\left[2x-\sigma\left(1+ y^{2}\sgn(V_{0})\sgn(\beta)\right)\right]\,.
		\end{eqnarray}
	\end{subequations}
 As in the power law case, one can notice that the dynamical system \eqref{dynsys_exp} is symmetric under reflection against the $y=0$ line, i.e. under the transformation $y\to-y$. This implies that it suffices to take into consideration only the $y>0$ region of the phase space, since the phase portrait in the $y<0$ region will just be a reflection of that against $y=0$. From the physical point of view, this means that the qualitative behavior of the model is independent of the signature of the potential and we can confine our attention to $\sgn(V_0)=1$. Therefore, the interpretation of the $y=0$ invariant submanifold is very clear in this case. As before, for the physical viability of the model, one requires the condition $\Omega_{m}\geq 0$. 
	The phase space of the system \eqref{dynsys_pow} is therefore constrained within the region given by
	\begin{eqnarray}
		&&\lbrace (x,y) \in \mathbb{R}^2: y \geq 0, x^2-y^2-\sgn(\beta)\geq 0\rbrace\,.  
	\end{eqnarray}
	
	%%%%%%%%%%%%%%%%%%%%%%%%%%
	% \begin{table}[H]
	% \begin{center}
	% \begin{tabular}{ |c|c|c|c| } 
	% \hline
	% Point & Co-ordinate $(\Bar{x},\Bar{y})$ &  Existence &  Physical viability $(\Omega_{m}\geq0)$\\ \hline
	% $A_{1+}$ & $\left(\frac{1}{\sqrt{2}},0\right)$ & $\beta >0$ & Always \\ \hline
	% $A_{1-}$ & $\left(-\frac{1}{\sqrt{2}},0\right)$ & $\beta >0$ & Always \\ \hline
	% $A_{2+}$ & $\left(\frac{2}{\sqrt{4+\sigma^{2}}},\sqrt{\frac{4+\sigma^{2}}{4+2\sigma^{2}}}\right)$ & $\beta<0$ and $\sigma<0$ & Always \\ \hline
	% $A_{2-}$ & $\left(-\frac{2}{\sqrt{4+\sigma^{2}}},\sqrt{\frac{4+\sigma^{2}}{4+2\sigma^{2}}}\right)$ & $\beta<0$ and $\sigma>0$ & Always \\ \hline
	% $A_{3+}$ & $\left(\frac{2}{\sqrt{4+\sigma^{2}}},\sqrt{1-\frac{\sigma^{2}}{4}}\right)$ & $\beta>0$ and $0<\sigma<2$ & Always \\ \hline
	% $A_{3-}$ & $\left(-\frac{2}{\sqrt{4+\sigma^{2}}},\sqrt{1-\frac{\sigma^{2}}{4}}\right)$ & $\beta>0$ and $-2<\sigma<0$ & Always  \\ \hline
	% $A_{4}$ & $\left(\frac{\sigma}{\sqrt{(1+\omega_{m})^{2}+\sigma^{2}}},\sqrt{\frac{{1-\omega_{m}}}{2}}\right)$ & $\beta>0$ & $|\sigma |\geq\sqrt{2(1+\omega_{m})}$ \\ \hline
	% \end{tabular}
	% \end{center}
	% \caption{All fixed points.}
	% \label{tab:finite_fixed_pts_exp}
	% \end{table}
	
	\begin{table}[H]
		\begin{center}
			\begin{tabular}{|*{4}{c|}} 
				\hline
				Point & Co-ordinate $(x,y)$ &  Existence &  Physical viability $(\Omega_{m}\geq0)$\\ \hline
				\parbox[c][0.6cm]{0.5cm}{$A_{1+}$} & $\left(1,0\right)$ & $\beta >0$ & Always \\ \hline
				\parbox[c][0.6cm]{0.5cm}{$A_{1-}$} & $\left(-1,0\right)$ & $\beta >0$ & Always \\ \hline
				\parbox[c][1cm]{0.5cm}{$A_{2+}$} & $\left(\frac{2}{|\sigma|},\frac{\sqrt{4+\sigma^{2}}}{|\sigma|}\right)$ & $\beta<0$ and $\sigma<0$ & Always \\ \hline
				\parbox[c][1cm]{0.5cm}{$A_{2-}$} & $\left(-\frac{2}{|\sigma|},\frac{\sqrt{4+\sigma^{2}}}{|\sigma|}\right)$ & $\beta<0$ and $\sigma>0$ & Always \\ \hline
				\parbox[c][1cm]{0.5cm}{$A_{3+}$} & $\left(\frac{2}{|\sigma|},\frac{\sqrt{4-\sigma^{2}}}{|\sigma|}\right)$ & $\beta>0$ and $0<\sigma<2$ & Always \\ \hline
				\parbox[c][1cm]{0.5cm}{$A_{3-}$} & $\left(-\frac{2}{|\sigma|},\frac{\sqrt{4-\sigma^{2}}}{|\sigma|}\right)$ & $\beta>0$ and $-2<\sigma<0$ & Always  \\ \hline
				\parbox[c][1cm]{0.5cm}{$A_{4}$} & $\left(\frac{\sigma}{(1+\omega_{m})},\sqrt{\frac{1-\omega_{m}}{1+\omega_{m}}}\right)$ & $\beta>0$ & $|\sigma |\geq\sqrt{2(1+\omega_{m})}$ \\ \hline
			\end{tabular}
		\end{center}
		\caption{Existence and physical viability conditions for finite fixed points for a scalar field with kinetic term $F(X)=\beta X$ and potential $V(\phi)=V_0 e^{-\lambda\phi/M_{Pl}}$, calculated from the system \eqref{dynsys_exp}.}
		\label{tab:finite_fixed_pts_exp}
	\end{table}
	
	%%%%%%%%%%%%%%%%%%%%%%%%%%
	
	\begin{table}[H]
		\begin{center}
			\begin{tabular}{|*{4}{c|}}
				\hline
				Point   & Co-ordinates $(x,y)$ & Stability & Cosmology \\ \hline
				\parbox[c][0.6cm]{0.5cm}{$A_{1+}$} & $\left(1,0\right)$ & unstable always & $a(t)=(t-t_{*})^{\frac{1}{3}}, ~t\geq t_{*}$\\ \hline
				\parbox[c][0.6cm]{0.5cm}{$A_{1-}$} & $\left(-1,0\right)$ & stable always & $a(t)=(t_{*}-t)^{\frac{1}{3}}, ~t\leq t_{*}$\\ \hline
				\parbox[c][1cm]{0.5cm}{$A_{2+}$} & $\left(\frac{2}{|\sigma|},\frac{\sqrt{4+\sigma^{2}}}{|\sigma|}\right)$ & N.H. always & $a(t)=\frac{1}{(t_{*}-t)^{\frac{4}{3\sigma^{2}}}}, ~t<t_{*}$ \\ \hline
				\parbox[c][1cm]{0.5cm}{$A_{2-}$} & $\left(-\frac{2}{|\sigma|},\frac{\sqrt{4+\sigma^{2}}}{|\sigma|}\right)$ & N.H. always & $a(t)=\frac{1}{(t-t_{*})^{\frac{4}{3\sigma^{2}}}}, ~t>t_{*}$ \\ \hline
				\parbox[c][1cm]{0.5cm}{$A_{3+}$} & $\left(\frac{2}{|\sigma|},\frac{\sqrt{4-\sigma^{2}}}{|\sigma|}\right)$ & N.H. always & $a(t)=(t-t_{*})^{\frac{3\sigma^{2}}{4}}, ~t\geq t_{*}$ \\ \hline
				\parbox[c][1cm]{0.5cm}{$A_{3-}$} & $\left(-\frac{2}{|\sigma|},\frac{\sqrt{4-\sigma^{2}}}{|\sigma|}\right)$ & N.H. always & $a(t)=(t_{*}-t)^{\frac{3\sigma^{2}}{4}}, ~t\leq t_{*}$ \\ \hline
				\parbox[c][1cm]{0.5cm}{$A_{4}$} & $\left(\frac{\sigma}{(1+\omega_{m})},\sqrt{\frac{1-\omega_{m}}{1+\omega_{m}}}\right)$ & N.H. always & $a(t)=(t-t_{*})^{\frac{2}{3(1+\omega_{m})}}, ~t\geq t_{*}$ \\ \hline
			\end{tabular}
		\end{center}
		\caption{Stability condition of physically viable fixed points given in Table \ref{tab:finite_fixed_pts_exp} along with their cosmological behavior. Stability of $A_{2\pm},\,A_{3\pm}$ and $A_4$ are determined from the Jacobian eigenvalues whereas the stability of $A_{1\pm}$ are determined by examining the stability of the invariant submanifolds $y=0$ and $\Omega_m=0$ (see appendix \ref{app:stab_inv_sub}).}
		\label{tab:stab_finite_fixed_pts_exp}
	\end{table}
	
	%%%%%%%%%%%%%%%%%%%%%%%%%%
	
	The system \eqref{dynsys_exp} contains seven finite fixed points viz. $A_{1\pm},\,A_{2\pm},\,A_{3\pm}$ and $A_{4}$ (see Table \ref{tab:finite_fixed_pts_exp}). The stability and cosmological evolution corresponding to each fixed point are listed in Table \ref{tab:stab_finite_fixed_pts_exp}. The fixed points $A_{1\pm}$ that exist only for the case of a canonical scalar field, correspond to decelerated expanding and contracting phases respectively, which we also obtained for the power law potential. The rest of the finite fixed points obtained for the exponential potential have no counterpart for the power law potential. For the case of a canonical scalar field, we get another pair of expanding and contracting solutions, $A_{3\pm}$, which can be accelerated or decelerated according to $\sigma^2>\frac{4}{3}$ or $\sigma^2<\frac{4}{3}$ respectively. $A_{3\pm}$ coincide with $A_{1\pm}$ in the limit $|\sigma|\rightarrow2$. Apart from that, for the canonical case, we also have an expanding power law solution $A_4$, which is accelerated or decelerated according to $1+3\omega_{m}>0$ or $1+3\omega_{m}<0$, i.e. whether the matter component satisfies the SEC or not. For the case of a phantom scalar field, we get two solutions $A_{2\pm}$, which represent finite time singularities in the past and the future respectively. In fact, $A_{2+}$ is a phantom-dominated phase that ends, as expected, in a big-rip singularity.

	\paragraph{Fixed Points at Infinity}
	
	To get a global picture of the 2-dimensional phase space we employ the compact dynamical variables $\Bar{x},\,\Bar{y}$ as in Eq.\eqref{3D_dynvar_comp_def}. For a constant $\sigma$ Eq.\eqref{dynsys_3d} reduces to
	\begin{subequations}\label{dynsys_2d}
		\begin{eqnarray}
			&& \frac{d\Bar{x}}{d\tau}=\frac{3}{2}(1-\Bar{x}^{2})^{\frac{3}{2}}\left[(\omega_{k}-\omega_{m})\left(\frac{\Bar{x}^{2}}{1-\Bar{x}^{2}}-\sgn(\beta)\right)+\left(1+\omega_{m}-\frac{\sigma\Bar{x}\sgn(\beta)}{\sqrt{1-\Bar{x}^{2}}}\right)\frac{\Bar{y}^{2}}{1-\Bar{y}^{2}}\right],\\    
			&& \frac{d\Bar{y}}{d\tau}=\frac{3}{2}\Bar{y}(1-\Bar{y}^{2})\left[-\sigma +\frac{(\omega_{k}+1)\Bar{x}}{\sqrt{1-\Bar{x}^{2}}}-\frac{\sigma\Bar{y}^{2}}{1-\Bar{y}^{2}}\sgn(\beta)\right] \,.
		\end{eqnarray}
	\end{subequations}
	%\begin{equation}
	%\frac{d\sigma}{dN}=-3\sigma^{2}(\Gamma-1)+\frac{3\sigma[2\Xi(\omega_{k}+1)+\omega_{k}-1]}{2(2\Xi+1)(\omega_{k}+1)}\left((\omega_{k}+1)\frac{\Bar{x}}{\sqrt{1-\Bar{x}^{2}}}-\frac{\sigma \Bar{y}^{2}}{1-\Bar{y}^{2}}\right)    
	%\end{equation}
	% One can notice that the dynamical system \eqref{dynsys_2d} is symmetric under reflection against the $\Bar{y}=0$ line, i.e. under the transformation $\Bar{y}\to-\Bar{y}$. This implies that it suffices to take into consideration only the $\Bar{y}>0$ region of the phase space, since the phase portrait in the $\Bar{y}<0$ region will just be a reflection of that against $\Bar{y}=0$. From the physical point of view it means that the qualitative behavior of the $\beta X - V_{0}e^{-\lambda \phi/M_{Pl}}$ model is independent of the signature of the potential. Henceforth, therefore, we take the potential to be positive, i.e. $V_0>0$.
	As in the previous case, to regularize the dynamical system  Eq.\eqref{dynsys_2d} we redefine the phase space time variable as
	\begin{equation}\label{time_redef_exp}
		d\tau \rightarrow d\Bar{\tau} = \frac{d\tau}{(1-\Bar{x}^{2})^{\frac{1}{2}}(1-\Bar{y}^{2})}.
	\end{equation}
	%\begin{equation*}
	%\frac{d\Bar{x}}{d\Bar{N}}=\frac{3}{2}(1-\Bar{x}^{2})\left[(\omega_{k}-\omega_{m})\left(\Bar{x}^{2}-(1-\Bar{x}^{2})\sgn(\beta)\right)(1-\Bar{y}^{2})+\left((1+\omega_{m})(1-\Bar{x}^{2})-\sigma\Bar{x}\sgn(\beta)\sqrt{1-\Bar{x}^{2}}\right)\Bar{y}^{2}\sgn(V_{0})\right]    
	%\end{equation*}
	With respect to this redefined time variable, the dynamical system \eqref{dynsys_2d} can be rewritten as
	\begin{subequations}\label{dynsys_2d_new}
		\begin{eqnarray}
			&& \frac{d\Bar{x}}{d\Bar{\tau}} = \frac{3}{2}(\omega_{k}-\omega_{m})\left(\Bar{x}^{2}-(1-\Bar{x}^{2})\sgn(\beta)\right)(1-\Bar{x}^{2})(1-\Bar{y}^{2}) + \frac{3}{2}\left((1+\omega_{m})(1-\Bar{x}^{2})-\sigma\Bar{x}\sgn(\beta)\sqrt{1-\Bar{x}^{2}}\right)(1-\Bar{x}^{2})\Bar{y}^{2},\nonumber\\
			&& \label{dynsys_2d_new_x}\\
			&& \frac{d\Bar{y}}{d\Bar{\tau}} = \frac{3}{2}\Bar{y}(1-\Bar{y}^{2})\left[(-\sigma\sqrt{1-\Bar{x}^{2}} +(\omega_{k}+1)\Bar{x})(1-\Bar{y}^{2})-\sigma\Bar{y}^{2}\sqrt{1-\Bar{x}^{2}}\sgn(\beta)\right]. \label{dynsys_2d_new_y}
		\end{eqnarray}
	\end{subequations}
	The constraints in Eqs. \eqref{physviab_can} and \eqref{physviab_noncan}, which specified the physically viable region of the phase space for power law potential, remain valid also for exponential potential. This implies we can only focus on phantom scalar field ($\beta<0$) in whatever follows if we want to achieve nonsingular bounces.
	
	%%%%%%%%%%%%%%%%%%%%%%
	
	\begin{table}[H]
		\begin{center}
			\begin{tabular}{|*{4}{c|}}
				\hline
				Point & Co-ordinate $(\Bar{x},\Bar{y})$ &  Existence &  Physical viability $(\Omega_{m}\geq0)$\\ \hline
				\parbox[c][0.6cm]{0.5cm}{$B_{1+}$} & $(1,1)$ & Always & Always \\ \hline
				\parbox[c][0.6cm]{0.5cm}{$B_{1-}$} & $(-1,1)$ & Always & Always \\ \hline
				\parbox[c][0.6cm]{0.5cm}{$B_{2+}$} & $(1,0)$ & Always & Always \\ \hline
				\parbox[c][0.6cm]{0.5cm}{$B_{2-}$} & $(-1,0)$ & Always & Always \\ \hline
				\parbox[c][1.2cm]{0.5cm}{$B_{3+}$} & $\left(\frac{1+\omega_{m}}{\sqrt{(1+\omega_{m})^{2}+\sigma^{2}}},1\right)$ & \begin{tabular}{@{}c@{}} $((1+\omega_{m})^{2}+\sigma^{2}>0)\wedge$\\ $\left((\beta<0)\wedge(\sigma\leq0)\right)\vee\left((\beta>0)\wedge(\sigma\geq0)\right)$ \end{tabular} & $\sigma=0$\\ \hline
				\parbox[c][1.2cm]{0.5cm}{$B_{3-}$} & $\left(-\frac{1+\omega_{m}}{\sqrt{(1+\omega_{m})^{2}+\sigma^{2}}},1\right)$ & \begin{tabular}{@{}c@{}}$((1+\omega_{m})^{2}+\sigma^{2}>0)\wedge$\\ $\left((\beta<0)\wedge(\sigma\geq0)\right)\vee\left((\beta>0)\wedge(\sigma\leq0)\right)$ \end{tabular} & $\sigma=0$ \\ \hline
			\end{tabular}
		\end{center}
		\caption{Existence and physical viability condition for fixed points at infinity for a scalar field with kinetic term $F(X)=\beta X$ and potential $V(\phi)=V_0 e^{-\lambda\phi/M_{Pl}}$, calculated from the system \eqref{dynsys_2d_new}. }
		\label{tab:infinite_fixed_pts_exp}
	\end{table}
	
%%%%%%%%%%%%%%%%%%%%%%
	
\begin{table}[H]
\begin{center}
			\begin{tabular}{|*{4}{c|}}
				\hline
				Point   & Co-ordinates $(\Bar{x},\Bar{y})$ & Stability & Cosmology \\ \hline
				\parbox[c][1.2cm]{0.5cm}{$B_{1+}$} & $(1,1)$ &  \begin{tabular}{@{}c@{}}stable for $\sigma=0 ~\text{or} ~\{\sigma \neq 0, \sgn(\beta)\neq \sgn(\sigma)\}$ \\ \& saddle for $\{\sigma \neq 0, \sgn(\beta)= \sgn(\sigma)\}$\end{tabular} & De Sitter\\ \hline
				\parbox[c][1.2cm]{0.5cm}{$B_{1-}$} & $(-1,1)$ & \begin{tabular}{@{}c@{}} unstable for $\sigma=0 ~\text{or} ~\{\sigma \neq 0, \sgn(\beta)= \sgn(\sigma)\}$  \\ \& saddle for $\{\sigma \neq 0, \sgn(\beta)\neq \sgn(\sigma)\}$\end{tabular} & De Sitter \\ \hline
				\parbox[c][0.8cm]{0.5cm}{$B_{2+}$}   & $(1,0)$ & saddle always & $a(t)=(t-t_{*})^{\frac{2}{3(1+\omega_{m})}}, ~t\geq t_{*}$\\ \hline
				\parbox[c][0.8cm]{0.5cm}{$B_{2-}$}   & $(-1,0)$ & saddle always & $a(t)=(t_{*}-t)^{\frac{2}{3(1+\omega_{m})}}, ~t\leq t_{*}$\\ \hline
			\end{tabular}
		\end{center}
		\caption{Stability condition of physically viable fixed points given in Table \ref{tab:infinite_fixed_pts_exp} along with their cosmological behavior. The stability of these fixed points is determined by investigating the stability of the invariant submanifolds $\Bar{x}=\pm1$ and $\Bar{y}=0,1$ (see appendix \ref{app:stab_inv_sub}).}
		\label{tab:stab_infinite_fixed_pts_exp}
	\end{table}

	Fixed points for the system \eqref{dynsys_2d_new} are listed in Table \ref{tab:infinite_fixed_pts_exp}, along with their stability conditions in Table \ref{tab:stab_infinite_fixed_pts_exp}. Although from Table \ref{tab:infinite_fixed_pts_exp} it might appear that there are three different pairs of fixed points at the infinity of the phase space, the pair $B_{3\pm}$ is physically viable only for $\sigma=0$, in which case it coincides with $B_{1\pm}$. Therefore there are only two different pairs of fixed points at infinity. $B_{1\pm}$ are De-Sitter solutions whereas $B_{2\pm}$ corresponds to cosmological phases dominated by the hydrodynamic matter component. For $\omega_{m}=0$ the fixed point $B_{2+}$ can be interpreted as the matter-dominated epoch since this is a saddle and therefore always represents an intermediate phase of evolution. It might also be noted that $A_{2\pm}$ and $A_{3\pm}$ merges with $B_{1\pm}$ at the limit $|\sigma|\rightarrow\infty$.
	
	Below we provide compact 2-dimensional phase portraits for various cases. In all the plots we also specify the physically viable region as calculated by the constraints \eqref{physviab_can} or \eqref{physviab_noncan} and the regions where the NEC and SEC are satisfied. The NEC and SEC are determined in terms of the ``effective'' equation of state parameter defined in Eq.\eqref{eq:weff}.
	
	%%%%%%%%%%%%%%%%%%%%%%%%%%%%%%%%%%%%%%%%%%%%%%%%%%%%%%%%%%%%%%%%%%%%%%%%%%%%%%%%%%%%%%%%%%%%%%%%%%%%%%%%%%%%
	\begin{figure}[H]
		\centering
		\minipage{0.35\textwidth}
		\includegraphics[width=\textwidth]{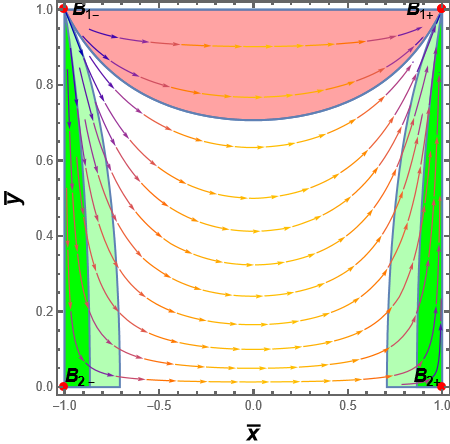}
		\endminipage
		\caption{The phase space portrait for $\omega_{m}=0$ with $\sigma=0$, $\beta<0$ and $V_{0}>0$. Satisfaction of the NEC is shown in light green, the satisfaction of the SEC in dark green and the red area denotes a non-physical part of the phase space.}\label{FIG.3}
	\end{figure}
	%%%%%%%%%%%%%%%%%%%%%%%%%%%%%%%%%%%%%%%%%%%%%%%%%%%%%%%%%%%%%%%
	%%%%%%%%%%%%%%%%%%%%%%%%%%%%%%%%%%%%%%%%%%%%%%%%%%%%%%%
 \begin{figure}[H]
     \centering
     \includegraphics[scale=0.6]{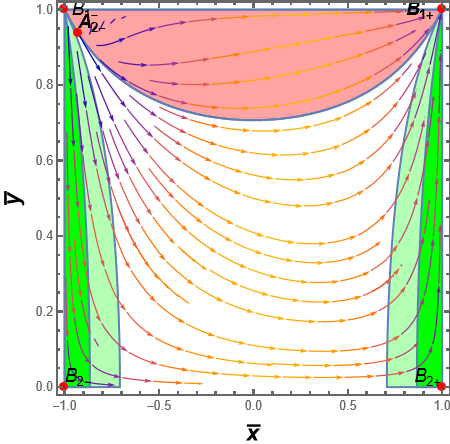}\hspace{1.5cm}
     \includegraphics[scale=0.6]{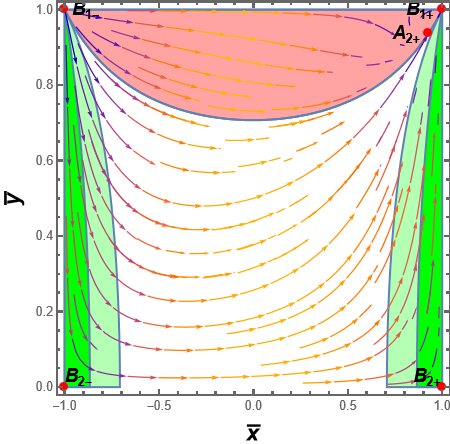}\vspace{0.8cm}
     \includegraphics[scale=0.6]{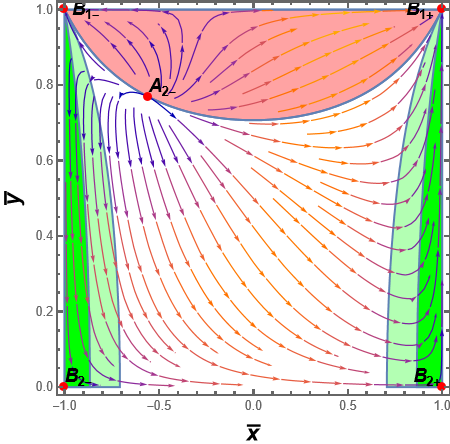}\hspace{1.5cm}
     \includegraphics[scale=0.6]{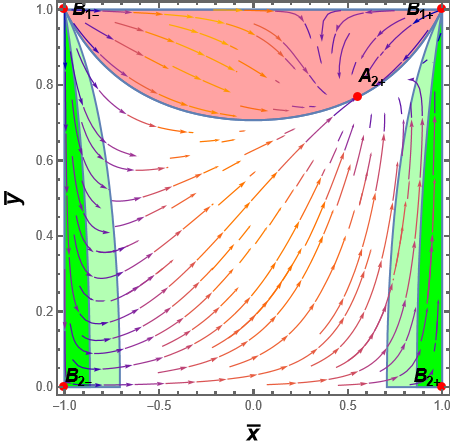}
     \caption{The phase space portrait for $\sigma=0.8$ (upper left panel) $\sigma=-0.8$ (upper right panel) $\sigma=3$ (lower left panel) $\sigma=-3$ (lower right panel) with $\omega_{m}=0$, $\beta<0$ and $V_{0}>0$. Satisfaction of the NEC is shown in light green, the satisfaction of the SEC in dark green and the red area denotes a non-physical part of the phase space.}\label{FIG.4}
 \end{figure}
\begin{figure}[H]
		\centering
		\minipage{0.35\textwidth}
		\includegraphics[width=\textwidth]{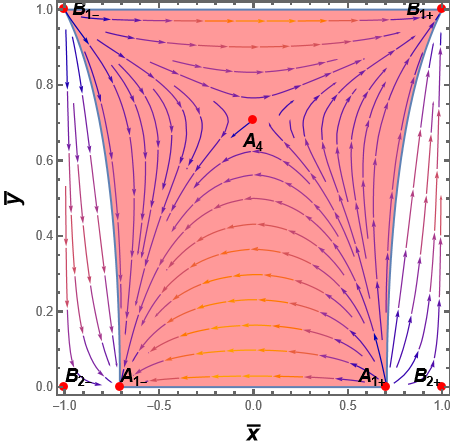}
		\endminipage
		\caption{The phase space portrait for $\omega_{m}=0$ with $\sigma=0$, $\beta>0$ and $V_{0}>0$. The shaded region represents the non-physical viable region.}\label{FIG.5}
	\end{figure}
		%%%%%%%%%%%%%%%%%%%%%%%%%%%%%%%%%%%%%%%%%%%%%%%%%%%%%%%%%%%%%%%
	%%%%%%%%%%%%%%%%%%%%%%%%%%%%%%%%%%%%%%%%%%%%%%%%%%%%%%%

 From Fig.\ref{FIG.3} and Fig.\ref{FIG.4}, it can be seen that for a phantom scalar field ($\beta<0$), all the phase trajectories within the physically viable region show a nonsingular bounce. Nonsingular bounce is a generic feature in this case. the figures also confirm the violation of both NEC and SEC to achieve a bounce. For the sake of completeness, we also show the phase portrait for a particular case for $\beta>0$ in Fig.\ref{FIG.5}, which shows that all the trajectories representing a nonsingular bounce fall within the physically non-viable region, as expected.    

%%%%%%%%%%%%%%%%%%%%%%%%%%%%%%%%%%%%%%%%%%%%%%%%%%%%%%%%%%%%%%%%%%%%%%%%%%%%%%%%%%%%%%%%%%%%%%%%%%%%%%%%%%%%%%%%%%%%%%%%%%%%%%%%%%%
    
\subsection{\bf{Specific Case II: Noncanonical scalar field $F(X)=\beta X^{m}$ \quad ($m\neq1$)} }\label{subsec:general}

In this section, we generalize our scalar field Lagrangian to the noncanonical form
\begin{equation}
\mathcal{L}(\phi,X) = \beta X^m - V(\phi)\,,  \label{langragian_2}  
\end{equation}
 Since $\rho_{\phi}+P_{\phi}=2XF_X=2m\beta X^m$ and $X\geq0$, it corresponds to a non-phantom scalar field when $m\beta>0$ and to a phantom scalar field when $m\beta<0$. Again, one can notice from equation \eqref{eq:Raychaudhuri} that for $\dot{H}>0$ near the bounce one must necessarily need $m\beta<0$, i.e. a phantom scalar field. We particularly concentrate on the case $m\neq1$, as the case $m=1$ has been considered previously. 
 
 One can calculate that, in this case,
\begin{equation}
  \rho_{k}=(2m-1)\beta X^{m}.  
\end{equation}
We keep on using the same compact and non-compact dynamical variables as introduced earlier. The physical viability of the model requires the condition $\Omega_{m}\geq0$.  In terms of the non-compact dynamical variables $(x,y)$, it can be written as
 \begin{equation}
    \lbrace (x,y,\sigma) \in \mathbb{R}^3: x^2-y^2-\sgn((2m-1)\beta)\geq 0\rbrace\,.\label{viable:gen}
 \end{equation}
In terms of the compact dynamical variables $(\Bar{x},\Bar{y})$ defined in \eqref{3D_dynvar_comp_def}, one can write
    \begin{equation}
        \frac{\bar{x}^{2}}{1-\bar{x}^{2}}-\frac{\bar{y}^{2}}{1-\bar{y}^{2}}-\sgn{((2m-1)\beta)}=\Omega_{m}.
    \end{equation}
Note that the physical viability conditions for $m\neq1$ differ from that for $m=1$ only in the fact that $sgn(\beta)$ is replaced by $sgn((2m-1)\beta)$. When $\sgn((2m-1)\beta)=1$ one needs
\begin{equation}
        \frac{\bar{x}^{2}}{1-\bar{x}^{2}}-\frac{\bar{y}^{2}}{1-\bar{y}^{2}}-1=\Omega_{m}\geq 0.
\end{equation}
This leads to the necessary and sufficient conditions for the physical viability respectively as 
\begin{equation}
         \Bar{y}^{2}\leq 2-\frac{1}{\Bar{x}^{2}}, \qquad \Bar{x}^{2}\geq \frac{1}{2},   
\end{equation}
which are the same conditions obtained in Eq.\eqref{physviab_can}. When $\sgn((2m-1)\beta)=-1$ one needs
\begin{equation}
        \frac{\bar{x}^{2}}{1-\bar{x}^{2}}-\frac{\bar{y}^{2}}{1-\bar{y}^{2}}+1=\Omega_{m}\geq 0.
\end{equation}
This leads to the necessary and sufficient conditions as
\begin{equation}
         \Bar{y}^{2}\leq \frac{1}{2-\Bar{x}^{2}},  
\end{equation}
which is the same condition as obtained in Eq.\eqref{physviab_noncan}. Just like one could conclude for the case of $F(X)=\beta X$ that a physically viable nonsingular bounce requires $\beta<0$, one can conclude here that for the generic case $F(X)=\beta X^m$, a physically viable nonsingular bounce requires $(2m-1)\beta<0$. This implies the following parameter range
\begin{equation}
    \{(m>1/2)\wedge(\beta<0)\} \vee \{(m<1/2)\wedge(\beta>0)\}\,. 
\end{equation}
As we have already seen from \eqref{eq:Raychaudhuri} that achieving a nonsingular bounce necessarily requires a phantom scalar field, i.e. $m\beta<0$, together with the condition $(2m-1)\beta<0$ this slightly constrain the parameter range
\begin{equation}\label{param_space}
    \{(m>1/2)\wedge(\beta<0)\} \vee \{(m<0)\wedge(\beta>0)\}\,. 
\end{equation}
In particular, the parameter range $0\leq m\leq 1/2$ is not allowed if we want to achieve a nonsingular bounce.

\subsubsection{\bf{Power Law Potential $V(\phi)=V_{0}\phi^{n}$}}\label{subsec:modelI_m}

 For the kinetic term $F(X)=\beta X^{m}$ and the potential $V(\phi)=V_{0}\phi^{n}$, we have
	\begin{equation}
		\Xi=m-1, \qquad \Gamma=1-\frac{1}{n}, \qquad \omega_{k}=\frac{1}{2m-1}, \qquad \rho_{k}=(2m-1)\beta X^{m}\,.
	\end{equation}

\paragraph{Finite Fixed Point Analysis}

In this case, the system \eqref{dynsys} reduces to
   \begin{subequations}\label{dynsys_pow_m}
		\begin{eqnarray}
			&& \frac{dx}{d\tau} = \frac{3}{2}x\left[\left(\frac{2m}{2m-1}\right)x - \sigma y^2 \sgn(V)\sgn[(2m-1)\beta]\right] - \frac{3}{2}\left[\left(\frac{1}{2m-1}-\omega_{m}\right)\sgn[(2m-1)\beta] + (1+\omega_{m})(x^{2}-y^2 \sgn(V))\right],\label{eq:ds_x_pow_m} \nonumber\\
            && \\
			&& \frac{dy}{d\tau} = \frac{3}{2}y\left[-\sigma + \left(\frac{2m}{2m-1}\right)x - \sigma y^2 \sgn(V)\sgn[(2m-1)\beta]\right],\label{eq:ds_y_pow_m}\\
			&& \frac{d\sigma}{d\tau} = \frac{3}{n}\sigma^{2}+3\sigma\frac{(2m-3)m+1}{(4m-2)m}\left(\left(\frac{2m}{2m-1}\right)x-\sigma y^{2}\sgn(V)\right)\,.\label{eq:ds_s_pow_m}
		\end{eqnarray}
       \end{subequations}	
	%%%%%%%%%%%%%%%%%%%%%%%%%%
 The dynamical system is independent of the parameter $\lambda$. The above system reduces to the system \eqref{dynsys_pow} for $m=1$. As in the case of $m=1$, the system is symmetric under reflection around $y=0$, which happens to be an invariant submanifold. Therefore it suffices to consider only the part of the phase space given by
\begin{equation}
    \lbrace (x,y,\sigma) \in \mathbb{R}^3: y \geq 0, x^2-y^2-\sgn((2m-1)\beta)\geq 0\rbrace\,.\label{viable:gen_m}
 \end{equation}

\begin{table}[H]
		\begin{center}
			\begin{tabular}{|*{4}{c|}} 
				\hline
				Point & Co-ordinate $(x,y,\sigma)$ &  Existence &  Physical viability $(\Omega_{m}\geq0)$\\ \hline
				\parbox[c][0.6cm]{0.5cm}{$A_{1+}$} & $(1,0,0)$ & $(2m-1)\beta>0\wedge m\neq0$ & Always \\ \hline
				\parbox[c][0.6cm]{0.5cm}{$A_{1-}$} & $(-1,0,0)$ & $(2m-1)\beta>0\wedge m\neq0$ & Always \\ \hline
			\end{tabular}
		\end{center}
		\caption{Existence and physical viability conditions for finite fixed points for a non-canonical scalar field with kinetic term $F(X)=\beta X^m$ ($m\neq 1$) and potential $V(\phi)=V_0 \phi^n$, calculated from the system \eqref{dynsys_pow_m}.}
		\label{T:ex_pow_nc_m}
	\end{table} 
 	\begin{table}[H]
		\begin{center}
			\begin{tabular}{|*{4}{c|}}
				\hline
				Point   & Co-ordinates $(x,y,\sigma)$ & Stability & Cosmology \\ \hline
				\parbox[c][1.6cm]{0.5cm}{$A_{1+}$} & $(1,0,0)$ &  \begin{tabular}{@{}c@{}} Stable for $0<m<\frac{1}{2}$ \&\\ unstable for $\frac{1}{2}<m<1$\\ saddle otherwise\end{tabular} & $a(t)=(t-t_{*})^{\frac{2m-1}{3m}}, ~t\geq t_{*}$ \\ \hline
				\parbox[c][1.6cm]{0.5cm}{$A_{1-}$} & $(-1,0,0)$ & \begin{tabular}{@{}c@{}} Stable for $\frac{1}{2}<m<1$ \&\\ unstable for $0< m<\frac{1}{2}$\\ saddle otherwise\end{tabular} & $a(t)=(t_{*}-t)^{\frac{2m-1}{3m}}, ~t\leq t_{*}$ \\ \hline
			\end{tabular}
		\end{center}
		\caption{Stability condition of physically viable critical points given in Table \ref{T:ex_pow_nc_m} along with their cosmological behaviour.}
  \label{T:st_pow_nc_m}
	\end{table}
The system \eqref{dynsys_pow_m} contains two finite fixed points $A_{1\pm}$, which exist only for non-phantom fields (see Table \ref{T:ex_pow_nc_m}). The stabilities and corresponding cosmologies are given in Table \ref{T:st_pow_nc_m}. Since these critical points exist only when $(2m-1)\beta>0$. When $(2m-1)\beta>0$, they cannot give rise to physically viable nonsingular bouncing trajectories.
       
\paragraph{Fixed Points at Infinity}
 %       \begin{subequations}\label{dynsys_3d_m}
	% 	\begin{eqnarray}
	% 		\frac{d\Bar{x}}{d\tau}&=&\frac{3}{2}(1-\Bar{x}^{2})^{\frac{3}{2}}\left[(\omega_{k}-\omega_{m})\left(\frac{\Bar{x}^{2}}{1-\Bar{x}^{2}}-\sgn[(2m-1)\beta]\right)+\left(1+\omega_{m}-\frac{\Bar{\sigma}\Bar{x}\sgn[(2m-1)\beta]}{\sqrt{1-\Bar{\sigma}^2}\sqrt{1-\Bar{x}^{2}}}\right)\frac{\Bar{y}^{2}\sgn(V)}{1-\Bar{y}^{2}}\right],\\    
	% 		\frac{d\Bar{y}}{d\tau}&=&\frac{3}{2}\Bar{y}(1-\Bar{y}^{2})\left[-\frac{\Bar{\sigma}}{\sqrt{1-\Bar{\sigma}^2}} +\frac{(\omega_{k}+1)\Bar{x}}{\sqrt{1-\Bar{x}^{2}}}-\frac{\Bar{\sigma}\Bar{y}^{2}\sgn(V)}{\sqrt{1-\Bar{\sigma}^2}(1-\Bar{y}^{2})}\sgn[(2m-1)\beta]\right] \,,\\
	% 		\frac{d\Bar{\sigma}}{d\tau}&=&\left(1-\Bar{\sigma}^{2}\right)^{\frac{3}{2}}\bigg[\frac{3}{n} \frac{\Bar{\sigma}^{2}}{(1-\Bar{\sigma}^2)}+\frac{2\Xi(\omega_{k}+1)+\omega_{k}-1}{(4\Xi+2)(\omega_{k}+1)}\frac{3\Bar{\sigma}}{\sqrt{1-\Bar{\sigma}^{2}}}\bigg(\frac{(\omega_{k}+1)\Bar{x}}{\sqrt{1-\Bar{x}^{2}}}-\frac{\Bar{\sigma}}{\sqrt{1-\Bar{\sigma}^{2}}}\frac{\Bar{y}^{2}}{1-\Bar{y}^{2}}\bigg)\bigg]\,. 
	% 	\end{eqnarray}
	% \end{subequations}

In terms of the compact dynamical variables $(\Bar{x},\Bar{y},\Bar{\sigma})$ defined in equation \eqref{3D_dynvar_comp_def} and the redefined time variable defined in equation \eqref{time_redef_pow}, the dynamical system for $F(X)=\beta X^{m},\,V=V_0\phi^n$ can be rewritten in the following form 
\begin{subequations}\label{dynsys_3d_pow_m}
\begin{eqnarray}
\frac{d\Bar{x}}{d\Bar{\tau}} &=& \frac{3}{2}\left(\frac{1}{2m-1}-\omega_{m}\right)\left(\Bar{x}^{2}-(1-\Bar{x}^{2})\sgn[(2m-1)\beta]\right) (1-\Bar{x}^{2})(1-\Bar{y}^{2})\sqrt{1-\Bar{\sigma}^2}+\frac{3}{2}\big((1+\omega_{m}) (1-\Bar{x}^{2}) \sqrt{1-\Bar{\sigma}^2}\nonumber\\
&&-\Bar{\sigma}\Bar{x}\sgn[(2m-1)\beta]\sqrt{1-\Bar{x}^{2}}\big)(1-\Bar{x}^{2})\Bar{y}^{2}, \label{dynsys_3d_pow_x_m}\\ 
\frac{d\Bar{y}}{d\Bar{\tau}}& =& \frac{3}{2}\Bar{y}(1-\Bar{y}^{2})\left[\left(-\Bar{\sigma}\sqrt{1-\Bar{x}^{2}}+\left(\frac{2m}{2m-1}\right)\Bar{x}\sqrt{1-\Bar{\sigma}^2}\right)(1-\Bar{y}^{2})-\Bar{\sigma}\Bar{y}^{2}\sqrt{1-\Bar{x}^{2}}\sgn[(2m-1)\beta]\right]\,, \label{dynsys_3d_pow_y_m}\\
\frac{d\Bar{\sigma}}{d\Bar{\tau}}&=&(1-\Bar{\sigma}^{2})\left[\frac{3}{n}\Bar{\sigma}^{2}\sqrt{1-\Bar{x}^{2}}(1-\Bar{y}^{2})+3\Bar{\sigma}\frac{(2m-3)m+1}{(4m-2)m}\left(\left(\frac{2m}{2m-1}\right)\Bar{x}\sqrt{1-\Bar{\sigma}^{2}}(1-\Bar{y}^{2})-\Bar{\sigma}\Bar{y}^{2}\sqrt{1-\Bar{x}^{2}}\right)\right]\,.\label{dynsys_3d_pow_s_m}
% \frac{d\Bar{x}}{d\Bar{\tau}} &=& \frac{3}{2}(\omega_{k}-\omega_{m})\left(\Bar{x}^{2}-(1-\Bar{x}^{2})\sgn[(2m-1)\beta]\right)(1-\Bar{x}^{2})(1-\Bar{y}^{2}) \sqrt{1-\Bar{\sigma}^2} + \frac{3}{2}\big((1+\omega_{m})(1-\Bar{x}^{2}) \sqrt{1-\Bar{\sigma}^2}\nonumber\\
% &&-\Bar{\sigma}\Bar{x}\sgn[(2m-1)\beta]\sqrt{1-\Bar{x}^{2}}\big)(1-\Bar{x}^{2})\Bar{y}^{2} ,\label{dynsys_3d_new_x_m}\\
% \frac{d\Bar{y}}{d\Bar{\tau}}& =& \frac{3}{2}\Bar{y}(1-\Bar{y}^{2})\left[(-\Bar{\sigma}\sqrt{1-\Bar{x}^{2}} +(\omega_{k}+1)\Bar{x}\sqrt{1-\Bar{\sigma}^2})(1-\Bar{y}^{2})-\Bar{\sigma}\Bar{y}^{2}\sqrt{1-\Bar{x}^{2}}\sgn[(2m-1)\beta]\right]\,, \label{dynsys_3d_new_y_m}\\
% \frac{d\Bar{\sigma}}{d\Bar{\tau}}&=&(1-\Bar{\sigma}^{2})\left[\frac{3}{n}\Bar{\sigma}^{2}\sqrt{1-\Bar{x}^{2}}(1-\Bar{y}^{2})+\frac{3\Bar{\sigma}[2\Xi(\omega_{k}+1)+\omega_{k}-1]}{(4\Xi+2)(\omega_{k}+1)}\left((\omega_{k}+1)\Bar{x}\sqrt{1-\Bar{\sigma}^{2}}(1-\Bar{y}^{2})-\Bar{\sigma}\Bar{y}^{2}\sqrt{1-\Bar{x}^{2}}\right)\right]\,.\label{dynsys_3d_new_s_m}
\end{eqnarray}
\end{subequations}	

	\begin{table}[H]
		\begin{center}
			\begin{tabular}{|*{4}{c|}}
				\hline
				Point & Co-ordinate $(\Bar{x},\Bar{y},\Bar{\sigma})$ &  Existence &  Physical viability $(\Omega_{m}\geq0)$\\ \hline
				\parbox[c][0.6cm]{0.5cm}{$B_{1+}$} & $(1,1,\Bar{\sigma})$ & Always & Always \\ \hline
				\parbox[c][0.6cm]{0.5cm}{$B_{1-}$} & $(-1,1,\Bar{\sigma})$ & Always & Always \\ \hline         
                \parbox[c][0.6cm]{0.5cm}{$B_{2+}^{a}$} & $(1,0,0)$ & Always & Always \\ \hline 
                \parbox[c][0.6cm]{0.5cm}{$B_{2-}^{a}$} & $(-1,0,0)$ & Always & Always \\ \hline
				\parbox[c][1.2cm]{0.5cm}{$B_{3+}$} & $\left(\Bar{x},0,1\right)$ & Always & \begin{tabular}{@{}c@{}}$\frac{1}{2}\leq \Bar{x}^{2}\leq 1$ if $\sgn((2m-1)\beta)=1$ \\ $\Bar{x}^{2}\leq 1$ if $\sgn((2m-1)\beta)=-1$\end{tabular}\\ \hline
				\parbox[c][1.2cm]{0.5cm}{$B_{3-}$} & $\left(\Bar{x},0,-1\right)$ & Always & \begin{tabular}{@{}c@{}}$\frac{1}{2}\leq \Bar{x}^{2}\leq 1$ if $\sgn((2m-1)\beta)=1$ \\ $\Bar{x}^{2}\leq 1$ if $\sgn((2m-1)\beta)=-1$\end{tabular}\\ \hline
				\parbox[c][1.2cm]{0.5cm}{$B_{4+}^{a}$} & $\left(\frac{(1+\omega_{m})\sqrt{1-\Bar{\sigma}^2}}{\sqrt{(1+\omega_{{m}})^2-\Bar{\sigma}^2\omega_{m}(\omega_{m}+2)}},1,1\right)$ & \begin{tabular}{@{}c@{}}$\Bar{\sigma}^2\leq \frac{(1+\omega_{m})^2}{\omega_{m}(\omega_{m}+2)}$ \\ \end{tabular} &$\Bar{\sigma}=0$\\ \hline
                \parbox[c][1.2cm]{0.5cm}{$B_{4+}^{b}$} & $\left(\frac{(1+\omega_{m})\sqrt{1-\Bar{\sigma}^2}}{\sqrt{(1+\omega_{{m}})^2-\Bar{\sigma}^2\omega_{m}(\omega_{m}+2)}},1,-1\right)$ & \begin{tabular}{@{}c@{}}$\Bar{\sigma}^2\leq \frac{(1+\omega_{m})^2}{\omega_{m}(\omega_{m}+2)}$ \\ \end{tabular} &$\Bar{\sigma}=0$\\ \hline
				\parbox[c][1.2cm]{0.5cm}{$B_{4+}^{c}$} & $\left(\frac{(1+\omega_{m})\sqrt{1-\Bar{\sigma}^2}}{\sqrt{(1+\omega_{{m}})^2-\Bar{\sigma}^2\omega_{m}(\omega_{m}+2)}},1,0\right)$ & \begin{tabular}{@{}c@{}}$\Bar{\sigma}^2\leq \frac{(1+\omega_{m})^2}{\omega_{m}(\omega_{m}+2)}$ \\  \end{tabular} &$\Bar{\sigma}=0$\\ \hline
				\parbox[c][1.2cm]{0.5cm}{$B_{4-}^{a}$} & $\left(-\frac{(1+\omega_{m})\sqrt{1-\Bar{\sigma}^2}}{\sqrt{(1+\omega_{{m}})^2-\Bar{\sigma}^2\omega_{m}(\omega_{m}+2)}},1,1\right)$ & \begin{tabular}{@{}c@{}}$\Bar{\sigma}^2\leq \frac{(1+\omega_{m})^2}{\omega_{m}(\omega_{m}+2)}$ \\ \end{tabular} &$\Bar{\sigma}=0$\\ \hline
                \parbox[c][1.2cm]{0.5cm}{$B_{4-}^{b}$} & $\left(-\frac{(1+\omega_{m})\sqrt{1-\Bar{\sigma}^2}}{\sqrt{(1+\omega_{{m}})^2-\Bar{\sigma}^2\omega_{m}(\omega_{m}+2)}},1,-1\right)$ & \begin{tabular}{@{}c@{}}$\Bar{\sigma}^2\leq \frac{(1+\omega_{m})^2}{\omega_{m}(\omega_{m}+2)}$ \\ \end{tabular} &$\Bar{\sigma}=0$\\ \hline
				\parbox[c][1.2cm]{0.5cm}{$B_{4-}^{c}$} & $\left(-\frac{(1+\omega_{m})\sqrt{1-\Bar{\sigma}^2}}{\sqrt{(1+\omega_{{m}})^2-\Bar{\sigma}^2\omega_{m}(\omega_{m}+2)}},1,0\right)$ & \begin{tabular}{@{}c@{}}$\Bar{\sigma}^2\leq \frac{(1+\omega_{m})^2}{\omega_{m}(\omega_{m}+2)}$ \\  \end{tabular} &$\Bar{\sigma}=0$\\ \hline
				\parbox[c][1.2cm]{0.5cm}{$B_{5+}$} & $\left(0,\frac{1}{\sqrt{2}},1\right)$ & $(2m-1)\beta<0$ & Always \\ \hline
				\parbox[c][1.2cm]{0.5cm}{$B_{5-}$} &$\left(0,\frac{1}{\sqrt{2}},-1\right)$ & $(2m-1)\beta<0$ & Always \\ \hline
			\end{tabular}
		\end{center}
		\caption{Existence and physical viability condition for fixed points at infinity for a noncanonical scalar field with kinetic term $F(X)=\beta X^m$ ($m\neq 1$) and potential $V(\phi)=V_0 \phi^n$, calculated from the system (\ref{dynsys_3d_pow_m}). }
  \label{T:ex_pow_com_m}
	\end{table}
	%%%%%%%%%%%%%%%%%%%%%%%
% {\color{red} Stability of the lines of fixed points $B_{1\pm}$ can be determined by investigating the stability of the invariant submanifolds $\Bar{x}=\pm1$ and $\Bar{y}=1$ (see appendix \ref{app:stab_inv_sub_general}). Fixed points $B_{2\pm}^{a}$ are on the intersection of the invariant submanifold $\Bar{x}=\pm1$ and $\Bar{y}=0$. One can notice that the vicinity of $\Bar{x}=1$ for $\Bar{y}=0$ shows attracting behavior for $m>1/2$ and the vicinity of $\Bar{y}=0$ for $\Bar{x}=1$ shows repelling behavior. similar kind of behaviour one can notice for $\Bar{x}=-1$ and $\Bar{y}=0$ i.e. opposite behaviour for $m>1/2$, hence the fixed point $B_{2\pm}^{a}$ always shows the saddle behaviour. Fixed points $B_{3\pm}$ lies on the intersection of $\Bar{y}=0$ and $\Bar{\sigma}=\pm 1$ stability can checked from the stability of invariant submanifold $\Bar{y}=0$ and $\Bar{\sigma}=\pm 1$. Stability of $B_{5\pm}$ is determined from the eigenvalues of the Jacobian matrix.}{\color{blue} I think this paragraph may not be necessary here} {\color{purple} after the stability table, I wanted to add this, but just to adjust the space I added here. You can remove it dada.} 

\begin{figure}[H]
		\centering
		\minipage{0.45\textwidth}
		\includegraphics[width=\textwidth]{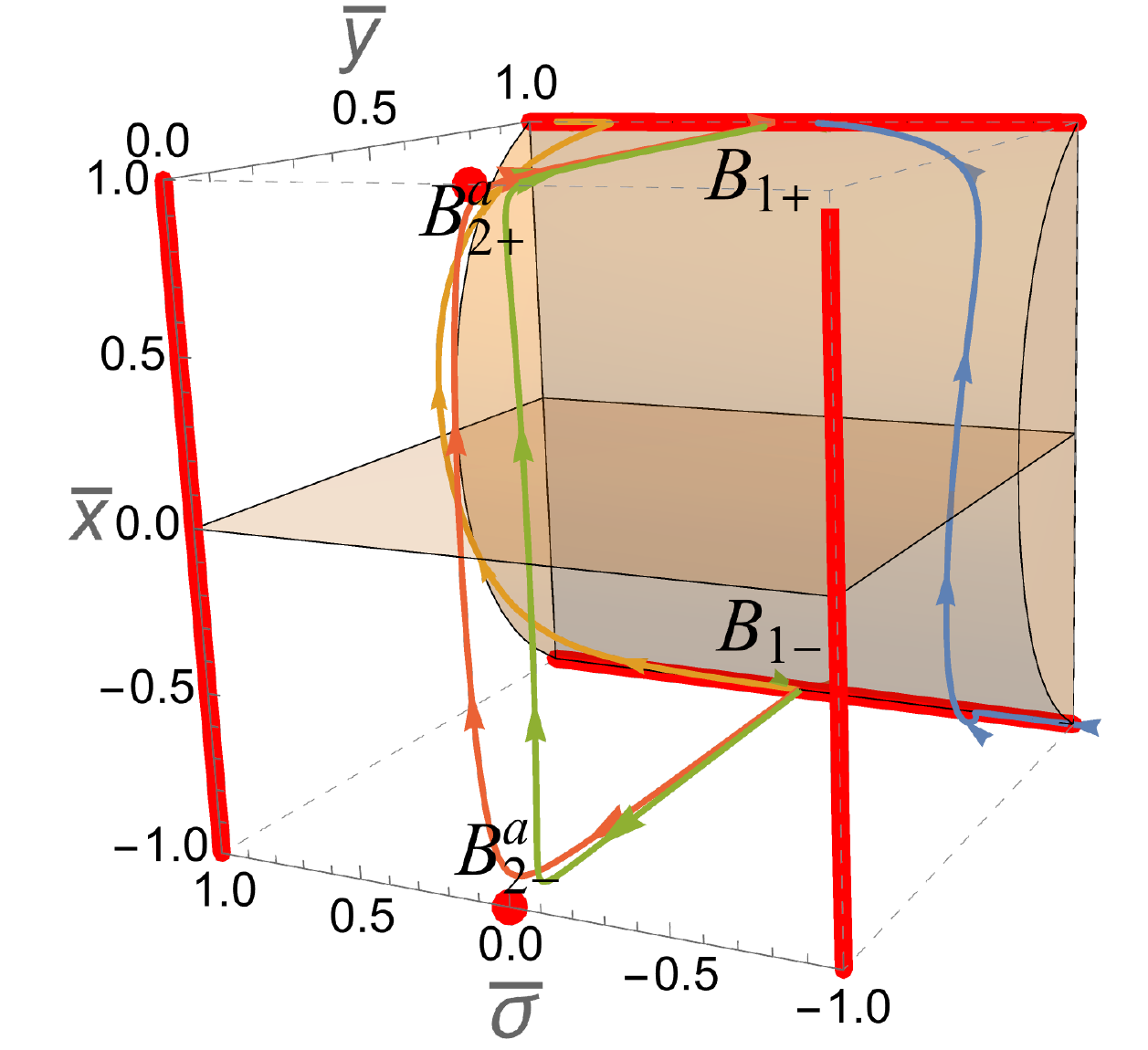}
		\endminipage
		\minipage{0.45\textwidth}
		\includegraphics[width=\textwidth]{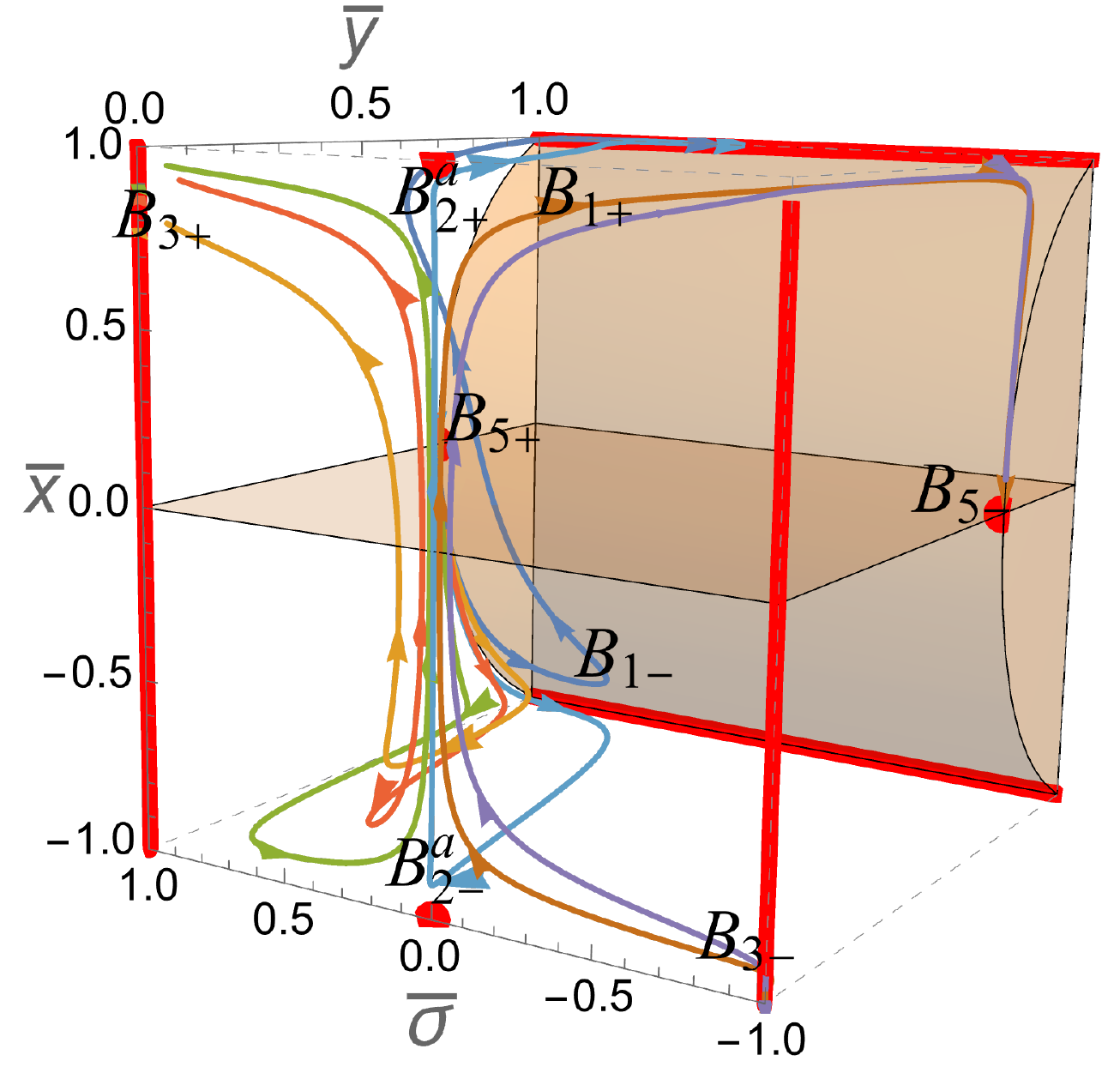}
		\endminipage
		\caption{Phase portrait of the system (\ref{dynsys_3d_pow_m}) with shaded region represents a non-physical region of the phase space. The left plot is for $n=-5, m=\frac{2}{3}$ and the right plot is for $n=4, m=3$ moreover $\sgn((2m-1)\beta)=-1$.}
        \label{FIG.6}
	\end{figure}

The system \eqref{dynsys_3d_pow_m} presents six invariant submanifolds $\bar{x}=\pm1,\,\bar{y}=0,1$ and $\bar{\sigma}=\pm1$. Their stability is calculated in appendix \ref{app:stab_inv_sub_general}). The fixed points for the system \eqref{dynsys_3d_pow_m} are presented in Table \ref{T:ex_pow_com_m}. In the presence of pressureless dust, their stability conditions and corresponding cosmologies are presented in Table \ref{T:st_pow_com_m}. The lines of fixed points $B_{1\pm},\,B_{3\pm}$ and the isolated fixed points $B_{5\pm}$ are the same ones that we obtained earlier for the case $m=1$ (see table \ref{tab:infinite_fixed_pts_pow}). However, unlike in the case of $m=1$, the entire lines $B_{2\pm}$ and $B_{4\pm}$ are not lines of fixed points when $m\neq1$. Instead, in this case, we only get two isolated fixed points $B^a_{2\pm}$ that lie on the line $B_{2\pm}$ respectively, and six isolated fixed points $B^{a,b,c}_{4\pm}$ that lie on the line $B_{4\pm}$ respectively. The points are listed in Table \ref{T:ex_pow_com_m}. Moreover, $B_{4\pm}^{a,b,c}$ are physically viable only for $\bar{\sigma}=0$, in which case they fall back into the lines of fixed points $B_{1\pm}$ respectively. Therefore at the infinity of the phase space there are only two lines of fixed points $B_{1\pm},\,B_{3\pm}$ and two pairs of isolated fixed points $B_{2\pm}^a$ and $B_{5\pm}$, whose nature of stability and the corresponding cosmology is listed in Table \ref{T:st_pow_com_m}. The cosmologies are the same as we have earlier obtained for $m=1$, as expected. Fig.\ref{FIG.6} presents the 3D phase portrait of the system \eqref{dynsys_3d_pow_m} in the compact space for two cases, showing different types of physically viable nonsingular bouncing trajectories.

 \begin{table}[H]
		\begin{center}
			\begin{tabular}{|*{4}{c|}} 
				\hline
				Point   & Co-ordinates $(\Bar{x},\Bar{y},\Bar{\sigma})$ & Stability & Cosmology \\ \hline
				\parbox[c][1.6cm]{0.5cm}{$B_{1+}$} & $(1,1,\Bar{\sigma})$ & \begin{tabular}{@{}c@{}} Stable for \\
                $\left(m>\frac{1}{2}\right)\wedge\left(\sgn(\Bar{\sigma})\neq\sgn\beta\right)$\\ 
                or $\left(m>\frac{1}{2}\right)\wedge(\Bar{\sigma}=0)$ \\ saddle otherwise \end{tabular} & De Sitter\\ \hline
				\parbox[c][1.6cm]{0.5cm}{$B_{1-}$} &  $(-1,1,\Bar{\sigma})$& \begin{tabular}{@{}c@{}} Unstable for \\ $\left(m>\frac{1}{2}\right)\wedge\left(\sgn(\Bar{\sigma})=\sgn\beta\right)$ \\or $\left(m>\frac{1}{2}\right)\wedge(\Bar{\sigma}=0)$ \\ saddle otherwise \end{tabular} & De Sitter \\ \hline
                \parbox[c][1cm]{0.5cm}{$B_{2+}^{a}$} & $(1,0,0)$ & Saddle always & $a(t)=(t-t_{*})^{\frac{2}{3}}, ~t\geq t_{*}$ \\ \hline      \parbox[c][1cm]{0.5cm}{$B_{2-}^{a}$} & $(-1,0,0)$ & Saddle always & $a(t)=(t-t_{*})^{\frac{2}{3}}, ~t\leq t_{*}$ \\ \hline
				\parbox[c][1.2cm]{0.5cm}{$B_{3+}$} & $\left(\Bar{x},0,1\right)$ & \begin{tabular}{@{}c@{}} Stable for $n>0$ \\ saddle otherwise \end{tabular} & Depending on $\Bar{x}$ and $\sgn(2m-1)\beta$ \\ \hline
				\parbox[c][1.2cm]{0.5cm}{$B_{3-}$} & $\left(\Bar{x},0,-1\right)$ & \begin{tabular}{@{}c@{}} Unstable for $n>0$ \\ saddle otherwise \end{tabular} & Depending on $\Bar{x}$ and $\sgn(2m-1)\beta$ \\ \hline
				\parbox[c][1.2cm]{0.5cm}{$B_{5+}$} & $\left(0,\frac{1}{\sqrt{2}},1\right)$ & \begin{tabular}{@{}c@{}} Unstable for $\frac{3(mn-n-2m)}{2mn}\geq 0$ \\ saddle otherwise \end{tabular} & $a(t)=$ constant\\ \hline
				\parbox[c][1.2cm]{0.5cm}{$B_{5-}$} & $\left(0,\frac{1}{\sqrt{2}},-1\right)$ & \begin{tabular}{@{}c@{}} Stable for $\frac{3(mn-n-2m)}{2mn}\geq 0$ \\ saddle otherwise \end{tabular} & $a(t)=$ constant\\ \hline
			\end{tabular}
		\end{center}
		\caption{Stability condition and the cosmology of the fixed points given in Table \ref{T:ex_pow_com_m} in presence of pressureless dust $(\omega_m=0)$. Stability of the lines of fixed points $B_{1\pm}$ and $B_{3\pm}$ can be determined by investigating the stability of the invariant submanifolds $\Bar{x}=\pm1$, $\Bar{y}=0,1$ and $\Bar{\sigma}\pm 1$ (see appendix \ref{app:stab_inv_sub_general}). }
  \label{T:st_pow_com_m}
	\end{table}

The phase trajectories in the right panel of Fig.\ref{FIG.6} shows nonsingular bouncing trajectories for the case $F(X)=\beta X^3\,(\beta<0),\,V(\phi)=V_{0}\phi^4$. Several different types of bouncing trajectories are possible, e.g. trajectories connecting a contracting De-Sitter phase to an expanding De-Sitter phase ($B_{1-}\rightarrow B_{1+}$), trajectories connecting a static and a De-Sitter phase ($B_{1-}\rightarrow B_{5-}$ and $B_{1+}\rightarrow B_{5+}$) and trajectories connecting $B_{3\pm}$ with $B_{1\pm}$ or $B_{5\pm}$. However, one cannot say this is the generic behaviour of phase trajectories for $F(X)=\beta X^3\,(\beta<0),\,V(\phi)=V_{0}\phi^4$, as $B_{1-}$ and $B_{1+}$ are not global repellers and attractors. For $m=3,\,n=4$, there are two other attractor/repeller pairs $B_{3\pm}$ and $B_{5\pm}$. Therefore, there can exist heteroclinic trajectories connecting them, which may or may not correspond to a nonsingular bouncing cosmology, depending on how the trajectories evolve.

The phase trajectories in the left panel of Fig.\ref{FIG.6}, which corresponds to $F(X)=\beta X^{2/3}\,(\beta<0),\,V(\phi)=V_{0}\phi^{-5}$, show nonsingular bouncing trajectories connecting two De-Sitter phases ($B_{1-} \rightarrow B_{1+}$). For $m=\frac{2}{3},\,n=-5$, $B_{1+}$ and  $B_{1-}$ are the only attractor/repeller pair possible, i.e., they are global attractors and repellers. The trajectories connecting them must necessarily undergo a nonsingular bounce and we have shown explicitly in the figure four such characteristic trajectories. In this case, one can confidently say that a nonsingular bounce is a generic behaviour of the phase trajectories.

 %%%%%%%%%%%%%%%%%%%%%%%%%%%%%%%%%%%%%%%%%%%%%%%%%%%%%%%%%%%%%%%%%%%%%%%%%%%%%%%%%%%%%%%%%%%%%%%%%%%%%%%%%%%%%%%%%%%%%%%%%%%%%%%%%%%%%%%%%%%%%%%%%%%%%%%%%%%%%%%%%%%%%%%%%%
 
 \subsubsection{\bf{Exponential Potential $V(\phi)=V_0 e^{-\lambda \phi/M_{Pl}}$}}\label{subsec:modelII_m}
 
 For the kinetic term $F(X)=\beta X^m$ and the potential $V(\phi)=V_0 e^{-\lambda \phi/M_{Pl}}$, we have
 \begin{equation}
 \Xi=m-1, \qquad \Gamma =1, \qquad \omega_{k}=\frac{1}{2m-1}, \qquad \rho_{k}=(2m-1)\beta X^{m}    
 \end{equation}

\paragraph{Finite Fixed Point Analysis}
 
In this case, the system \eqref{dynsys} reduces to
	 \begin{subequations}\label{dynsys_exp_m}
		\begin{eqnarray}
			&& \frac{dx}{d\tau} = \frac{3x}{2}\left[\left(\frac{2m}{2m-1}\right)x - \sigma y^2\sgn[(2m-1)\beta]\right] - \frac{3}{2}\left[\left(\frac{1}{2m-1}-\omega_{m}\right)\sgn[(2m-1)\beta] + (1+\omega_{m})(x^{2}-y^2)\right],\label{eq:ds_x_exp_m}\\
			&& \frac{dy}{d\tau} = \frac{3}{2}y\left[-\sigma + \left(\frac{2m}{2m-1}\right)x - \sigma y^2\sgn[(2m-1)\beta]\right],\label{eq:ds_y_exp_m}\\
			&& \frac{d\sigma}{d\tau} = 3\sigma\frac{(2m-3)m+1}{(4m-2)m}\left[\left(\frac{2m}{2m-1}\right)x-\sigma y^{2}\right]\,.\label{eq:ds_s_exp_m}
		\end{eqnarray}
       \end{subequations}
%%%%%%%%%%%%%%%%%%%%%%%%%%
For $m=1$ the dynamical system for the exponential potential was 2D (\eqref{dynsys_exp}). For $m\neq1$, the dynamical system for the exponential potential becomes 3D. The dynamical system is independent of the parameter $\lambda$. The above system reduces to the system \eqref{dynsys_exp} for $m=1$. The system is symmetric under reflection around $y=0$, which happens to be an invariant submanifold. Therefore it suffices to consider only the part of the phase space given by
\begin{equation}
    \lbrace (x,y,\sigma) \in \mathbb{R}^3: y \geq 0, x^2-y^2-\sgn((2m-1)\beta)\geq 0\rbrace\,,
 \end{equation}
which corresponds to taking $V_0>0$.
	
	\begin{table}[H]
		\begin{center}
			\begin{tabular}{|*{4}{c|}}
				\hline
				Point & Co-ordinate $(x,y,\sigma)$ &  Existence &  Physical viability $(\Omega_{m}\geq0)$\\ \hline
				\parbox[c][0.6cm]{0.5cm}{$A_{1+}$} & $(1,0,0)$ & $(2m-1)\beta>0$ & Always \\ \hline
				\parbox[c][0.6cm]{0.5cm}{$A_{1-}$} & $(-1,0,0)$ & $(2m-1)\beta>0$ & Always \\ \hline
			%  $A_{2+}$ & $\left(\frac{2}{|\sigma|},\frac{\sqrt{4+\sigma^{2}}}{|\sigma|},\sigma\right)$ & $(\beta<0)\wedge (m=1) \wedge (\sigma\leq 0)$ & Always \\ \hline
			% 	$A_{2-}$ & $\left(-\frac{2}{|\sigma|},\frac{\sqrt{4+\sigma^{2}}}{|\sigma|},\sigma\right)$ & $(\beta<0)\wedge (m=1) \wedge (\sigma\geq 0)$ & Always \\ \hline  
			% 	$A_{3+}$ & $\left(\frac{2}{|\sigma|},\frac{\sqrt{4-\sigma^{2}}}{|\sigma|},\sigma\right)$ & $(\beta>0)\wedge (m=1) \wedge (\sigma\geq 0)$ & Always \\ \hline
			% 	$A_{3-}$ & $\left(-\frac{2}{|\sigma|},\frac{\sqrt{4-\sigma^{2}}}{|\sigma|},\sigma\right)$ & $(\beta>0)\wedge (m=1) \wedge (\sigma\leq 0)$ & Always \\ \hline  
			% 	$A_{4}$ & $\left(\frac{\sigma}{1+\omega_{m}},\sqrt{\frac{1-\omega_{m}}{1+\omega_{m}}},\sigma\right)$ & $\beta>0\wedge m=1$ & $|\sigma|\geq \sqrt{2(1+\omega_{m})}$ \\ \hline
		  \end{tabular}
		\end{center}
		\caption{Existence and the physical viability condition for finite fixed points for a noncanonical scalar field with kinetic term $F(X)=\beta X^m$ ($m\neq1$) and potential $V(\phi)=V_{0}e^{-\lambda\phi/M_{Pl}}$, calculated from the system (\ref{dynsys_exp_m}).}
  \label{T:ex_nc_m}
	\end{table}
	\begin{table}[H]
		\begin{center}
			\begin{tabular}{|*{4}{c|}} 
				\hline
				Point & Co-ordinate $(x,y,\sigma)$ &  Stability &  Cosmology \\ \hline
				\parbox[c][1cm]{0.5cm}{$A_{1+}$} & $(1,0,0)$ &\begin{tabular}{@{}c@{}} Unstable if $m> 1$\\ saddle otherwise \end{tabular}& $a(t)=(t-t_{*})^{\frac{2m-1}{3m}}, ~t\geq t_{*}$ \\ \hline
				\parbox[c][1cm]{0.5cm}{$A_{1-}$} & $(-1,0,0)$ &\begin{tabular}{@{}c@{}}Stable if $m>1$\\ saddle otherwise \end{tabular}& $a(t)=(t-t_{*})^{\frac{2m-1}{3m}}, ~t\leq t_{*}$ \\ \hline
				\end{tabular}
		\end{center}
		\caption{Stability and cosmological behavior of the physically viable fixed points given in Table \ref{T:ex_nc_m}.} 
    \label{T:st_exp_nc_m}
	\end{table}
 The system \eqref{dynsys_exp_m} contains two finite fixed points $A_{1\pm}$, which exist only for non-phantom fields (see Table \ref{T:ex_nc_m}). The stabilities and corresponding cosmologies are given in Table \ref{T:st_exp_nc_m}. Since these critical points exist only when $(2m-1)\beta>0$. When $(2m-1)\beta>0$, they cannot give rise to physically viable nonsingular bouncing trajectories. The finite fixed points $A_{2\pm},\,A_{3\pm}$ and $A_4$ that appeared in Table \ref{tab:finite_fixed_pts_exp}, are specific to the case $m=1$ and does not arise for $m\neq1$.

\paragraph{Fixed Points at Infinity}
 %       \begin{subequations}
	% 	\begin{eqnarray}
	% 		\frac{d\Bar{x}}{d\tau}&=&\frac{3}{2}(1-\Bar{x}^{2})^{\frac{3}{2}}\left[(\omega_{k}-\omega_{m})\left(\frac{\Bar{x}^{2}}{1-\Bar{x}^{2}}-\sgn[(2m-1)\beta]\right)+\left(1+\omega_{m}-\frac{\Bar{\sigma}\Bar{x}\sgn[(2m-1)\beta]}{\sqrt{1-\Bar{\sigma}^2}\sqrt{1-\Bar{x}^{2}}}\right)\frac{\Bar{y}^{2}\sgn(V)}{1-\Bar{y}^{2}}\right],\\    
	% 		\frac{d\Bar{y}}{d\tau}&=&\frac{3}{2}\Bar{y}(1-\Bar{y}^{2})\left[-\frac{\Bar{\sigma}}{\sqrt{1-\Bar{\sigma}^2}} +\frac{(\omega_{k}+1)\Bar{x}}{\sqrt{1-\Bar{x}^{2}}}-\frac{\Bar{\sigma}\Bar{y}^{2}\sgn(V)}{\sqrt{1-\Bar{\sigma}^2}(1-\Bar{y}^{2})}\sgn[(2m-1)\beta]\right] \,,\\
	% 		\frac{d\Bar{\sigma}}{d\tau}&=&3\Bar{\sigma}\left(1-\Bar{\sigma}^{2}\right)\left[\frac{2\Xi(\omega_{k}+1)+\omega_{k}-1}{(4\Xi+2)(\omega_{k}+1)}
 %            \left(\frac{(\omega_{k}+1)\Bar{x}}{\sqrt{1-\Bar{x}^{2}}}-\frac{\Bar{\sigma}}{\sqrt{1-\Bar{\sigma}^{2}}}\frac{\Bar{y}^{2}}{1-\Bar{y}^{2}}\right)\right]\,. 
	% 	\end{eqnarray}
	% \end{subequations}
In terms of the compact dynamical variables $(\Bar{x},\Bar{y},\Bar{\sigma})$ defined in equation \eqref{3D_dynvar_comp_def} and the redefined time variable defined in equation \eqref{time_redef_pow}, the dynamical system for $F(X)=\beta X^{m},\,V=V_0 e^{-\lambda\phi/M_{Pl}}$ can be rewritten in the following form 
\begin{subequations}\label{dynsys_3d_exp_m}
\begin{eqnarray}
\frac{d\Bar{x}}{d\Bar{\tau}} &=& \frac{3}{2}\left(\frac{1}{2m-1}-\omega_{m}\right)\left(\Bar{x}^{2}-(1-\Bar{x}^{2})\sgn[(2m-1)\beta]\right) (1-\Bar{x}^{2})(1-\Bar{y}^{2})\sqrt{1-\Bar{\sigma}^2}+\frac{3}{2}\big((1+\omega_{m}) (1-\Bar{x}^{2}) \sqrt{1-\Bar{\sigma}^2}\nonumber\\
&&-\Bar{\sigma}\Bar{x}\sgn[(2m-1)\beta]\sqrt{1-\Bar{x}^{2}}\big)(1-\Bar{x}^{2})\Bar{y}^{2}, \label{dynsys_3d_exp_x_m}\\ 
\frac{d\Bar{y}}{d\Bar{\tau}}& =& \frac{3}{2}\Bar{y}(1-\Bar{y}^{2})\left[\left(-\Bar{\sigma}\sqrt{1-\Bar{x}^{2}}+\left(\frac{2m}{2m-1}\right)\Bar{x}\sqrt{1-\Bar{\sigma}^2}\right)(1-\Bar{y}^{2})-\Bar{\sigma}\Bar{y}^{2}\sqrt{1-\Bar{x}^{2}}\sgn[(2m-1)\beta]\right]\,, \label{dynsys_3d_exp_y_m}\\ 
%\frac{d\Bar{x}}{d\Bar{\tau}} &=& \frac{3}{2}(\omega_{k}-\omega_{m})\left[\Bar{x}^{2}-(1-\Bar{x}^{2})\sgn[(2m-1)\beta]\right](1-\Bar{x}^{2})(1-\Bar{y}^{2}) \sqrt{1-\Bar{\sigma}^2} + \frac{3}{2}\big[(1+\omega_{m})(1-\Bar{x}^{2}) \sqrt{1-\Bar{\sigma}^2}\nonumber\\
%&&-\Bar{\sigma}\Bar{x}\sgn[(2m-1)\beta]\sqrt{1-\Bar{x}^{2}}\big](1-\Bar{x}^{2})\Bar{y}^{2} ,\label{dynsys_3d_exp_x_m}\\
%\frac{d\Bar{y}}{d\Bar{\tau}}& =& \frac{3}{2}\Bar{y}(1-\Bar{y}^{2})\left[(-\Bar{\sigma}\sqrt{1-\Bar{x}^{2}} +(\omega_{k}+1)\Bar{x}\sqrt{1-\Bar{\sigma}^2})(1-\Bar{y}^{2})-\Bar{\sigma}\Bar{y}^{2}\sqrt{1-\Bar{x}^{2}}\sgn[(2m-1)\beta]\right]\,, \label{dynsys_3d_exp_y_m}\\
\frac{d\Bar{\sigma}}{d\Bar{\tau}}&=&3\Bar{\sigma}(1-\Bar{\sigma}^{2})
\left[\frac{(2m-3)m+1}{(4m-2)m}
\left(\left(\frac{2m}{2m-1}\right)\Bar{x}\sqrt{1-\Bar{\sigma}^{2}}(1-\Bar{y}^{2})-\Bar{\sigma}\Bar{y}^{2}\sqrt{1-\Bar{x}^{2}}\right)\right] \,.\label{dynsys_3d_exp_s_m}
\end{eqnarray}
\end{subequations}

\begin{figure}[H]
		\centering
		\minipage{0.45\textwidth}
		\includegraphics[width=\textwidth]{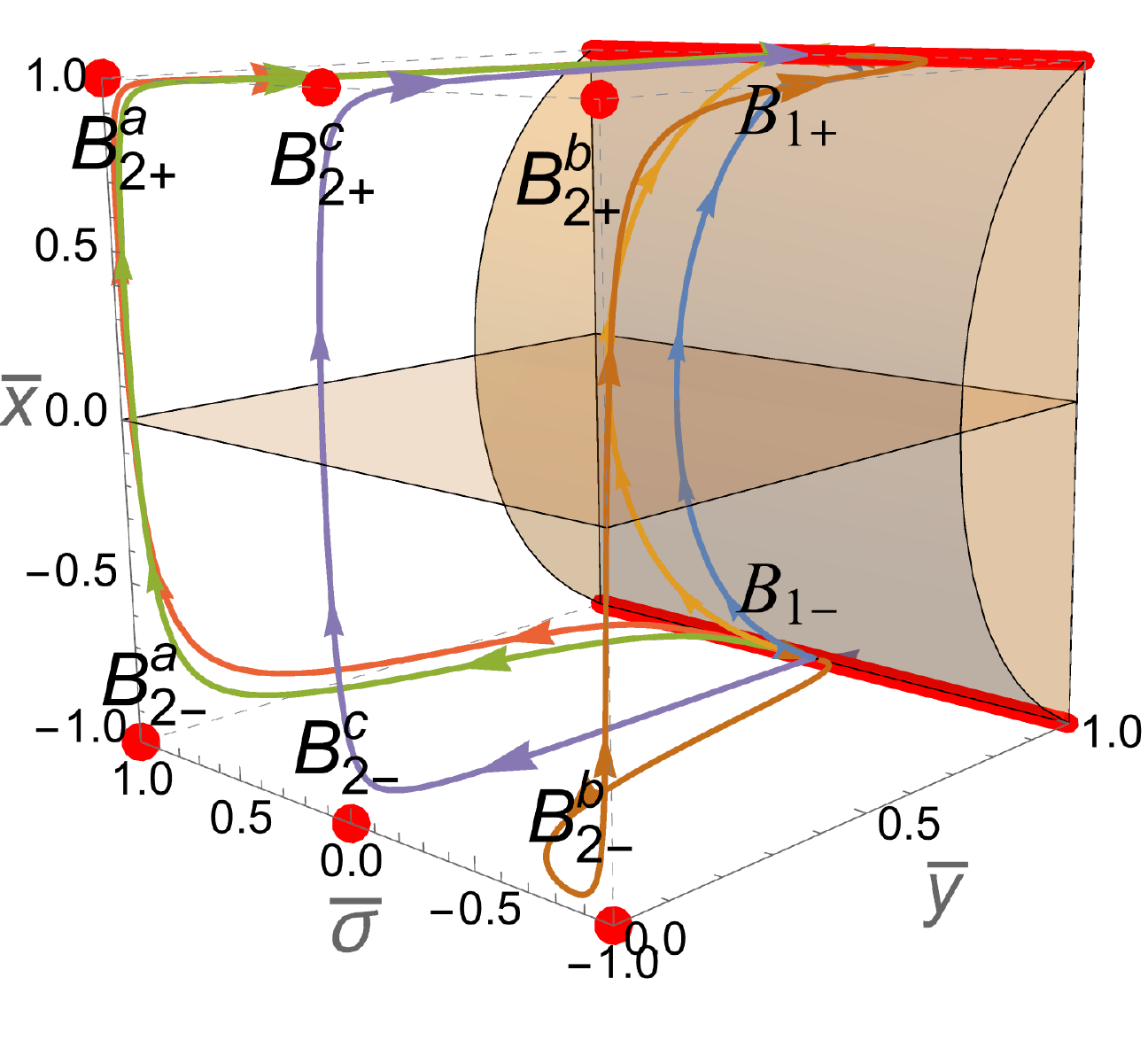}
		\endminipage
		\minipage{0.45\textwidth}
		\includegraphics[width=\textwidth]{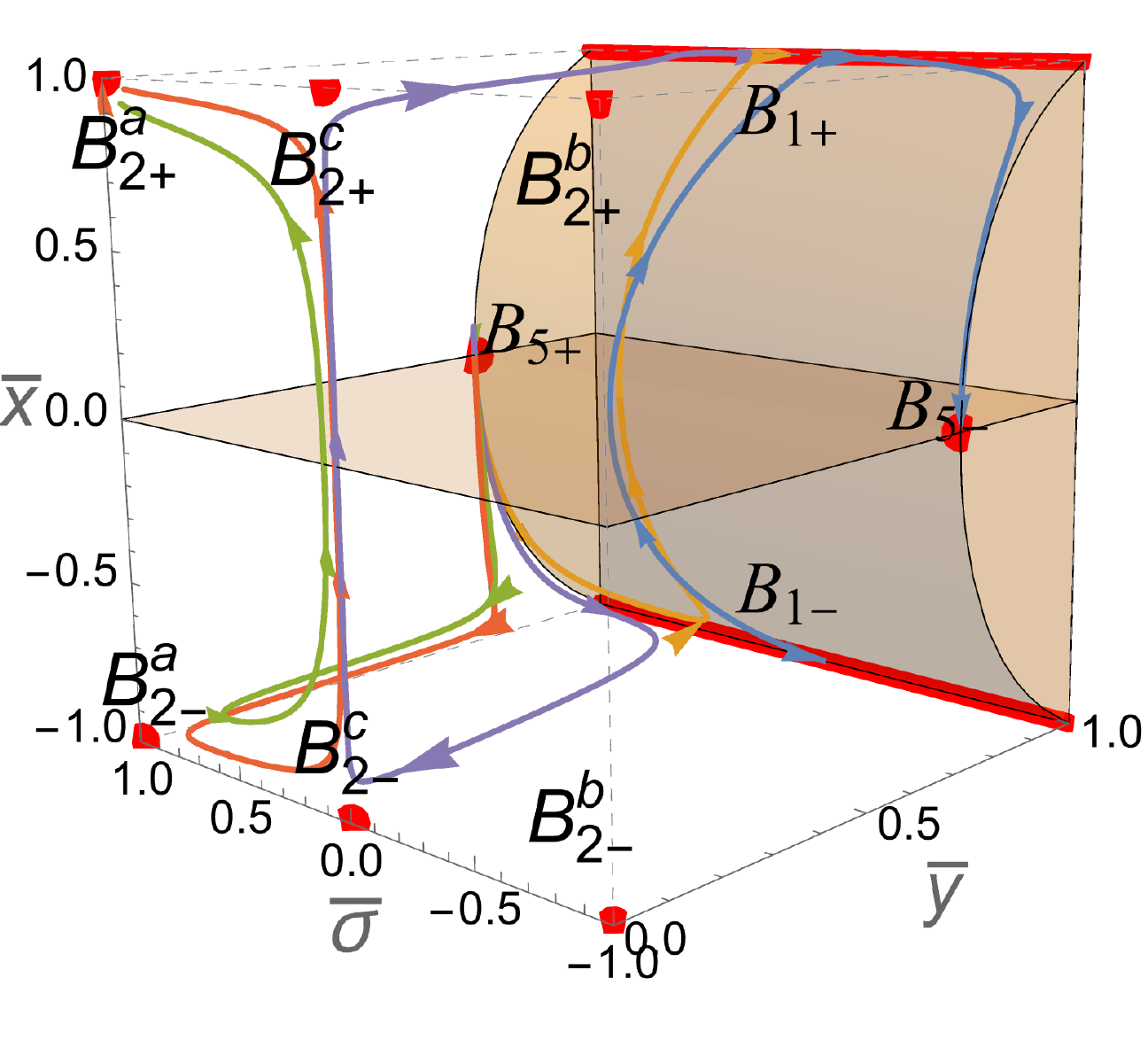}
		\endminipage
		\caption{Phase portrait of the system (\ref{dynsys_3d_exp_m}) with shaded region represents a non-physical region of the phase space. The plots are for $m=2/3$ (left panel) and $m=3$ (right panel) moreover $\sgn((2m-1)\beta)=-1$.}
   \label{FIG.7}
	\end{figure}

	\begin{table}[H]
		\begin{center}
			\begin{tabular}{|*{4}{c|}}
				\hline
				Point & Co-ordinate $(\Bar{x},\Bar{y},\Bar{\sigma})$ &  Existence &  Physical viability $(\Omega_{m}\geq0)$\\ \hline
				\parbox[c][0.5cm]{0.5cm}{$B_{1+}$} & $(1,1,\Bar{\sigma})$ & Always & Always \\ \hline
				\parbox[c][0.5cm]{0.5cm}{$B_{1-}$} & $(-1,1,\Bar{\sigma})$ & Always & Always \\ \hline
				\parbox[c][0.5cm]{0.5cm}{$B_{2+}^{a}$} & $(1,0,1)$ & Always & Always \\ \hline
                \parbox[c][0.5cm]{0.5cm}{$B_{2+}^{b}$} & $(1,0,-1)$ & Always & Always \\ \hline
				\parbox[c][0.5cm]{0.5cm}{$B_{2+}^{c}$} & $(1,0,0)$ & Always & Always \\ \hline
				\parbox[c][0.5cm]{0.5cm}{$B_{2-}^{a}$} & $(-1,0,1)$ & Always & Always \\ \hline
                \parbox[c][0.5cm]{0.5cm}{$B_{2-}^{b}$} & $(-1,0,-1)$ & Always & Always \\ \hline
				\parbox[c][0.5cm]{0.5cm}{$B_{2-}^{c}$} & $(-1,0,0)$ & Always & Always \\ \hline
				\parbox[c][1cm]{0.5cm}{$B_{3+}^{a}$} & $\left(\frac{(1+\omega_{m})\sqrt{1-\Bar{\sigma}^2}}{\sqrt{(1+\omega_{{m}})^2(1-\Bar{\sigma}^{2})+\Bar{\sigma}^2}},1,1\right)$ & \begin{tabular}{@{}c@{}} $(1+\omega_{{m}})^2(1-\Bar{\sigma}^{2})+\Bar{\sigma}^2\geq 0$\\ \end{tabular} & $\Bar{\sigma}=0$ \\ \hline
                \parbox[c][1cm]{0.5cm}{$B_{3+}^{b}$} & $\left(\frac{(1+\omega_{m})\sqrt{1-\Bar{\sigma}^2}}{\sqrt{(1+\omega_{{m}})^2(1-\Bar{\sigma}^{2})+\Bar{\sigma}^2}},1,-1\right)$ & \begin{tabular}{@{}c@{}} $(1+\omega_{{m}})^2(1-\Bar{\sigma}^{2})+\Bar{\sigma}^2\geq 0$ \\  \end{tabular} & $\Bar{\sigma}=0$ \\ \hline
				\parbox[c][1cm]{0.5cm}{$B_{3+}^{c}$} & $\left(\frac{(1+\omega_{m})\sqrt{1-\Bar{\sigma}^2}}{\sqrt{(1+\omega_{{m}})^2(1-\Bar{\sigma}^{2})+\Bar{\sigma}^2}},1,0\right)$ & \begin{tabular}{@{}c@{}} $(1+\omega_{{m}})^2(1-\Bar{\sigma}^{2})+\Bar{\sigma}^2\geq 0$\end{tabular} & $\Bar{\sigma}=0$ \\ \hline	\parbox[c][1cm]{0.5cm}{$B_{3-}^{a}$} & $\left(-\frac{(1+\omega_{m})\sqrt{1-\Bar{\sigma}^2}}{\sqrt{(1+\omega_{{m}})^2(1-\Bar{\sigma}^{2})+\Bar{\sigma}^2}},1,1\right)$ & \begin{tabular}{@{}c@{}} $(1+\omega_{{m}})^2(1-\Bar{\sigma}^{2})+\Bar{\sigma}^2\geq 0$ \\ \end{tabular} & $\Bar{\sigma}=0$ \\ \hline	
				\parbox[c][1cm]{0.5cm}{$B_{3-}^{b}$} & $\left(-\frac{(1+\omega_{m})\sqrt{1-\Bar{\sigma}^2}}{\sqrt{(1+\omega_{{m}})^2(1-\Bar{\sigma}^{2})+\Bar{\sigma}^2}},1,-1\right)$ & \begin{tabular}{@{}c@{}} $(1+\omega_{{m}})^2(1-\Bar{\sigma}^{2})+\Bar{\sigma}^2\geq 0$\\ \end{tabular} & $\Bar{\sigma}=0$ \\ \hline		
				\parbox[c][1cm]{0.5cm}{$B_{3-}^{c}$} & $\left(-\frac{(1+\omega_{m})\sqrt{1-\Bar{\sigma}^2}}{\sqrt{(1+\omega_{{m}})^2(1-\Bar{\sigma}^{2})+\Bar{\sigma}^2}},1,0\right)$ & \begin{tabular}{@{}c@{}} $(1+\omega_{{m}})^2(1-\Bar{\sigma}^{2})+\Bar{\sigma}^2\geq 0$\end{tabular} & $\Bar{\sigma}=0$ \\ \hline
                    \parbox[c][1.2cm]{0.5cm}{$B_{5+}$} & $\left(0,\frac{1}{\sqrt{2}},1\right)$ & $(2m-1)\beta<0$ & Always \\ \hline
				\parbox[c][1.2cm]{0.5cm}{$B_{5-}$} &$\left(0,\frac{1}{\sqrt{2}},-1\right)$ & $(2m-1)\beta<0$ & Always \\ \hline
				\end{tabular}
		\end{center}
		\caption{Existence and the physical viability condition for fixed points at infinity for a noncanonical scalar field with kinetic term $F(X)=\beta X^m$ ($m\neq1$) and $V(\phi)=V_{0}e^{-\lambda\phi/M_{Pl}}$, calculated from the system (\ref{dynsys_3d_exp_m}).}
  \label{T:ex_exp_com_m}
	\end{table}
 	%%%%%%%%%%%%%%%%%%%%%%%
The system \eqref{dynsys_3d_exp_m} presents six invariant submanifolds $\bar{x}=\pm1,\,\bar{y}=0,1$ and $\bar{\sigma}=\pm1$, same as in the power law case. Their stability is calculated in appendix \ref{app:stab_inv_sub_general}. The fixed points for the system \eqref{dynsys_3d_exp_m} are presented in Table \ref{T:ex_exp_com_m}. In the presence of pressureless dust, their stability conditions and corresponding cosmologies are presented in Table \ref{T:st_exp_com_m}. The six isolated fixed points $B_{3\pm}^{a,b,c}$ exist only for $\Bar{\sigma}=0$, for which they fall back on the lines of fixed points $B_{1\pm}$. The $\{x,y\}$ coordinates of the fixed points $B_{1\pm}$ and $B_{2\pm}^c$ are the same as the fixed points $B_{1\pm}$ and $B_{2\pm}$ of the 2D phase space for the case $m=1$ (see Table \ref{tab:infinite_fixed_pts_exp}). Since for the case $m\neq1$, $\sigma$ is a dynamical variable, we get the fixed points $B_{2\pm}^{a,b}$ and $B_{5\pm}$ at the boundary $\sigma\rightarrow\pm\infty$. We did not get any corresponding point for the case $m=1$ because in that case, $\sigma$ was a parameter and we considered it to be finite. In Fig.\ref{FIG.7} we present the 3D phase portrait of the system \eqref{dynsys_3d_exp_m} in the compact phase space for two cases, showing different types of physically viable nonsingular bouncing trajectories. 
  
  % Stability of the lines of fixed points $B_{1\pm}$ can be determined by investigating the stability of the invariant submanifolds $\Bar{x}=\pm1$ and $\Bar{y}=1$ (see appendix \ref{app:stab_inv_sub_general}). Fixed points $B_{2\pm}^{a}$, $B_{2\pm}^{b}$ and $B_{2\pm}^{c}$ are on the intersection of the invariant submanifold $\Bar{x}=\pm1$ and $\Bar{y}=0$. One can notice that the vicinity of $\Bar{x}=1$ for $\Bar{y}=0$ shows attracting behavior for $m>1/2$ and the vicinity of $\Bar{y}=0$ for $\Bar{x}=1$ shows repelling behavior. similar kind of behaviour one can notice for $\Bar{x}=-1$ and $\Bar{y}=0$ i.e. opposite behaviour for $m>1/2$, hence the fixed point $B_{2\pm}^{a}$, $B_{2\pm}^{b}$ and $B_{2\pm}^{c}$ always shows the saddle bbehaviour.

   	\begin{table}[H]
		\begin{center}
			\begin{tabular}{|*{4}{c|}} 
				\hline
				Point   & Co-ordinates $(\Bar{x},\Bar{y},\Bar{\sigma})$ & Stability & Cosmology \\ \hline
				\parbox[c][1.6cm]{0.5cm}{$B_{1+}$} & $(1,1,\Bar{\sigma})$ & \begin{tabular}{@{}c@{}} Stable for \\
                $\left(m>\frac{1}{2}\right)\wedge\left(\sgn(\Bar{\sigma})\neq\sgn\beta\right)$\\ 
                or $\left(m>\frac{1}{2}\right)\wedge(\Bar{\sigma}=0)$ \\ saddle otherwise \end{tabular}  & De Sitter\\ \hline
				\parbox[c][1.6cm]{0.5cm}{$B_{1-}$} &  $(-1,1,\Bar{\sigma})$ & \begin{tabular}{@{}c@{}} Unstable for \\ $\left(m>\frac{1}{2}\right)\wedge\left(\sgn(\Bar{\sigma})=\sgn\beta\right)$ \\or $\left(m>\frac{1}{2}\right)\wedge(\Bar{\sigma}=0)$ \\ saddle otherwise \end{tabular}  & De Sitter \\ \hline
		        \parbox[c][0.8cm]{0.5cm}{$B_{2+}^{a}$} & $(1,0,1)$ & Saddle always & $a(t)=(t-t_{*})^{\frac{2}{3}}, t\geq t_{*}$ \\ \hline
                \parbox[c][0.8cm]{0.5cm}{$B_{2+}^{b}$} & $(1,0,-1)$ & Saddle always & $a(t)=(t-t_{*})^{\frac{2}{3}}, t\geq t_{*}$ \\ \hline
				\parbox[c][0.8cm]{0.5cm}{$B_{2+}^{c}$} & $(1,0,0)$ & Saddle always & $a(t)=(t-t_{*})^{\frac{2}{3}}, t\geq t_{*}$ \\ \hline
				\parbox[c][0.8cm]{0.5cm}{$B_{2-}^{a}$} & $(-1,0,1)$ & Saddle always & $a(t)=(t-t_{*})^{\frac{2}{3}}, t\leq t_{*}$ \\ \hline
                \parbox[c][0.8cm]{0.5cm}{$B_{2-}^{b}$} & $(-1,0,-1)$ & Saddle always & $a(t)=(t-t_{*})^{\frac{2}{3}}, t\leq t_{*}$ \\ \hline
				\parbox[c][0.8cm]{0.5cm}{$B_{2-}^{c}$} & $(-1,0,0)$ & Saddle always & $a(t)=(t-t_{*})^{\frac{2}{3}}, t\leq t_{*}$ \\ \hline
                    \parbox[c][1.2cm]{0.5cm}{$B_{5+}$} & $\left(0,\frac{1}{\sqrt{2}},1\right)$ & \begin{tabular}{@{}c@{}} Unstable for $m<0$ or $m>1$ \\ saddle otherwise \end{tabular} & $a(t)=$ constant\\ \hline
				\parbox[c][1.2cm]{0.5cm}{$B_{5-}$} & $\left(0,\frac{1}{\sqrt{2}},-1\right)$ & \begin{tabular}{@{}c@{}} Stable for $m<0$ or $m>1$ \\ saddle otherwise \end{tabular} & $a(t)=$ constant\\ \hline
				\end{tabular}
		\end{center}
		\caption{Stability and the cosmological behavior of the physically viable fixed points defined in Table \ref{T:ex_exp_com_m} in presence of pressureless dust ($\omega_m=0$). Stability of the fixed points (or the line of fixed points) $B_{1\pm}$, $B_{2\pm}^{a,b,c}$ can be determined by investigating the stability of the invariant submanifolds $\Bar{x}=\pm1$ and $\Bar{y}=0,1$ (see appendix \ref{app:stab_inv_sub_general}).}
  \label{T:st_exp_com_m}
	\end{table}

The phase trajectories in the right panel of Fig.\ref{FIG.7} shows nonsingular bouncing trajectories for the case $F(X)=\beta X^3\,(\beta<0),\,V(\phi)=V_0 e^{-\lambda\phi/M_{pl}}$. In this case, two types of bouncing trajectories are possible, e.g. trajectories connecting a contracting De-Sitter phase to an expanding De-Sitter phase ($B_{1-}\rightarrow B_{1+}$), trajectories connecting a static and a De-Sitter phase ($B_{1-}\rightarrow B_{5-}$ and $B_{5+}\rightarrow B_{1+}$). However, one cannot say this is the generic behaviour of phase trajectories for $F(X)=\beta X^3\,(\beta<0),\,V(\phi)=V_0 e^{-\lambda\phi/M_{pl}}$, as $B_{1-}$ and $B_{1+}$ are not global repellers and attractors. For $m=3$, there are another attractor/repeller pair $B_{5\pm}$. Therefore, there can exist heteroclinic trajectories connecting them, which may or may not correspond to a nonsingular bouncing cosmology, depending on how the trajectories evolve.

The phase trajectories in the left panel of Fig.\ref{FIG.7}, which corresponds to $F(X)=\beta X^{2/3}\,(\beta<0),\,V(\phi)=V_0 e^{-\lambda\phi/M_{pl}}$, show nonsingular bouncing trajectories connecting two De-Sitter phases ($B_{1-} \rightarrow B_{1+}$). For $m=\frac{2}{3}$, $B_{1+}$ and  $B_{1-}$ are the only attractor/repeller pair possible, i.e., they are global attractors and repellers. The trajectories connecting them must necessarily undergo a nonsingular bounce and we have shown explicitly in the figure five such characteristic trajectories. In this case, one can confidently say that a nonsingular bounce is a generic behaviour of the phase trajectories.

 %%%%%%%%%%%%%%%%%%%%%%%%%%%%%%%%%%%%%%%%%%%%%%%%%%%%%%%%%%%%%%%%%%%%%%%%%%%%%%%%%%%%%%%%%%%%%%%%%%%%%%%%%%%%%%%%%%%%%%%%%%%%%%%%%%%%%%%%%%%%%%%%%%%%%%%%%%%%%%%%%%%%%%%%%%%%%%%%%%%%%%%%%%%%%%%%
	\section{Summary}\label{sec:summary}
	
	A careful analysis of the phase space structure of the models considered reveals some important physical aspects of the models. In this section, we summarize some of the main points of our phase space analysis.
	\begin{itemize}
		\item \textbf{Matter dominated epoch:} At the fixed points $B_{2\pm}$, $(x,y)\rightarrow(\pm\infty,0)$ and one can write from Eqs.\eqref{dynvar_def} and \eqref{constr}
		\begin{equation}
			\frac{\Omega_m}{x^2} = \frac{\rho_m}{3M_{Pl}^{2}H^2} \rightarrow 1.
		\end{equation}
		Therefore these two points are matter-dominated fixed points having a power law evolution (see Tables \ref{tab:stab_infinite_fixed_pts_pow}, \ref{tab:stab_infinite_fixed_pts_exp}, \ref{T:st_pow_com_m}, \ref{T:st_exp_com_m}). In all the cases considered above $B_{2\pm}$ are saddles. The saddle fixed point $B_{2+}$ can be interpreted as the intermediate matter-dominated epoch during the expanding phase of the universe.

		\item \textbf{Genericity of nonsingular bounces:} By genericity of nonsingular bouncing solutions, what we mean is, no matter what initial state one may choose during the contracting phase (i.e. whatever phase space point one chooses in the region $\Bar{x}<0$), one always ends up with a nonsingular bounce. If there is a global repeller in the region $\Bar{x}<0$ and a global attractor in the region $\Bar{x}>0$, then all the heteroclinic trajectories must necessarily represent a nonsingular bounce and one can say that nonsingular bounce is generic. In this case, one can also say that the bouncing solutions are stable in both past and future directions. No perturbation of initial conditions alters the past and future asymptotic of the evolution.  
  
        Nonsingular bounces are only possible for phantom scalar fields within GR. When the kinetic term in the Lagrangian of the scalar field is $F(X)=\beta X$ ($\beta<0$), nonsingular bounce is generic for an exponential potential but not for a power law potential, as we have discussed in Sec.\ref{subsec:modelI} and Sec.\ref{subsec:modelII} and as is also clear from Figs.\ref{fig:infinte_pow},\ref{FIG.3},\ref{FIG.4}. This is because the fixed points $B_{1\pm}$ are global attractors/repellers for an exponential potential but not for a power law potential. 
  
        When the kinetic term in the Lagrangian of the scalar field is $F(X)=\beta X^m$ with $\beta,\,m$ belonging to the range specified in \eqref{param_space}, nonsingular bounce is not generic for either power law or exponential potentials, as we have discussed in Sec.\ref{subsec:modelI_m} and Sec.\ref{subsec:modelII_m} and as is also clear from the examples presented in Figs.\ref{FIG.6},\ref{FIG.7}. This is because the fixed points $B_{1\pm}$ are not global attractors/repellers for either case. It is possible for heteroclinic trajectories to exist which do not undergo any bounce. However, there exist specific ranges for the model parameters for which nonsingular bounce is generic. For $F(X)=\beta X^m$ ($m\neq1$) and $V(\phi)=V_0 \phi^n$, when $\{m,\,n\}$ is within the range
        \begin{equation}
            \frac{1}{2} < m < 1\,, \qquad n < \frac{2m}{m-1}\,,
        \end{equation}
        the points $B_{3\pm}$ and $B_{5\pm}$ are saddles and $B_{1\pm}$ becomes global attractors and repellers (see Table \ref{T:st_pow_com_m}). For $F(X)=\beta X^m$ ($m\neq1$) and $V(\phi)=V_0 e^{-\lambda\phi/M_Pl}$, when 
        \begin{equation}
            \frac{1}{2} < m < 1\,, 
        \end{equation}
        the points $B_{5\pm}$ are saddles and $B_{1\pm}$ becomes global attractors and repellers (see Table \ref{T:st_exp_com_m}). In these cases, nonsingular bounce becomes generic, as can be seen in the examples presented in the left panels of Figs.\ref{FIG.6} and \ref{FIG.7}.

        We must mention here that, for cases when the bounce is not generic, one can of course still get a bounce for a set of initial conditions. However, unlike the case when the bounce is generic, for a chosen initial condition that leads to a bounce, it cannot be guaranteed that a random slight change of it will still lead to a bounce. In this sense, the cases for which bounce is generic are more interesting than the cases when they are not, as long as constructing a bouncing model is concerned.

        \item \textbf{Cosmic future when the bounce is generic:} It is interesting to investigate from the phase space the cosmic future of the nonsingular bouncing cosmologies when they are generic. 
        
        For $F(X)=\beta X$ ($\beta<0$) and an exponential potential, although mathematically the fixed points $A_{2\pm}$ are non-hyperbolic, from Fig.\ref{FIG.4} one can see that they are actually global future and past attractors. The end state of a nonsingular bouncing cosmology is a future attractor which can be either a De-Sitter phase or a big-rip cosmology, depending on the choice of the model parameters. If $\sigma=\frac{\sqrt{{2}/{3}}\lambda}{\sqrt{-\beta}}\geq0$, the future attractor is a De-Sitter phase given by the point $B_{1+}$. In this case, the cosmology can be matched with the $\Lambda$CDM at the asymptotic future. If, however, $\sigma=\frac{\sqrt{{2}/{3}}\lambda}{\sqrt{-\beta}}<0$, then the future attractor is a big-rip singularity given by the point $A_{2+}$.

        On the contrary, when $F(X)=\beta X^m$ for either a power law or an exponential potential, and in cases when the bounce is generic, the only end state that is possible is the De-Sitter phases represented by the point $B_{1+}$. In these cases, the cosmology can be matched with $\Lambda$CDM at the asymptotic future.  
	\end{itemize}
	
	%%%%%%%%%%%%%%%%%%%%%%%%%%%%%%%%%%%%%%%%%%
	
	\section{Conclusion}\label{sec:concl}
	
	Nonsingular bouncing solutions are important candidates for early universe cosmology and it is necessary to investigate different aspects of them. This article deals with investigating them from the phase space point of view. For our analysis, we have taken nonsingular bouncing models in $F(X)-V(\phi)$ theory, considering two simple choices for the potential, namely power law and exponential potential. Our main motivation behind doing this exercise is to find how generic nonsingular bouncing solutions are. More precisely, even if for a theory a bouncing solution may exist, whether or not it arises from a large number of initial conditions. In the phase space picture, this question can be recast as to whether phase trajectories representing nonsingular bouncing solutions come from a large area of the phase space or only from some small patches. 
	
	For the purpose of a dynamical system analysis we have used the formulation of references \cite{De-Santiago:2012ibi,Panda:2015wya}. The formulation employs density-normalized dimensionless dynamical variables instead of the usual Hubble-normalized dynamical variables. This is because the Hubble normalized dynamical variables diverge at the bounce. We extend the phase space analysis of \cite{De-Santiago:2012ibi} by compactifying the phase space. Compactification of the phase space helps us visualize its global structure and answer questions regarding past and future asymptotic of a cosmological model. In our case, the compact phase space analysis helps us investigate the genericity of solutions as well as answer questions about their past and future asymptotic dynamics.
	
	For both potentials, we prove the existence of intermediate matter-dominated epochs which arise as saddle fixed points in the phase space. We recover the result that for a nonsingular bounce to happen in an $F(X)$-$V(\phi)$ type scalar field Lagrangian in GR, the scalar field needs to be phantom. \footnote{If one goes beyond a scalar field with an $F(X)$-$V(\phi)$ type Lagrangians and GR, it is possible to achieve a nonsingular bounce even without non-phantom scalar fields e.g. ghost condensate or Galileon models} This result is also expected out of the fact that nonsingular bounce in spatially flat FLRW cosmology in GR requires violation of NEC \cite{Xue:2013iqy,Battefeld:2014uga}. We showed that, in general, nonsingular bounce in these models is not generic due to the non-existence of global past or future attractors. However, we identify the parameter range for which nonsingular bounce can be generic. For the kinetic term $F(X)=\beta X^m$ ($\beta<0$), these ranges are as follows:
    \begin{itemize}
        \item For a power law potential $V(\phi) = V_0 \phi^n$, the range is $\left\lbrace\frac{1}{2}<m<1,\,n<\frac{2m}{m-1}\right\rbrace$.
        \item For an exponential potential $V(\phi) = V_0 e^{-\lambda\phi/M_{Pl}}$, the range is $\left\lbrace\frac{1}{2}<m\leq1\right\rbrace$. 
    \end{itemize}
    When the model parameters fall outside this range, one can of course still obtain a bouncing solution by choosing a certain set of initial conditions and evolving the system numerically. However, it is totally possible that the initial condition is fine-tuned. An arbitrary slight change of initial conditions takes one to a phase trajectory that does not represent a bounce. On the contrary, when the model parameters are in the aforementioned range, any arbitrary change of the initial conditions would still give us a bounce; the initial conditions are not fine-tuned.
    For the special case when $F(X) = \beta X$ ($\beta<0$) and $V(\phi) = V_0 e^{-\lambda\phi/M_{Pl}}$, the asymptotic future of the bouncing cosmology can be either De-Sitter or a big-rip. In all the other cases when the bounce is generic, the asymptotic future is definitely De-Sitter.

    In spite of the fact that there are issues arising at the level of inhomogeneous cosmological perturbations in $F(X)$-$V(\phi)$ model, we have worked with this simpler model in order to showcase the importance of investigating the genericity of bouncing solutions (i.e. stability with respect to perturbations in initial conditions) and how one can use the compact phase space formulation to address this question. We argue that, for a theory under consideration, the cases for which bounce is generic are more interesting than the cases when they are not, as long as constructing a bouncing model in that theory is concerned. Also, a compact phase space analysis allows us to investigate the cosmic future of the nonsingular bouncing cosmology when it is generic. There are a plethora of nonsingular bouncing models in the literature, both in GR and in modified gravity theories (see the reviews \cite{Battefeld:2014uga,Novello:2008ra}; for some recent works see \cite{Cai:2017dyi,Ilyas:2020qja,Vikman_2005}). Since compact phase space analysis catches the global dynamic of the system, it would be interesting to consider the compact phase space for such theories, identify the cases when the bounce is generic and find out the cosmic future in such cases. We leave these ideas for future projects.

    % {\color{blue} We demonstrated that the bounce can only be explained if the field violates the null energy constraint, as shown in section \ref{subsec:general}. This is a well-known result that may have effects on the stability of the field. Although we did not discuss inhomogeneous perturbations in this paper, it has been suggested that these Lagrangians have issues with both classical and quantum stability when they violate the NEC \cite{Kallosh_2008,Vikman_2005}. All of the aforementioned justifications lead us to the conclusion that even fields as straightforward as $F(X)-V(\phi)$ are not likely to violate NEC and consequently lead to a bounce. Even while it could be necessary to analyze different kinds of fields, doing so goes beyond the scope of the current paper, which exclusively focused on the homogeneous dynamics of the fields.}
	
	%%%%%%%%%%%%%%%%%%%%%%%%%%%%%%%%%%%%%%%%%%%
	
	\begin{appendix}
		
		%%%%%%%%%%%%%%%%%%%%%%%%%%%%%%%%%%%%%%%%%%

  \section{Center manifold dynamics for point \texorpdfstring{$A_{1-}$}{} of model \texorpdfstring{$F(X)=\beta X, V(\phi)=V_0 \phi^n$}{}}\label{app:cmt}
		In this appendix, we apply the center manifold theory to study the dynamics of the system \eqref{dynsys_pow} near a point $A_{1-}(-1,0,0)$. For detailed mathematical background on center manifold theory, we refer the reader to Ref. \cite{Perko}.
		
		Firstly, we translate the point $(-1,0,0)$ to $(0,0,0)$ via a transformation $x\rightarrow x-1$, $y\rightarrow y$, $\sigma \rightarrow \sigma$ under which
		the system \eqref{dynsys_pow} becomes
		\begin{subequations}\label{dynsys_pow_cmt}
			\begin{eqnarray}
				&& \frac{dx}{d\tau} = \frac{3}{2}(x-1)\left[2(x-1) - \sigma y^2 \right] - \frac{3}{2}\left[(1-\omega_{m}) + (1+\omega_{m})((x-1)^{2}-y^2 )\right],\label{eq:x_cmt}\\
				&& \frac{dy}{d\tau} = \frac{3}{2}y\left[-\sigma + 2(x-1) - \sigma y^2 \right],\label{eq:y_cmt}\\
				&& \frac{d\sigma}{d\tau} = \frac{3}{n}\sigma^{2}\,,\label{eq:s_cmt}
			\end{eqnarray}
		\end{subequations}
		which can be written as
		\[\left(\begin{array}{c}
			\frac{dx}{d\tau} \\
			\frac{dy}{d\tau}\\
			\frac{d\sigma}{d\tau} \end{array} \right)=\left(\begin{array}{ccc}
			-3(1-\omega_{m})  & 0 & 0  \\
			0  & -3& 0\\
			0  & 0 & 0   \end{array} \right) \left(\begin{array}{c}
			x\\
			y\\
			\sigma \end{array} \right)+\left(\begin{array}{c}
			g_1\\
			g_2\\
			f \end{array} \right)\,,\]
		where 
		\begin{eqnarray}
			g_1 &=& -\frac{3}{2}\,\sigma\,x{y}^{2}+\frac{3}{2}\,{x}^{2}+\frac{3}{2}\,\sigma\,{y}^{2}-\frac{3}{2}{x}^{2}\omega_{{m}}+\frac{3}{2}\,{y}^{2}\omega_{{m}}+\frac{3}{2}\,{y}^{2}\,,\nonumber\\
			g_2&=&-\frac{3}{2}\,\sigma\,{y}^{3}-\frac{3}{2}\sigma\,y+3\,xy\,,\nonumber\\
			f&=&\frac{3}{n}\sigma^{2}\,.\nonumber
		\end{eqnarray}
		
		The local center manifold is given by
		\begin{equation}
			\lbrace \mathbf{z}=\mathbf{h}(\sigma): \mathbf{h}(0)=\mathbf{0}, D\mathbf{h}(\mathbf{0})=\mathbf{0}\rbrace\,,
		\end{equation}
		where  
		\[\mathbf{h}=\left(\begin{array}{c}
			h_1\\
			h_2\ \end{array} \right)\,,\quad \mathbf{z}=\left(\begin{array}{c}
			x\\
			y\ \end{array} \right)\,.
		\] 
		The function $\mathbf{h}$  can be approximated   in terms of a power series as
		\begin{eqnarray}
			h_1(\sigma)&=&a_2 \sigma^2+a_3 \sigma^3+\mathcal{O}(\sigma^4),\\
			h_2(\sigma)&=&b_2 \sigma^2+b_3 \sigma^3+\mathcal{O}(\sigma^4)\,,
		\end{eqnarray}
		which is determined by the quasilinear partial differential equation \cite{Perko}
		\begin{equation}\label{quasi_C}
			D \mathbf{h}(\sigma)\left[A \sigma+\mathbf{F}(\sigma,\mathbf{h}(\sigma))\right]-B \mathbf{h}(\sigma)-\mathbf{g}(\sigma,\mathbf{h}(\sigma))=\mathbf{0} \,,
		\end{equation}
		with \[\mathbf{g}=\left(\begin{array}{c}
			g_1\\
			g_2 \end{array} \right),~~~~~ \mathbf{F}=f, ~~~~~B= \left(\begin{array}{cc}
			-3 (1-\omega_{m}) & 0 \\
			0 &-3 \end{array} \right),~~~~~ A=0. \]
		
		In order to solve the Eq.~(\ref{quasi_C}), we substitute $A$, $\textbf{h}$, $\mathbf{F}$, $B$, $\mathbf{g}$ in to it. On comparing like powers of $\sigma$ from Eq.~(\ref{quasi_C}) we obtain that constants $a_2=0$, $a_3=0$, $b_2=0$, $b_3=0$. Thus, the local center manifold is the $\sigma$-axis which coincides with the center subspace (a subspace generated by the eigenvectors corresponds to a vanishing eigenvalue of the corresponding Jacobian matrix).
		
		Finally, the dynamics in a local center manifold is determined by the equation
		\begin{equation}
			\frac{d\sigma}{d\tau} =A\,\sigma+\mathbf{F}(\sigma,\mathbf{h}(\sigma)),
		\end{equation}
		which is simply
		\begin{align}
			\frac{d\sigma}{d\tau} =\frac{3}{n}\sigma^2\,.
		\end{align}
		Hence, point $A_{1-}$ is always a saddle as we obtain an even-parity order term.
		
		\section{Stability at invariant sub-manifold}\label{app:stab_inv_sub}
		
		It is important to investigate the stability of invariant submanifolds because that helps us infer the nature of stability of the fixed points that lie at the intersection of the invariant submanifolds. If a fixed point is located at the intersection of N invariant suibmanifolds in an N-dimensional phase space, then the point is
		\begin{itemize}
			\item Stable if all the invariant submanifolds are of attracting nature.
			\item Unstable if all the invariant submanifolds are of repelling nature.
			\item Saddle if some of the invariant submanifolds are attracting and some are repelling.
		\end{itemize}
		For model II considered in Sec.\ref{subsec:modelII}, the fixed points $A_{1\pm}$ are at the intersection of the invariant submanifolds $y=0,\,\Omega=0$. For both the models, the fixed points (or the lines of fixed points) $B_{1\pm}$ are at the intersection of the invariant submanifolds $\Bar{x}=\pm1,\,\Bar{y}=1$ and the fixed points $B_{2\pm}$ are at the intersection of the invariant submanifolds $\Bar{x}=\pm1,\,\Bar{y}=0$.
		
		Below we investigate the stability of different invariant submanifolds.
		%%%%%%%%%%%%%%%%%%%%%%%%%%%%%%%%%%%%%%%%%%
		
		\subsection{\texorpdfstring{$\Bar{x}=\pm 1$}{}}
		
		Consider $\Bar{x}$ in the vicinity of $+1$ or $-1$, i.e. $0<\epsilon\equiv(1-\Bar{x}^{2})\ll1$. One can then rewrite Eq.\eqref{dynsys_2d_new_x} as
		\begin{equation}
			\frac{d\Bar{x}}{d\Bar{\tau}} = \frac{3}{2}(\omega_{k}-\omega_{m})\left(1-\epsilon-\epsilon \sgn(\beta)\right)\epsilon(1-\Bar{y}^{2}) + \frac{3}{2}\left((1+\omega_{m})\epsilon-\sigma\Bar{x}\sgn(\beta)\sqrt{\epsilon}\right)\epsilon\Bar{y}^{2}.   
		\end{equation}
		Since $\omega_{k}=1$, to the lowest order of $\epsilon$ one can write
		\begin{eqnarray}
			\frac{d\Bar{x}}{d\Bar{\tau}} \simeq \frac{3}{2}(1-\omega_{m})(1-\Bar{y}^{2})\epsilon.
		\end{eqnarray}
		Since all quantities on the right-hand side are positive. The flow will always be in positive $x$-direction in the vicinity of $\Bar{x}=\pm1$. Therefore the invariant submanifold $\Bar{x}=1$ is attracting while the invariant submanifold $\Bar{x}=-1$ is repelling. 
		
		%%%%%%%%%%%%%%%%%%%%%%%%%%%%%%%%%%%%%%%%%
		
		\subsection{\texorpdfstring{$\Bar{y}=1$}{}}
		
		Consider $\Bar{y}$ in the vicinity of $1$, i.e. $0<\epsilon\equiv(1-\Bar{y}^{2})\ll1$. We will consider the cases $\sigma=0$ and $\sigma\neq0$ separately.
		
		%%%%%%%%%%%%%%%%%%%%%%%%%%%%%%%%%%%%%%%%%%
		
		\subsubsection{\texorpdfstring{$\sigma=0$}{}}
		
		One can rewrite Eq.\eqref{dynsys_2d_new_y} as
		\begin{equation}
			\frac{d\Bar{y}}{d\Bar{\tau}} = \frac{3}{2}\Bar{y}(1-\Bar{y}^{2})^{2}(\omega_{k}+1)\Bar{x} = \frac{3}{2}(\omega_{k}+1)\Bar{x}\epsilon^{2}\sqrt{1-\epsilon},   
		\end{equation}
		Since $\omega_k=1$, to the lowest order of $\epsilon$ one can write
		\begin{equation}
			\frac{d\Bar{y}}{d\Bar{\tau}} \simeq 3\Bar{x}\epsilon^{2}.
		\end{equation}
		From the above one can conclude that
		\begin{itemize}
			\item $\frac{d\Bar{y}}{d\Bar{\tau}}>0$ in the vicinity of $\Bar{y}=1$ i.e. the invariant submanifold $\Bar{y}=1$ is attracting for $\Bar{x}>0$.
			\item $\frac{d\Bar{y}}{d\Bar{\tau}}<0$ in the vicinity of $\Bar{y}=1$ i.e. the invariant submanifold $\Bar{y}=1$ is repelling for $\Bar{x}<0$.
		\end{itemize}
		
		%%%%%%%%%%%%%%%%%%%%%%%%%%%%%%%%%%%%%%%%%
		
		\subsubsection{\texorpdfstring{$\sigma\neq0$}{}}
		
		One can rewrite Eq.\eqref{dynsys_2d_new_y} as
		\begin{eqnarray}
			\frac{d\Bar{y}}{d\Bar{\tau}} &=& \frac{3}{2}\Bar{y}(1-\Bar{y}^{2})\left[(-\sigma\sqrt{1-\Bar{x}^{2}} +(\omega_{k}+1)\Bar{x})(1-\Bar{y}^{2}) - \sigma\Bar{y}^{2}\sqrt{1-\Bar{x}^{2}}\sgn(\beta)\right] \nonumber\\
			&=& \frac{3}{2}\epsilon\sqrt{1-\epsilon}\left[(-\sigma\sqrt{1-\Bar{x}^{2}}+(\omega_{k}+1)\Bar{x})\epsilon - \sigma(1-\epsilon)\sqrt{1-\Bar{x}^{2}}\sgn(\beta)\right],
		\end{eqnarray}
		Since $\omega_k=1$, to the lowest order of $\epsilon$ one can write
		\begin{equation}
			\frac{d\Bar{y}}{d\Bar{\tau}} \simeq -\frac{3}{2}\epsilon\sigma\sqrt{1-\Bar{x}^{2}}\sgn(\beta).   
		\end{equation}
		From the above one can conclude that
		\begin{itemize}
			\item $\frac{d\Bar{y}}{d\Bar{\tau}}>0$ in the vicinity of $\Bar{y}=1$ i.e. the invariant submanifold $\Bar{y}=1$ is attracting if $\sgn(\beta)\neq \sgn(\sigma)$.
			\item $\frac{d\Bar{y}}{d\Bar{\tau}}<0$ in the vicinity of $\Bar{y}=1$ i.e. the invariant submanifold $\Bar{y}=1$ is repelling if $\sgn(\beta)=\sgn(\sigma)$.
		\end{itemize}
		
		%%%%%%%%%%%%%%%%%%%%%%%%%%%%%%%%%%%%%%%%%
		
		\subsection{\texorpdfstring{$\Bar{y}=0$}{}}
		
		Consider $\Bar{y}$ in the vicinity of $\Bar{y}=0$, i.e. $0<\Bar{y}\ll1$. We will consider the cases $\sigma=0$, $\sigma>0$ and $\sigma<0$ separately.
		
		\subsubsection{\texorpdfstring{$\sigma=0$}{}}
		
		One can rewrite Eq.\eqref{dynsys_2d_new_y} as
		\begin{equation}
			\frac{d\Bar{y}}{d\Bar{\tau}} = \frac{3}{2}\Bar{y}(1-\Bar{y}^{2})^{2}(\omega_{k}+1)\Bar{x}.   
		\end{equation}
		Since $\omega_k=1$, to leading order of $\Bar{y}$ one can write
		\begin{equation}
			\frac{d\Bar{y}}{d\Bar{\tau}} \simeq 3\Bar{x}\Bar{y}.   
		\end{equation}
		From the above one can conclude that
		\begin{itemize}
			\item $\frac{d\Bar{y}}{d\Bar{\tau}}>0$ in the vicinity of $\Bar{y}=0$ i.e. the invariant submanifold $\Bar{y}=0$ is repelling for $\Bar{x}>0$.
			\item $\frac{d\Bar{y}}{d\Bar{\tau}}<0$ in the vicinity of $\Bar{y}=0$ i.e. the invariant submanifold $\Bar{y}=0$ is attracting for $\Bar{x}<0$.
		\end{itemize}
		
		%%%%%%%%%%%%%%%%%%%%%%%%%%%%%%%%%%%%%%%%%
		
		\subsubsection{\texorpdfstring{$\sigma\neq0$}{}}
		
		One can rewrite Eq.\eqref{dynsys_2d_new_y} as
		\begin{equation}
			\frac{d\Bar{y}}{d\Bar{\tau}} = \frac{3}{2}\Bar{y}(1-\Bar{y}^{2})\left[(-\sigma\sqrt{1-\Bar{x}^{2}} +(\omega_{k}+1)\Bar{x})(1-\Bar{y}^{2}) - \sigma\Bar{y}^{2}\sqrt{1-\Bar{x}^{2}}\sgn(\beta)\right].    
		\end{equation}
		Since $\omega_k=1$, to the leading order of $\Bar{y}$ one can write
		\begin{equation}
			\frac{d\Bar{y}}{d\Bar{\tau}} \simeq \frac{3}{2}\Bar{y}\left[ -\sigma \sqrt{1-\Bar{x}^{2}}+2\Bar{x} \right].   
		\end{equation}
		From the above one can conclude that
		\begin{itemize}
			\item $\frac{d\Bar{y}}{d\Bar{\tau}}>0$ in the vicinity of $\Bar{y}=0$ i.e. the invariant submanifold $\Bar{y}=0$ is repelling for $\Bar{x}>\frac{\sigma}{\sqrt{4+\sigma^{2}}}$.
			\item $\frac{d\Bar{y}}{d\Bar{\tau}}<0$ in the vicinity of $\Bar{y}=0$ i.e. the invariant submanifold $\Bar{y}=0$ is attracting for $\Bar{x}<\frac{\sigma}{\sqrt{4+\sigma^{2}}}$. 
		\end{itemize}
		
		%%%%%%%%%%%%%%%%%%%%%%%%%%%%%%%%%%%%%%%%%
		
		% \subsubsection{$\sigma <0$}
		% \begin{eqnarray}
		% \frac{d\Bar{y}}{d\Bar{\tau}}&=&\frac{3}{2}\Bar{y}(1-\Bar{y}^{2})\left[ -\sigma \sqrt{1-\Bar{x}^{2}}+(\omega_{k}+1)\Bar{x} \right](1-\Bar{y}^{2})\nonumber \\
		% \frac{d\Bar{y}}{d\Bar{\tau}}&=&\frac{3}{2}\epsilon\left[ -\sigma \sqrt{1-\Bar{x}^{2}}+2\Bar{x} \right]
		% \end{eqnarray}
		
		% It is attracting for $\Bar{x}<\frac{\sigma}{\sqrt{4+\sigma^{2}}}$ and repelling for $\Bar{x}>-\frac{\sigma}{\sqrt{4+\sigma^{2}}}$.
		
		%%%%%%%%%%%%%%%%%%%%%%%%%%%%%%%%%%%%%%%%%
		
		\subsection{\texorpdfstring{$\Omega_{m}=0$}{}}
		
		Apart from $\Bar{x}=\pm1$ and $\Bar{y}=0,1$, there is another invariant submanifold given by $\Omega_m=0$. The existence of this invariant submanifold is not apparent from the dynamical system \eqref{dynsys_2d_new}. One could have guessed the existence of this submanifold from the physical argument that, if cosmology is initially a vacuum, it remains so as there is no mechanism for matter creation in classical physics. That $\Omega_m=0$ is an invariant submanifold can be explicitly shown if one tries to write a dynamical equation for $\Omega_m$ in terms of $\Bar{\tau}$ using the dynamical equations \eqref{dynsys} and the definitions \eqref{constr_comp}, \eqref{time_redef_exp}
		\begin{equation}
			\frac{d\Omega_{m}}{d\Bar{\tau}}=3\Omega_{m}[(1-\omega_{m})\Bar{x}(1-\Bar{y}^{2})-\sigma \Bar{y}^{2}\sqrt{1-\Bar{x}^{2}}\sgn(\beta)] .   
		\end{equation}
		From the above one can conclude that the invariant submanifold $\Omega_{m}=0$ is
		\begin{itemize}
			\item attracting for $(1-\omega_{m})\Bar{x}(1-\Bar{y}^{2})-\sigma \Bar{y}^{2}\sqrt{1-\Bar{x}^{2}}\sgn(\beta)<0$.
			\item repelling for $(1-\omega_{m})\Bar{x}(1-\Bar{y}^{2})-\sigma \Bar{y}^{2}\sqrt{1-\Bar{x}^{2}}\sgn(\beta)>0$.
		\end{itemize}

	\section{Stability at invariant sub-manifold of \texorpdfstring{$F(X)=\beta X^{m}$}{}}\label{app:stab_inv_sub_general}	
 
 As we have seen the stability through the invariant submanifold, the fixed points (or the lines of fixed points) $B_{1\pm}$ are at the intersection of the invariant submanifolds $\Bar{x}=\pm1,\,\Bar{y}=1$, the fixed points $B_{2\pm}^{a},\, B_{2\pm}^{b},\, B_{2\pm}^{c}$ are at the intersection of the invariant submanifolds $\Bar{x}=\pm1,\,\Bar{y}=0$ and the fixed points $B_{3\pm}$ are at the intersection of the invariant submanifolds $\Bar{y}=0,\,\Bar{\sigma}=\pm 1$.
% \begin{subequations} 
%  \begin{eqnarray}
% \frac{d\Bar{x}}{d\Bar{\tau}} &=& \frac{3}{2}\left(\frac{1}{2m-1}-\omega_{m}\right)      \left(\Bar{x}^{2}-(1-\Bar{x}^{2})\sgn[(2m-1)\beta]\right) (1-\Bar{x}^{2})(1-\Bar{y}^{2})\sqrt{1-\Bar{\sigma}^2}+\frac{3}{2}\big((1+\omega_{m})(1-\Bar{x}^{2}) \sqrt{1-\Bar{\sigma}^2}\nonumber\\
% &&-\Bar{\sigma}\Bar{x}\sgn[(2m-1)\beta]\sqrt{1-\Bar{x}^{2}}\big)(1-\Bar{x}^{2})\Bar{y}^{2}\\ 
% \frac{d\Bar{y}}{d\Bar{\tau}}& =& \frac{3}{2}\Bar{y}(1-\Bar{y}^{2})\left[\left(-\Bar{\sigma}\sqrt{1-\Bar{x}^{2}}+\left(\frac{2m}{2m-1}\right)\Bar{x}\sqrt{1-\Bar{\sigma}^2}\right)(1-\Bar{y}^{2})-\Bar{\sigma}\Bar{y}^{2}\sqrt{1-\Bar{x}^{2}}\sgn[(2m-1)\beta]\right]\\  
% \frac{d\Bar{\sigma}}{d\Bar{\tau}}&=&3\Bar{\sigma}(1-\Bar{\sigma}^{2})\left[\Bar{\sigma}\sqrt{1-\Bar{x}^{2}}(1-\Bar{y}^{2})(1-\Gamma)+\frac{(2m -3)m+1}{(4m-2)m}\left(\left(\frac{2m}{2m-1}\right) \Bar{x}\sqrt{1-\Bar{\sigma}^{2}}(1-\Bar{y}^{2})-\sqrt{1-\Bar{x}^{2}}\Bar{\sigma}\Bar{y}^{2}\right)\right]    
% \end{eqnarray}   
% \end{subequations}
\subsection{\texorpdfstring{$\Bar{x}=\pm 1$}{}}
Consider $\Bar{x}$ in the vicinity of $+1$ or $-1$, i.e. $0<\epsilon\equiv (1-\Bar{x}^{2})\ll 1$.
\begin{equation}
 \frac{d\Bar{x}}{d\Bar{\tau}}=\frac{3}{2}\epsilon\left(\frac{1}{2m-1}-\omega_{m}\right)(1-\Bar{y}^{2})\sqrt{1-\Bar{\sigma}^{2}}   
\end{equation}
Since all the quantities $\epsilon,(1-\Bar{y}^{2}), \sqrt{1-\Bar{\sigma}^{2}}$ are positive in the right hand side. If $\omega_{m}=0$, the quantity $\frac{1}{2m-1}$ will decide the attracting or repelling behaviour of the $\Bar{x}=\pm 1$.

\begin{itemize}
\item $\frac{d\Bar{x}}{d\Bar{\tau}}>0$ if $m>\frac{1}{2}$ the invariant submanifold $\Bar{x}=1$ is attracting and the invariant submanifold $\Bar{x}=-1$ is repelling.

\item $\frac{d\Bar{x}}{d\Bar{\tau}}<0$ if $m<\frac{1}{2}$ the invariant submanifold $\Bar{x}=1$ is repelling and the invariant submanifold $\Bar{x}=-1$ is attracting.

\end{itemize}
\subsection{\texorpdfstring{$\Bar{y}= 1$}{}}
\subsubsection{\texorpdfstring{$\Bar{\sigma}\neq 0$}{}}
Consider $\Bar{y}$ in the vicinity of 1, i.e. $0<\epsilon\equiv (1-\Bar{y}^{2})\ll 1$.
\begin{equation}
\frac{d\Bar{y}}{d\Bar{\tau}}=-\frac{3}{2}\epsilon \Bar{\sigma}\sqrt{1-\Bar{x}^{2}}\sgn((2m-1)\beta)    
\end{equation}
\begin{itemize}
    \item $\frac{d\Bar{y}}{d\Bar{\tau}}>0$ in the vicinity of $\Bar{y}=1$ i.e. the invariant submanifold $\Bar{y}=1$ is attracting if $\sgn((2m-1)\beta)\neq \sgn(\Bar{\sigma})$.
    \item $\frac{d\Bar{y}}{d\Bar{\tau}}<0$ in the vicinity of $\Bar{y}=1$ i.e. the invariant submanifold $\Bar{y}=1$ is repelling if $\sgn((2m-1)\beta)= \sgn(\Bar{\sigma})$.
\end{itemize}
\subsubsection{\texorpdfstring{$\Bar{\sigma}= 0$}{}}
\begin{equation}
 \frac{d\Bar{y}}{d\Bar{\tau}}= \left(\frac{3m}{2m-1}\right)\bar{x}\epsilon^{2} 
\end{equation}
\begin{itemize}
    \item $\frac{d\Bar{y}}{d\Bar{\tau}}>0$ in the vicinity of $\Bar{y}=1$ i.e. the invariant submanifold $\Bar{y}=1$ is attracting if $(\Bar{x}>0)\wedge\left((m<0)\vee(m>\frac{1}{2})\right)$ or $(\Bar{x}<0)\wedge(0<m<\frac{1}{2})$.
    \item $\frac{d\Bar{y}}{d\Bar{\tau}}<0$ in the vicinity of $\Bar{y}=1$ i.e. the invariant submanifold $\Bar{y}=1$ is repelling if $(\Bar{x}<0)\wedge\left((m<0)\vee(m>\frac{1}{2})\right)$ or $(\Bar{x}>0)\wedge(0<m<\frac{1}{2})$.
\end{itemize}

\subsection{\texorpdfstring{$\Bar{y}=0$}{}}
\subsubsection{\texorpdfstring{$\Bar{\sigma}\neq 0$}{}}
Consider $\Bar{y}$ in the vicinity of 0, i.e. $0<\bar{y}\ll 1$.
\begin{equation}
 \frac{d\Bar{y}}{d\Bar{\tau}}= \frac{3}{2}\Bar{y}\left(-\Bar{\sigma}\sqrt{1-\bar{x}^{2}}+\left(\frac{2m}{2m-1}\right)\bar{x}\sqrt{1-\bar{\sigma}^{2}}\right)  
\end{equation}
\begin{itemize}
\item $\frac{d\Bar{y}}{d\Bar{\tau}}>0$ in the vicinity of $\Bar{y}=0$ i.e. the invariant submanifold $\Bar{y}=0$ is repelling if $\Bar{\sigma}=-1$.

\item $\frac{d\Bar{y}}{d\Bar{\tau}}<0$ in the vicinity of $\Bar{y}=0$ i.e. the invariant submanifold $\Bar{y}=0$ is attracting if $\Bar{\sigma}=1$.

\item $\frac{d\Bar{y}}{d\Bar{\tau}}>0$ in the vicinity of $\Bar{y}=0$ i.e. the invariant submanifold $\Bar{y}=0$ is repelling if $(\Bar{x}=1)\wedge\left((m<0)\vee (m>\frac{1}{2})\right)$ or $(\Bar{x}=-1)\wedge(0<m<\frac{1}{2})$.

\item $\frac{d\Bar{y}}{d\Bar{\tau}}<0$ in the vicinity of $\Bar{y}=0$ i.e. the invariant submanifold $\Bar{y}=0$ is attracting if $(\Bar{x}=-1)\wedge\left((m<0)\vee (m>\frac{1}{2})\right)$ or $(\Bar{x}=1)\wedge(0<m<\frac{1}{2})$.
\end{itemize}
\subsubsection{\texorpdfstring{$\Bar{\sigma}= 0$}{}}
\begin{equation}
 \frac{d\Bar{y}}{d\Bar{\tau}}= \left(\frac{3m}{2m-1}\right)\bar{x}\Bar{y} 
\end{equation}
\begin{itemize}
    \item $\frac{d\Bar{y}}{d\Bar{\tau}}>0$ in the vicinity of $\Bar{y}=0$ i.e. the invariant submanifold $\Bar{y}=0$ is repelling if $(\Bar{x}>0)\wedge\left((m<0)\vee(m>\frac{1}{2})\right)$ or $(\Bar{x}<0)\wedge(0<m<\frac{1}{2})$.
    \item $\frac{d\Bar{y}}{d\Bar{\tau}}<0$ in the vicinity of $\Bar{y}=0$ i.e. the invariant submanifold $\Bar{y}=0$ is attracting if $(\Bar{x}<0)\wedge\left((m<0)\vee(m>\frac{1}{2})\right)$ or $(\Bar{x}>0)\wedge(0<m<\frac{1}{2})$.
\end{itemize}

\subsection{\texorpdfstring{$\Bar{\sigma}=\pm 1$}{}}
\begin{equation*}
\frac{d\Bar{\sigma}}{d\Bar{\tau}}=-3\epsilon\sqrt{1-\Bar{x}^{2}}\left((1-\Bar{y}^{2})(\Gamma -1)+\frac{(2m -3)m+1}{(4m-2)m}\Bar{y}^{2}\right)    
\end{equation*} 
 \begin{equation}
\frac{d\Bar{\sigma}}{d\Bar{\tau}}=\begin{cases}
			-3\epsilon\sqrt{1-\Bar{x}^{2}}\left(-\frac{(1-\Bar{y}^{2})}{n}+\frac{(2m -3)m+1}{(4m-2)m}\Bar{y}^{2}\right), & \text{For power law potential}\\
                -\frac{3((2m -3)m+1)}{(4m-2)m}\epsilon\sqrt{1-\Bar{x}^{2}}\Bar{y}^{2}, & \text{For exponential law potential}
		 \end{cases}    
\end{equation}
For the power law potential
\subsubsection{\texorpdfstring{$\Bar{y}\neq 0$}{}}
\begin{itemize}
\item $\frac{d\Bar{\sigma}}{d\bar{\tau}}>0$ if $-\frac{1}{n}+\left(\frac{(2m -3)m+1}{(4m-2)m}+\frac{1}{n}\right)\Bar{y}^{2}<0$ the invariant submanifold $\Bar{\sigma}=-1$ is repelling and $\Bar{\sigma}=1$ is attracting. 

\item  $\frac{d\Bar{\sigma}}{d\bar{\tau}}<0$ if $-\frac{1}{n}+\left(\frac{(2m -3)m+1}{(4m-2)m}+\frac{1}{n}\right)\Bar{y}^{2}>0$ the invariant submanifold $\Bar{\sigma}=-1$ is attracting and $\Bar{\sigma}=1$ is repelling.
\end{itemize}
\subsubsection{\texorpdfstring{$\Bar{y}= 0$}{}}
\begin{itemize}
\item $\frac{d\Bar{\sigma}}{d\bar{\tau}}>0$ if $\frac{1}{n}>0$ the invariant submanifold $\Bar{\sigma}=-1$ is repelling and $\Bar{\sigma}=1$ is attracting.

\item  $\frac{d\Bar{\sigma}}{d\bar{\tau}}<0$ if $\frac{1}{n}<0$ the invariant submanifold $\Bar{\sigma}=-1$ is attracting and $\Bar{\sigma}=1$ is repelling.
\end{itemize}

For the exponential law potential 
\begin{itemize}
\item $\frac{d\Bar{\sigma}}{d\bar{\tau}}>0$ if $\frac{(2m -3)m+1}{(4m-2)m}<0$ i.e. $(0<m<1)$ the invariant submanifold $\Bar{\sigma}=-1$ is repeling and $\Bar{\sigma}=1$ is attracting.

\item $\frac{d\Bar{\sigma}}{d\bar{\tau}}<0$ if $\frac{(2m -3)m+1}{(4m-2)m}>0$ i.e. $(m<0)\vee(m>1)$ the invariant submanifold $\Bar{\sigma}=-1$ is attracting and $\Bar{\sigma}=1$ is repelling. 
\end{itemize}

	\end{appendix}
	
	%%%%%%%%%%%%%%%%%%%%%%%%%%%%%%%%%%%%%%%%%%%%%%%%%%%%%%%%%%%%%%%
	%%%%%%%%%%%%%%%%%%%%%%%%%%%%%%%%%%%%%%%%%%
	\section*{Acknowledgement} ASA acknowledges the financial support provided by University Grants Commission (UGC) through Senior Research Fellowship (File No. 16-9 (June 2017)/2018 (NET/CSIR)), to carry out the research work. SC acknowledges funding support from the NSRF via the Program Management Unit for Human Resources and Institutional Development, Research and Innovation [grant number B01F650006]. BM acknowledges the support of IUCAA, Pune (India) through the visiting associateship program. JD was supported by the Core Research Grant of SERB, Department of Science and Technology India (File No. CRG $\slash 2018 \slash 001035$) and the Associate program of IUCAA.

	\section*{References}
	\bibliography{refs}

\begin{thebibliography}{26}%
\makeatletter
\providecommand \@ifxundefined [1]{%
 \@ifx{#1\undefined}
}%
\providecommand \@ifnum [1]{%
 \ifnum #1\expandafter \@firstoftwo
 \else \expandafter \@secondoftwo
 \fi
}%
\providecommand \@ifx [1]{%
 \ifx #1\expandafter \@firstoftwo
 \else \expandafter \@secondoftwo
 \fi
}%
\providecommand \natexlab [1]{#1}%
\providecommand \enquote  [1]{``#1''}%
\providecommand \bibnamefont  [1]{#1}%
\providecommand \bibfnamefont [1]{#1}%
\providecommand \citenamefont [1]{#1}%
\providecommand \href@noop [0]{\@secondoftwo}%
\providecommand \href [0]{\begingroup \@sanitize@url \@href}%
\providecommand \@href[1]{\@@startlink{#1}\@@href}%
\providecommand \@@href[1]{\endgroup#1\@@endlink}%
\providecommand \@sanitize@url [0]{\catcode `\\12\catcode `\$12\catcode
  `\&12\catcode `\#12\catcode `\^12\catcode `\_12\catcode `\%12\relax}%
\providecommand \@@startlink[1]{}%
\providecommand \@@endlink[0]{}%
\providecommand \url  [0]{\begingroup\@sanitize@url \@url }%
\providecommand \@url [1]{\endgroup\@href {#1}{\urlprefix }}%
\providecommand \urlprefix  [0]{URL }%
\providecommand \Eprint [0]{\href }%
\providecommand \doibase [0]{https://doi.org/}%
\providecommand \selectlanguage [0]{\@gobble}%
\providecommand \bibinfo  [0]{\@secondoftwo}%
\providecommand \bibfield  [0]{\@secondoftwo}%
\providecommand \translation [1]{[#1]}%
\providecommand \BibitemOpen [0]{}%
\providecommand \bibitemStop [0]{}%
\providecommand \bibitemNoStop [0]{.\EOS\space}%
\providecommand \EOS [0]{\spacefactor3000\relax}%
\providecommand \BibitemShut  [1]{\csname bibitem#1\endcsname}%
\let\auto@bib@innerbib\@empty
%</preamble>
\bibitem [{\citenamefont {Borde}\ \emph {et~al.}(2003)\citenamefont {Borde},
  \citenamefont {Guth},\ and\ \citenamefont {Vilenkin}}]{Borde:2001nh}%
  \BibitemOpen
  \bibfield  {author} {\bibinfo {author} {\bibfnamefont {A.}~\bibnamefont
  {Borde}}, \bibinfo {author} {\bibfnamefont {A.~H.}\ \bibnamefont {Guth}},\
  and\ \bibinfo {author} {\bibfnamefont {A.}~\bibnamefont {Vilenkin}},\
  }\bibfield  {title} {\bibinfo {title} {{Inflationary space-times are
  incompletein past directions}},\ }\href
  {https://doi.org/10.1103/PhysRevLett.90.151301} {\bibfield  {journal}
  {\bibinfo  {journal} {Phys. Rev. Lett.}\ }\textbf {\bibinfo {volume} {90}},\
  \bibinfo {pages} {151301} (\bibinfo {year} {2003})},\ \Eprint
  {https://arxiv.org/abs/gr-qc/0110012} {arXiv:gr-qc/0110012} \BibitemShut
  {NoStop}%
\bibitem [{\citenamefont {Xue}(2013)}]{Xue:2013iqy}%
  \BibitemOpen
  \bibfield  {author} {\bibinfo {author} {\bibfnamefont {B.}~\bibnamefont
  {Xue}},\ }\emph {\bibinfo {title} {{Nonsingular Bouncing Cosmology}}},\ \href
  {http://arks.princeton.edu/ark:/88435/dsp01n583xv101} {Ph.D. thesis},\
  \bibinfo  {school} {Princeton U.} (\bibinfo {year} {2013})\BibitemShut
  {NoStop}%
\bibitem [{\citenamefont {Battefeld}\ and\ \citenamefont
  {Peter}(2015)}]{Battefeld:2014uga}%
  \BibitemOpen
  \bibfield  {author} {\bibinfo {author} {\bibfnamefont {D.}~\bibnamefont
  {Battefeld}}\ and\ \bibinfo {author} {\bibfnamefont {P.}~\bibnamefont
  {Peter}},\ }\bibfield  {title} {\bibinfo {title} {{A Critical Review of
  Classical Bouncing Cosmologies}},\ }\href
  {https://doi.org/10.1016/j.physrep.2014.12.004} {\bibfield  {journal}
  {\bibinfo  {journal} {Phys. Rept.}\ }\textbf {\bibinfo {volume} {571}},\
  \bibinfo {pages} {1} (\bibinfo {year} {2015})},\ \Eprint
  {https://arxiv.org/abs/1406.2790} {arXiv:1406.2790 [astro-ph.CO]}
  \BibitemShut {NoStop}%
\bibitem [{\citenamefont {Cai}\ \emph {et~al.}(2012)\citenamefont {Cai},
  \citenamefont {Easson},\ and\ \citenamefont {Brandenberger}}]{Cai:2012va}%
  \BibitemOpen
  \bibfield  {author} {\bibinfo {author} {\bibfnamefont {Y.-F.}\ \bibnamefont
  {Cai}}, \bibinfo {author} {\bibfnamefont {D.~A.}\ \bibnamefont {Easson}},\
  and\ \bibinfo {author} {\bibfnamefont {R.}~\bibnamefont {Brandenberger}},\
  }\bibfield  {title} {\bibinfo {title} {{Towards a Nonsingular Bouncing
  Cosmology}},\ }\href {https://doi.org/10.1088/1475-7516/2012/08/020}
  {\bibfield  {journal} {\bibinfo  {journal} {JCAP}\ }\textbf {\bibinfo
  {volume} {08}},\ \bibinfo {pages} {020}},\ \Eprint
  {https://arxiv.org/abs/1206.2382} {arXiv:1206.2382 [hep-th]} \BibitemShut
  {NoStop}%
\bibitem [{\citenamefont {Cai}\ \emph {et~al.}(2013)\citenamefont {Cai},
  \citenamefont {McDonough}, \citenamefont {Duplessis},\ and\ \citenamefont
  {Brandenberger}}]{Cai:2013kja}%
  \BibitemOpen
  \bibfield  {author} {\bibinfo {author} {\bibfnamefont {Y.-F.}\ \bibnamefont
  {Cai}}, \bibinfo {author} {\bibfnamefont {E.}~\bibnamefont {McDonough}},
  \bibinfo {author} {\bibfnamefont {F.}~\bibnamefont {Duplessis}},\ and\
  \bibinfo {author} {\bibfnamefont {R.~H.}\ \bibnamefont {Brandenberger}},\
  }\bibfield  {title} {\bibinfo {title} {{Two Field Matter Bounce Cosmology}},\
  }\href {https://doi.org/10.1088/1475-7516/2013/10/024} {\bibfield  {journal}
  {\bibinfo  {journal} {JCAP}\ }\textbf {\bibinfo {volume} {10}},\ \bibinfo
  {pages} {024}},\ \Eprint {https://arxiv.org/abs/1305.5259} {arXiv:1305.5259
  [hep-th]} \BibitemShut {NoStop}%
\bibitem [{\citenamefont {Novello}\ and\ \citenamefont
  {Bergliaffa}(2008)}]{Novello:2008ra}%
  \BibitemOpen
  \bibfield  {author} {\bibinfo {author} {\bibfnamefont {M.}~\bibnamefont
  {Novello}}\ and\ \bibinfo {author} {\bibfnamefont {S.~E.~P.}\ \bibnamefont
  {Bergliaffa}},\ }\bibfield  {title} {\bibinfo {title} {{Bouncing
  Cosmologies}},\ }\href {https://doi.org/10.1016/j.physrep.2008.04.006}
  {\bibfield  {journal} {\bibinfo  {journal} {Phys. Rept.}\ }\textbf {\bibinfo
  {volume} {463}},\ \bibinfo {pages} {127} (\bibinfo {year} {2008})},\ \Eprint
  {https://arxiv.org/abs/0802.1634} {arXiv:0802.1634 [astro-ph]} \BibitemShut
  {NoStop}%
\bibitem [{\citenamefont {Carroll}\ \emph {et~al.}(2003)\citenamefont
  {Carroll}, \citenamefont {Hoffman},\ and\ \citenamefont
  {Trodden}}]{Carroll_2003}%
  \BibitemOpen
  \bibfield  {author} {\bibinfo {author} {\bibfnamefont {S.~M.}\ \bibnamefont
  {Carroll}}, \bibinfo {author} {\bibfnamefont {M.}~\bibnamefont {Hoffman}},\
  and\ \bibinfo {author} {\bibfnamefont {M.}~\bibnamefont {Trodden}},\
  }\bibfield  {title} {\bibinfo {title} {Can the dark energy equation-of-state
  parameter w be less than $\ensuremath{-}1?$},\ }\href
  {https://doi.org/10.1103/PhysRevD.68.023509} {\bibfield  {journal} {\bibinfo
  {journal} {Phys. Rev. D}\ }\textbf {\bibinfo {volume} {68}},\ \bibinfo
  {pages} {023509} (\bibinfo {year} {2003})},\ \Eprint
  {https://arxiv.org/abs/astro-ph/0301273} {arXiv:astro-ph/0301273}
  \BibitemShut {NoStop}%
\bibitem [{\citenamefont {Garriga}\ and\ \citenamefont
  {Vilenkin}(2013)}]{Garriga_2013}%
  \BibitemOpen
  \bibfield  {author} {\bibinfo {author} {\bibfnamefont {J.}~\bibnamefont
  {Garriga}}\ and\ \bibinfo {author} {\bibfnamefont {A.}~\bibnamefont
  {Vilenkin}},\ }\bibfield  {title} {\bibinfo {title} {Living with ghosts in
  lorentz invariant theories},\ }\href
  {https://doi.org/10.1088/1475-7516/2013/01/036} {\bibfield  {journal}
  {\bibinfo  {journal} {JCAP}\ }\textbf {\bibinfo {volume} {2013}}\bibfield
  {number} {\bibinfo  {number} { (01)},\ \bibinfo {pages} {036}},\ }\Eprint
  {https://arxiv.org/abs/1202.1239} {arXiv:1202.1239} \BibitemShut {NoStop}%
\bibitem [{\citenamefont {Sawicki}\ and\ \citenamefont
  {Vikman}(2013)}]{Sawicki_2013}%
  \BibitemOpen
  \bibfield  {author} {\bibinfo {author} {\bibfnamefont {I.}~\bibnamefont
  {Sawicki}}\ and\ \bibinfo {author} {\bibfnamefont {A.}~\bibnamefont
  {Vikman}},\ }\bibfield  {title} {\bibinfo {title} {Hidden negative energies
  in strongly accelerated universes},\ }\bibfield  {journal} {\bibinfo
  {journal} {Physical Review D}\ }\textbf {\bibinfo {volume} {87}},\ \href
  {https://doi.org/10.1103/physrevd.87.067301} {10.1103/physrevd.87.067301}
  (\bibinfo {year} {2013}),\ \Eprint {https://arxiv.org/abs/1209.2961}
  {arXiv:1209.2961} \BibitemShut {NoStop}%
\bibitem [{\citenamefont {Vikman}(2005)}]{Vikman_2005}%
  \BibitemOpen
  \bibfield  {author} {\bibinfo {author} {\bibfnamefont {A.}~\bibnamefont
  {Vikman}},\ }\bibfield  {title} {\bibinfo {title} {Can dark energy evolve to
  the phantom?},\ }\href {https://doi.org/10.1103/PhysRevD.71.023515}
  {\bibfield  {journal} {\bibinfo  {journal} {Phys. Rev. D}\ }\textbf {\bibinfo
  {volume} {71}},\ \bibinfo {pages} {023515} (\bibinfo {year} {2005})},\
  \Eprint {https://arxiv.org/abs/astro-ph/0407107} {arXiv:astro-ph/0407107}
  \BibitemShut {NoStop}%
\bibitem [{\citenamefont {Buchbinder}\ \emph {et~al.}(2007)\citenamefont
  {Buchbinder}, \citenamefont {Khoury},\ and\ \citenamefont
  {Ovrut}}]{Buchbinder_2007}%
  \BibitemOpen
  \bibfield  {author} {\bibinfo {author} {\bibfnamefont {E.~I.}\ \bibnamefont
  {Buchbinder}}, \bibinfo {author} {\bibfnamefont {J.}~\bibnamefont {Khoury}},\
  and\ \bibinfo {author} {\bibfnamefont {B.~A.}\ \bibnamefont {Ovrut}},\
  }\bibfield  {title} {\bibinfo {title} {New ekpyrotic cosmology},\ }\href
  {https://doi.org/10.1103/PhysRevD.76.123503} {\bibfield  {journal} {\bibinfo
  {journal} {Phys. Rev. D}\ }\textbf {\bibinfo {volume} {76}},\ \bibinfo
  {pages} {123503} (\bibinfo {year} {2007})},\ \Eprint
  {https://arxiv.org/abs/hep-th/0702154} {arXiv:hep-th/0702154} \BibitemShut
  {NoStop}%
\bibitem [{\citenamefont {Arkani-Hamed}\ \emph {et~al.}(2004)\citenamefont
  {Arkani-Hamed}, \citenamefont {Cheng}, \citenamefont {Luty},\ and\
  \citenamefont {Mukohyama}}]{Arkani-Hamed_2004}%
  \BibitemOpen
  \bibfield  {author} {\bibinfo {author} {\bibfnamefont {N.}~\bibnamefont
  {Arkani-Hamed}}, \bibinfo {author} {\bibfnamefont {H.-C.}\ \bibnamefont
  {Cheng}}, \bibinfo {author} {\bibfnamefont {M.~A.}\ \bibnamefont {Luty}},\
  and\ \bibinfo {author} {\bibfnamefont {S.}~\bibnamefont {Mukohyama}},\
  }\bibfield  {title} {\bibinfo {title} {Ghost condensation and a consistent
  infrared modification of gravity},\ }\href
  {https://doi.org/10.1088/1126-6708/2004/05/074} {\bibfield  {journal}
  {\bibinfo  {journal} {JHEP}\ }\textbf {\bibinfo {volume} {2004}}\bibfield
  {number} {\bibinfo  {number} { (05)},\ \bibinfo {pages} {074}},\ }\Eprint
  {https://arxiv.org/abs/hep-th/0312099} {arXiv:hep-th/0312099} \BibitemShut
  {NoStop}%
\bibitem [{\citenamefont {Creminelli}\ \emph {et~al.}(2010)\citenamefont
  {Creminelli}, \citenamefont {Nicolis},\ and\ \citenamefont
  {Trincherini}}]{Creminelli_2010_2010}%
  \BibitemOpen
  \bibfield  {author} {\bibinfo {author} {\bibfnamefont {P.}~\bibnamefont
  {Creminelli}}, \bibinfo {author} {\bibfnamefont {A.}~\bibnamefont
  {Nicolis}},\ and\ \bibinfo {author} {\bibfnamefont {E.}~\bibnamefont
  {Trincherini}},\ }\bibfield  {title} {\bibinfo {title} {Galilean genesis: an
  alternative to inflation},\ }\href
  {https://doi.org/10.1088/1475-7516/2010/11/021} {\bibfield  {journal}
  {\bibinfo  {journal} {JCAP}\ }\textbf {\bibinfo {volume} {2010}}\bibfield
  {number} {\bibinfo  {number} { (11)},\ \bibinfo {pages} {021}},\ }\Eprint
  {https://arxiv.org/abs/1007.0027} {arXiv:1007.0027 [hep-th]} \BibitemShut
  {NoStop}%
\bibitem [{\citenamefont {Easson}\ \emph {et~al.}(2011)\citenamefont {Easson},
  \citenamefont {Sawicki},\ and\ \citenamefont {Vikman}}]{Easson_2011}%
  \BibitemOpen
  \bibfield  {author} {\bibinfo {author} {\bibfnamefont {D.~A.}\ \bibnamefont
  {Easson}}, \bibinfo {author} {\bibfnamefont {I.}~\bibnamefont {Sawicki}},\
  and\ \bibinfo {author} {\bibfnamefont {A.}~\bibnamefont {Vikman}},\
  }\bibfield  {title} {\bibinfo {title} {G-bounce},\ }\href
  {https://doi.org/10.1088/1475-7516/2011/11/021} {\bibfield  {journal}
  {\bibinfo  {journal} {JCAP}\ }\textbf {\bibinfo {volume} {2011}}\bibfield
  {number} {\bibinfo  {number} { (11)},\ \bibinfo {pages} {021}},\ }\Eprint
  {https://arxiv.org/abs/1109.1047} {arXiv:1109.1047} \BibitemShut {NoStop}%
\bibitem [{\citenamefont {Qiu}\ \emph {et~al.}(2011)\citenamefont {Qiu},
  \citenamefont {Evslin}, \citenamefont {Cai}, \citenamefont {Li},\ and\
  \citenamefont {Zhang}}]{Qiu_2011_2011}%
  \BibitemOpen
  \bibfield  {author} {\bibinfo {author} {\bibfnamefont {T.}~\bibnamefont
  {Qiu}}, \bibinfo {author} {\bibfnamefont {J.}~\bibnamefont {Evslin}},
  \bibinfo {author} {\bibfnamefont {Y.-F.}\ \bibnamefont {Cai}}, \bibinfo
  {author} {\bibfnamefont {M.}~\bibnamefont {Li}},\ and\ \bibinfo {author}
  {\bibfnamefont {X.}~\bibnamefont {Zhang}},\ }\bibfield  {title} {\bibinfo
  {title} {Bouncing galileon cosmologies},\ }\href
  {https://doi.org/10.1088/1475-7516/2011/10/036} {\bibfield  {journal}
  {\bibinfo  {journal} {JCAP}\ }\textbf {\bibinfo {volume} {2011}}\bibfield
  {number} {\bibinfo  {number} { (10)},\ \bibinfo {pages} {036}},\ }\Eprint
  {https://arxiv.org/abs/1108.0593} {arXiv:1108.0593 [hep-th]} \BibitemShut
  {NoStop}%
\bibitem [{\citenamefont {Fang}\ \emph {et~al.}(2007)\citenamefont {Fang},
  \citenamefont {Lu},\ and\ \citenamefont {Huang}}]{Fang_2007}%
  \BibitemOpen
  \bibfield  {author} {\bibinfo {author} {\bibfnamefont {W.}~\bibnamefont
  {Fang}}, \bibinfo {author} {\bibfnamefont {H.~Q.}\ \bibnamefont {Lu}},\ and\
  \bibinfo {author} {\bibfnamefont {Z.~G.}\ \bibnamefont {Huang}},\ }\bibfield
  {title} {\bibinfo {title} {{Cosmologies with a general non-canonical scalar
  field}},\ }\href {https://doi.org/10.1088/0264-9381/24/15/002} {\bibfield
  {journal} {\bibinfo  {journal} {CQG}\ }\textbf {\bibinfo {volume} {24}},\
  \bibinfo {pages} {3799} (\bibinfo {year} {2007})},\ \Eprint
  {https://arxiv.org/abs/0610188} {arXiv:0610188 [hep-th]} \BibitemShut
  {NoStop}%
\bibitem [{\citenamefont {Unnikrishnan}\ \emph {et~al.}(2012)\citenamefont
  {Unnikrishnan}, \citenamefont {Sahni},\ and\ \citenamefont
  {Toporensky}}]{Unnikrishnan_2012}%
  \BibitemOpen
  \bibfield  {author} {\bibinfo {author} {\bibfnamefont {S.}~\bibnamefont
  {Unnikrishnan}}, \bibinfo {author} {\bibfnamefont {V.}~\bibnamefont
  {Sahni}},\ and\ \bibinfo {author} {\bibfnamefont {A.}~\bibnamefont
  {Toporensky}},\ }\bibfield  {title} {\bibinfo {title} {{Refining inflation
  using non-canonical scalars}},\ }\href
  {https://doi.org/10.1088/1475-7516/2012/08/018} {\bibfield  {journal}
  {\bibinfo  {journal} {JCAP}\ }\textbf {\bibinfo {volume} {2012}},\ \bibinfo
  {pages} {018}},\ \Eprint {https://arxiv.org/abs/1205.0786} {arXiv:1205.0786
  [astro-ph.CO]} \BibitemShut {NoStop}%
\bibitem [{\citenamefont {Wainwright}\ and\ \citenamefont
  {Ellis}(1997)}]{Ellis}%
  \BibitemOpen
  \bibfield  {author} {\bibinfo {author} {\bibfnamefont {J.}~\bibnamefont
  {Wainwright}}\ and\ \bibinfo {author} {\bibfnamefont {G.~F.~R.}\ \bibnamefont
  {Ellis}},\ }\href {https://doi.org/10.1017/CBO9780511524660} {\emph {\bibinfo
  {title} {{Dynamical Systems in Cosmology}}}}\ (\bibinfo  {publisher}
  {Cambridge University Press},\ \bibinfo {year} {1997})\BibitemShut {NoStop}%
\bibitem [{\citenamefont {Coley}(2003)}]{Coley:2003mj}%
  \BibitemOpen
  \bibfield  {author} {\bibinfo {author} {\bibfnamefont {A.~A.}\ \bibnamefont
  {Coley}},\ }\href {https://doi.org/10.1007/978-94-017-0327-7} {\emph
  {\bibinfo {title} {{Dynamical systems and cosmology}}}}\ (\bibinfo
  {publisher} {Kluwer},\ \bibinfo {address} {Dordrecht, Netherlands},\ \bibinfo
  {year} {2003})\BibitemShut {NoStop}%
\bibitem [{\citenamefont {Bahamonde}\ \emph {et~al.}(2018)\citenamefont
  {Bahamonde}, \citenamefont {B\"ohmer}, \citenamefont {Carloni}, \citenamefont
  {Copeland}, \citenamefont {Fang},\ and\ \citenamefont
  {Tamanini}}]{Bahamonde:2017ize}%
  \BibitemOpen
  \bibfield  {author} {\bibinfo {author} {\bibfnamefont {S.}~\bibnamefont
  {Bahamonde}}, \bibinfo {author} {\bibfnamefont {C.~G.}\ \bibnamefont
  {B\"ohmer}}, \bibinfo {author} {\bibfnamefont {S.}~\bibnamefont {Carloni}},
  \bibinfo {author} {\bibfnamefont {E.~J.}\ \bibnamefont {Copeland}}, \bibinfo
  {author} {\bibfnamefont {W.}~\bibnamefont {Fang}},\ and\ \bibinfo {author}
  {\bibfnamefont {N.}~\bibnamefont {Tamanini}},\ }\bibfield  {title} {\bibinfo
  {title} {{Dynamical systems applied to cosmology: dark energy and modified
  gravity}},\ }\href {https://doi.org/10.1016/j.physrep.2018.09.001} {\bibfield
   {journal} {\bibinfo  {journal} {Phys. Rept.}\ }\textbf {\bibinfo {volume}
  {775-777}},\ \bibinfo {pages} {1} (\bibinfo {year} {2018})},\ \Eprint
  {https://arxiv.org/abs/1712.03107} {arXiv:1712.03107 [gr-qc]} \BibitemShut
  {NoStop}%
\bibitem [{\citenamefont {De-Santiago}\ \emph {et~al.}(2013)\citenamefont
  {De-Santiago}, \citenamefont {Cervantes-Cota},\ and\ \citenamefont
  {Wands}}]{De-Santiago:2012ibi}%
  \BibitemOpen
  \bibfield  {author} {\bibinfo {author} {\bibfnamefont {J.}~\bibnamefont
  {De-Santiago}}, \bibinfo {author} {\bibfnamefont {J.~L.}\ \bibnamefont
  {Cervantes-Cota}},\ and\ \bibinfo {author} {\bibfnamefont {D.}~\bibnamefont
  {Wands}},\ }\bibfield  {title} {\bibinfo {title} {{Cosmological phase space
  analysis of the F(X) - V($\phi$) scalar field and bouncing solutions}},\
  }\href {https://doi.org/10.1103/PhysRevD.87.023502} {\bibfield  {journal}
  {\bibinfo  {journal} {Phys. Rev. D}\ }\textbf {\bibinfo {volume} {87}},\
  \bibinfo {pages} {023502} (\bibinfo {year} {2013})},\ \Eprint
  {https://arxiv.org/abs/1204.3631} {arXiv:1204.3631 [gr-qc]} \BibitemShut
  {NoStop}%
\bibitem [{\citenamefont {Panda}\ and\ \citenamefont
  {Sharma}(2016)}]{Panda:2015wya}%
  \BibitemOpen
  \bibfield  {author} {\bibinfo {author} {\bibfnamefont {S.}~\bibnamefont
  {Panda}}\ and\ \bibinfo {author} {\bibfnamefont {M.}~\bibnamefont {Sharma}},\
  }\bibfield  {title} {\bibinfo {title} {{Anisotropic Bouncing Scenario in
  $F(X)-V(\phi)$ model}},\ }\href {https://doi.org/10.1007/s10509-015-2594-y}
  {\bibfield  {journal} {\bibinfo  {journal} {Astrophys. Space Sci.}\ }\textbf
  {\bibinfo {volume} {361}},\ \bibinfo {pages} {87} (\bibinfo {year} {2016})},\
  \Eprint {https://arxiv.org/abs/1501.01765} {arXiv:1501.01765 [gr-qc]}
  \BibitemShut {NoStop}%
\bibitem [{\citenamefont {Bouhmadi-L\'opez}\ \emph {et~al.}(2017)\citenamefont
  {Bouhmadi-L\'opez}, \citenamefont {Marto}, \citenamefont {Morais},\ and\
  \citenamefont {Silva}}]{Bouhmadi-Lopez:2016dzw}%
  \BibitemOpen
  \bibfield  {author} {\bibinfo {author} {\bibfnamefont {M.}~\bibnamefont
  {Bouhmadi-L\'opez}}, \bibinfo {author} {\bibfnamefont {J.}~\bibnamefont
  {Marto}}, \bibinfo {author} {\bibfnamefont {J.}~\bibnamefont {Morais}},\ and\
  \bibinfo {author} {\bibfnamefont {C.~M.}\ \bibnamefont {Silva}},\ }\bibfield
  {title} {\bibinfo {title} {{Cosmic infinity: A dynamical system approach}},\
  }\href {https://doi.org/10.1088/1475-7516/2017/03/042} {\bibfield  {journal}
  {\bibinfo  {journal} {JCAP}\ }\textbf {\bibinfo {volume} {03}},\ \bibinfo
  {pages} {042}},\ \Eprint {https://arxiv.org/abs/1611.03100} {arXiv:1611.03100
  [gr-qc]} \BibitemShut {NoStop}%
\bibitem [{\citenamefont {Cai}\ and\ \citenamefont {Piao}(2017)}]{Cai:2017dyi}%
  \BibitemOpen
  \bibfield  {author} {\bibinfo {author} {\bibfnamefont {Y.}~\bibnamefont
  {Cai}}\ and\ \bibinfo {author} {\bibfnamefont {Y.-S.}\ \bibnamefont {Piao}},\
  }\bibfield  {title} {\bibinfo {title} {{A covariant Lagrangian for stable
  nonsingular bounce}},\ }\href {https://doi.org/10.1007/JHEP09(2017)027}
  {\bibfield  {journal} {\bibinfo  {journal} {JHEP}\ }\textbf {\bibinfo
  {volume} {09}},\ \bibinfo {pages} {027}},\ \Eprint
  {https://arxiv.org/abs/1705.03401} {arXiv:1705.03401 [gr-qc]} \BibitemShut
  {NoStop}%
\bibitem [{\citenamefont {Ilyas}\ \emph {et~al.}(2020)\citenamefont {Ilyas},
  \citenamefont {Zhu}, \citenamefont {Zheng}, \citenamefont {Cai},\ and\
  \citenamefont {Saridakis}}]{Ilyas:2020qja}%
  \BibitemOpen
  \bibfield  {author} {\bibinfo {author} {\bibfnamefont {A.}~\bibnamefont
  {Ilyas}}, \bibinfo {author} {\bibfnamefont {M.}~\bibnamefont {Zhu}}, \bibinfo
  {author} {\bibfnamefont {Y.}~\bibnamefont {Zheng}}, \bibinfo {author}
  {\bibfnamefont {Y.-F.}\ \bibnamefont {Cai}},\ and\ \bibinfo {author}
  {\bibfnamefont {E.~N.}\ \bibnamefont {Saridakis}},\ }\bibfield  {title}
  {\bibinfo {title} {{DHOST Bounce}},\ }\href
  {https://doi.org/10.1088/1475-7516/2020/09/002} {\bibfield  {journal}
  {\bibinfo  {journal} {JCAP}\ }\textbf {\bibinfo {volume} {09}},\ \bibinfo
  {pages} {002}},\ \Eprint {https://arxiv.org/abs/2002.08269} {arXiv:2002.08269
  [gr-qc]} \BibitemShut {NoStop}%
\bibitem [{\citenamefont {Perko}(2001)}]{Perko}%
  \BibitemOpen
  \bibfield  {author} {\bibinfo {author} {\bibfnamefont {L.}~\bibnamefont
  {Perko}},\ }\href {https://doi.org/10.1007/978-1-4613-0003-8} {\emph
  {\bibinfo {title} {{Differential Equations and Dynamical Systems}}}}\
  (\bibinfo  {publisher} {Springer New York, NY},\ \bibinfo {year}
  {2001})\BibitemShut {NoStop}%
\end{thebibliography}%

\end{document}